\documentclass[traditabstract]{aa} 
\usepackage{graphicx}
\usepackage{siunitx}
\usepackage{txfonts}
\usepackage{graphicx}
\usepackage{caption}
\usepackage{natbib}
\usepackage{subfigure} 
\usepackage{hyperref}
\usepackage[utf8]{inputenc}
\usepackage{placeins}
\usepackage{scrextend}
\usepackage{rotating} 
\usepackage{multirow} 
\usepackage{booktabs} 

\newcommand{\bd}{\begin{displaymath}}
\newcommand{\ed}{\end{displaymath}}
\newcommand{\be}{\begin{equation}}
\newcommand{\ee}{\end{equation}}
\newcommand{\beaa}{\begin{eqnarray*}}
\newcommand{\eeaa}{\end{eqnarray*}}
\newcommand{\bea}{\begin{eqnarray}}
\newcommand{\eea}{\end{eqnarray}}

\newcommand{\erg}{\mathrm{erg}}

\DeclareSIUnit\parsec{pc}
\DeclareSIUnit\lightyear{ly}
\DeclareSIUnit\year{yr}
\DeclareSIUnit\erg{erg}
\DeclareSIUnit\ster{ster}
\DeclareSIUnit\arcsec{arcsec}
\DeclareSIUnit\rad{rad}
\DeclareSIUnit\mag{mag}

\usepackage{color} 
\definecolor{gruen}{cmyk}{0.35,0.01,0.80,0.1}

\definecolor{blue}{rgb}{0.0,0.0,1.0}
\definecolor{magenta}{rgb}{1.0,0.0,1.0}
\definecolor{orange}{rgb}{1.0,0.5,0.0}

\def\kmsMpc {$\mathrm{km\,s^{-1}\,Mpc^{-1}}$}

\begin{document}

   \title{HOLISMOKES - V. Microlensing of type II supernovae and time-delay inference through spectroscopic phase retrieval}

  \titlerunning{HOLISMOKES - V. Microlensing of SNe II and spectroscopic time-delay inference}

   \author{J. Bayer\inst{1,2}
                 \and 
                 S. Huber \inst{1,2}  
                  \and 
             C. Vogl\inst{1,2,3}
             \and    
          S. H. Suyu\inst{1,2,4}
                    \and
             S. Taubenberger\inst{1}
             \and 
             D. Sluse\inst{5}
             \and 
             J. H. H. Chan\inst{6}
             \and
             W.~E.~Kerzendorf\inst{7,8}
          }

   \institute{Max-Planck-Institut f\"ur Astrophysik, Karl-Schwarzschild Str. 1, 85748 Garching, Germany\\
              \email{jana@MPA-Garching.MPG.DE}
         \and 
           Physik-Department, Technische Universit\"at M\"unchen, James-Franck-Stra\ss{}e~1, 85748 Garching, Germany
                        \and 
           Exzellenzcluster ORIGINS, Boltzmannstr. 2, 85748 Garching, Germany
           \and
           Institute of Astronomy and Astrophysics, Academia Sinica, 11F of ASMAB, No.1, Section 4, Roosevelt Road, Taipei 10617, Taiwan
\and
           STAR Institute, Quartier Agora – Allée du six Août, 19c, 4000 Liège, Belgium
\and
Institute of Physics, Laboratory of Astrophysics, Ecole Polytechnique Fédérale de Lausanne(EPFL), Observatoire de Sauverny, 1290 Versoix, Switzerland
\and
Department of Physics and Astronomy, Michigan State University, East Lansing, MI 48824, USA
\and
Department of Computational Mathematics, Science, and Engineering, Michigan State University, East Lansing, MI 48824, USA
   }

 
  \abstract
{We investigate strongly gravitationally lensed type II supernovae (LSNe
II) for time-delay cosmography, incorporating microlensing effects; this expands on previous microlensing studies of type Ia supernovae
(SNe Ia). We use the radiative-transfer code \textsc{tardis} to recreate five spectra of the prototypical SN 1999em at different times within the plateau phase of the light curve. The microlensing-induced
deformations of the spectra and light curves are calculated by placing
the SN into magnification maps generated with the code
\textsc{gerlumph}. We study the impact of microlensing on the color curves and find that there is no strong influence on them
during the investigated time interval of the plateau phase. 
The color curves are only weakly affected by microlensing
due to the almost achromatic behavior of the intensity profiles.  However,
the lack of nonlinear structure in the color curves during the plateau phase of type II-plateau supernovae makes time-delay measurements more challenging compared to SN Ia color curves, given the
 possible presence of differential dust
extinction.
Therefore, we further investigate SN phase
inference through spectral absorption lines
under the influence of microlensing and Gaussian noise.
As the spectral features shift to longer wavelengths
with progressing time after explosion, the measured wavelength of a
specific absorption line provides information on the epoch of the
SN. The comparison between retrieved epochs of two observed lensing
images then gives the time delay of the images.
We find that the phase retrieval method that uses spectral features yields accurate delays with uncertainties of $\lesssim$2\,days, making it a promising approach.}


   \maketitle


\section{Introduction}

Over the last few years, several new and independent techniques for measuring
the Hubble constant, $H_{0}$, have been presented to address the problem
of the $>$4$\sigma$ discrepancy \citep{Verde2019} between the
value measured from observations of the cosmic microwave background
\citep{Planck2020} and distance ladder measurements from
the supernova $H_{0}$ for the equa-
tion of state (SH0ES) program (\citealt{Riess2016, Riess2018, Riess2019}; see also
\citealt{ Freedman2019, Freedman2020}).
One of these techniques is to conduct time-delay cosmography on
strongly lensed variable objects \citep{Refsdal1964} to infer $H_{0}$
through time-delay measurements, lens mass modeling, and
reconstructing the line-of-sight mass distributions. The calculated
time-delay distance is inversely
proportional to $H_{0}$.

The method can be applied to different variable sources. One
possibility is to use quasars, which has already been done in the
H0LiCOW program \citep{Suyu2017, Birrer2018, Sluse2019, Chen2019,
  Wong2019, Rusu2019} in cooperation with the COSmological MOnitoring of GRAvItational Lenses \citep[COSMOGRAIL;][]{Eigenbrod2005, Courbin2017, Bonvin2018}, the Strong lensing at High Angular Resolution Program \citep[SHARP;][]{Chen2019}, and the STRong-lensing Insights into the Dark Energy Survey (STRIDES) collaboration \citep{Shajib2020}. 
The newly established Time-Delay COSMOgraphy (TDCOSMO) organization is further testing systematic effects and
expanding the lensed quasar sample for future studies
\citep{Millon2020a, Millon2020, Gilman2020, Birrer2020}.
A second way to conduct time-delay measurements is through strongly
lensed supernovae (LSNe), which was first envisioned by
\cite{Refsdal1964} when he introduced the concept of time-delay
cosmography.

Until today, only two LSNe
with multiple resolved images have been observed. The first, called
``SN Refsdal,'' is a type II supernova (SN II), discovered by \cite{Kelly2016a,
  Kelly2016b}, which was lensed by the galaxy cluster MACS
J1149.5+222.3. 
 The second LSN, iPTF16geu, was discovered by
\cite{Goobar2017} and is a type Ia supernova (SN Ia) at redshift 0.409 lensed by a galaxy
at a redshift of 0.216.
Several lensing studies on SN Refsdal have been performed \citep[e.g.,][]{Rodney2016, Kelly2016b, Treu2016, PierelRodney19, Grillo2016, Grillo2020, Baklanov2020};
iPTF16geu has been studied extensively \citep{Goobar2017, More2017,
  Yahalomi2017, Dhawan2018, Johansson2020} and shows strong signs of
microlensing. Further studies on SNe Ia and their usability to measure time delays have been conducted by \cite{Goldstein2018}, \cite{Foxley2018}, \cite{Huber2019}, and  \cite{Huber2020}. 
The main focus of those papers has been the characterization of the perturbations introduced by microlensing.

Microlensing of SNe
results from the additional lensing of strongly lensed SNe induced by the compact masses
that the lens galaxy is composed of (e.g.,~stars). The influence of
microlensing is independent for each observed image of the SN. This
causes the spectrum to be magnified to an unknown amount and makes
time-delay measurements difficult. \cite{Goldstein2018} and
\cite{Huber2020} show that microlensing of LSNe Ia is mostly achromatic.  In three
of five color curves, \cite{Huber2020} find characteristic minima or maxima within
the achromatic phase of SNe Ia, making time-delay measurements with
LSNe Ia feasible.

With this work, which is part of Highly Optimised Lensing Investigations of Supernovae, Microlensing Objects, and Kinematics of Ellipticals and Spirals \citep[HOLISMOKES;][]{Suyu2020}, we take a closer look at SNe II
for time-delay measurements that incorporate
microlensing effects. Characteristic features in the light curves and spectra of SNe II could be
used to conduct time-delay measurements. With the upcoming Rubin Observatory Legacy Survey
of Space and Time \citep[LSST;][]{LSSTScienceCollaboration2009} we
expect to detect over 1000 LSNe II \citep{Wojtak2019, Goldstein2018, Goldstein2016}. This
number is larger than the number of LSNe Ia expected to be observed, which is estimated to be around 500 to 900 \citep{Oguri2010, Wojtak2019}. Therefore, it would be a missed opportunity to not include SNe II for time-delay cosmography.

We use radiative-transfer models to investigate the potential of LSNe II for
cosmographic measurements. In recent years computational
capabilities have grown, allowing us to conduct more and more complex calculations
for modeling SNe of different types. The study of LSNe Ia by
\cite{Huber2019} used the Applied Radiative Transfer In Supernovae (\textsc{artis}) code, which was developed by \cite{Kromer2009}.

For this study of LSNe II, we use the Monte Carlo radiative
transfer code \textsc{tardis} \citep{Kerzendorf2014}, as modified by \cite{Vogl2019}. The spatial specific-intensity profiles for each wavelength are
microlensed by magnification maps generated with the
inverse ray-tracing code \textsc{gerlumph} \citep{Vernardos2015}, and deformed spectra are constructed. We
then evaluate the microlensed spectra, light curves, and color curves
on their usability for time-delay measurements. In doing so, we
investigate a new approach to inferring phases on the basis of the spectral evolution of the SN. 
Throughout the paper, we assume a radiation-free flat Lambda cold dark matter ($\mathrm{\Lambda}$CDM) cosmology with $H_{0} = 72$ \kmsMpc \citep{Bonvin2017}, $\Omega_{\mathrm{m}}$ = 0.26, and $\Omega_{\mathrm{\Lambda}}$ = 0.74, as assumed by \cite{Oguri2010}.

The structure of the paper is as follows. First, we discuss the \textsc{tardis} simulations and present the synthetic SN II spectra in Sect. \ref{sec: Type II
  Supernova models}. In Sect. \ref{sec: Microlensing on Type II
  Supernovae} we explain our application of microlensing to those spectra and show the resulting light curves and
color curves. Further, we show the impact of microlensing on the
absorption features. In Sect. \ref{sec: SN phase inference from
  spectra} we investigate the phase inference of SNe from the
absorption wavelengths. Finally, we discuss our results in
Sect. \ref{sec: Discussion} and conclude in Sect. \ref{sec:
  Conclusion}.
  \begin{figure}[hbt!]
\centering
{\includegraphics[width=0.489\textwidth]{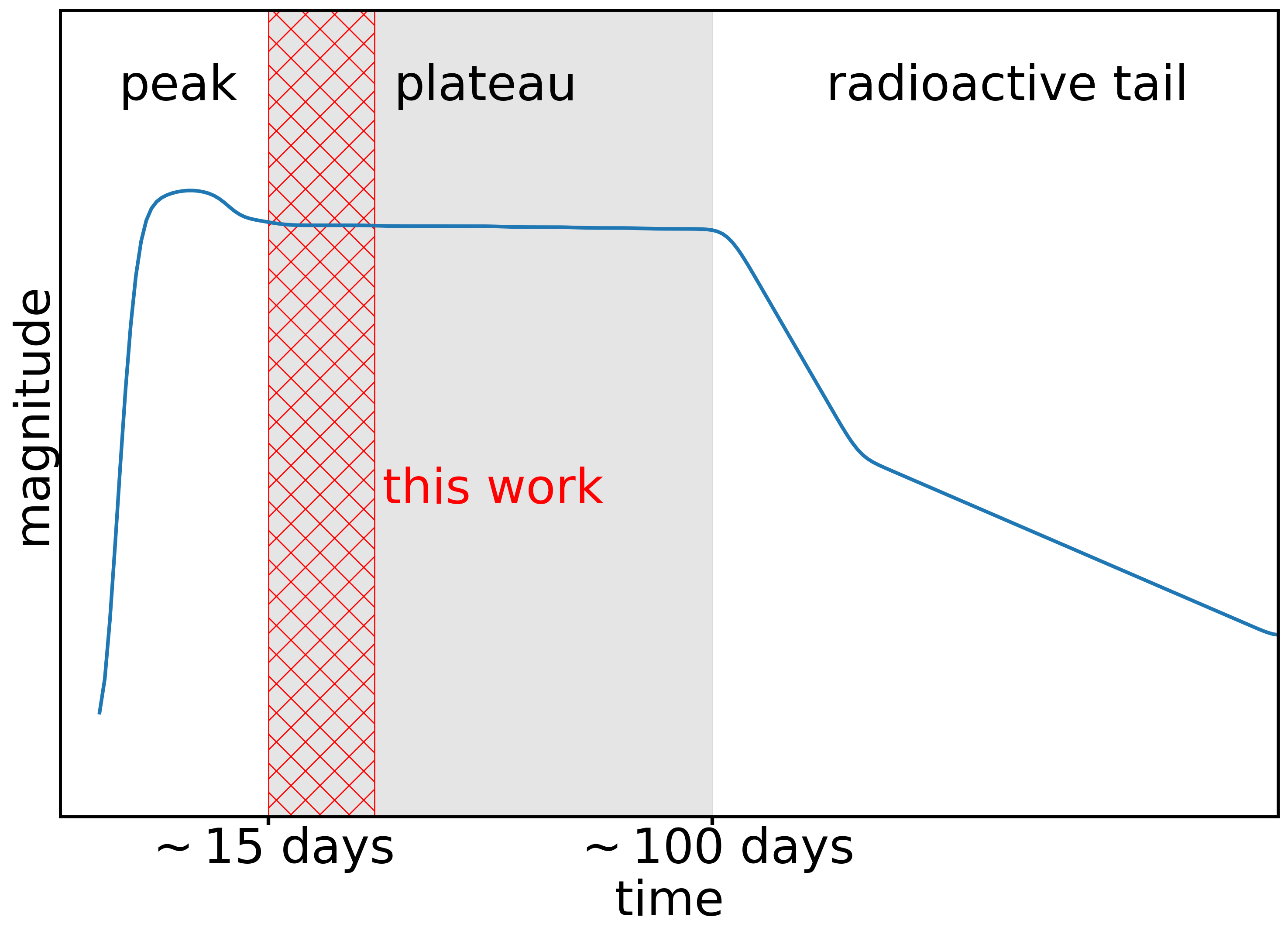}}\
\caption{\label{light_curve} Schematic light curve of a typical SN
  II-P. The red hatched area is the region of interest for the
  models in this paper and is part of the plateau phase marked by the
  gray area.}
\end{figure}

\section{Type II supernova models}
\label{sec: Type II Supernova models}

In this section we describe the creation of the SN II models we use
for the microlensing investigation in Sect. \ref{sec: Microlensing
  on Type II Supernovae}. We performed the calculations with the
extended version of the radiative-transfer code \textsc{tardis} from
\cite{Vogl2019} and applied it to the prototypical type II-plateau supernova (SN II-P) 1999em. The
reconstructed spectra are part of the plateau phase. The
characteristic temporal evolution of the light curve of a SN II-P is
shown in Fig. \ref{light_curve} with the region of interest marked
with a red crisscross pattern \citep{Barbon1979, Anderson2014, Valenti2015, Gall2015, Utrobin2017, Hicken2017}.

\subsection{\textsc{tardis} simulations}
\label{sec: TARDIS simulations}

\textsc{tardis} is a Monte Carlo based radiative transfer code based on the indivisible energy packet scheme of \cite{Lucy1999b,
Lucy2002, Lucy2003}, originally written for spectral modeling of SNe Ia
\citep{Kerzendorf2014}. \cite{Vogl2019} modified the code to 
perform spectral synthesis for SNe II and specifically applied it to
spectra of SN 1999em \citep{Vogl2020}.

The original \textsc{tardis} code considers Thomson scattering and
bound-bound line interactions as an approximation for SNe Ia
\citep{Barna2017, Boyle2017, Magee2016, Magee2017}. To account for the
hydrogen-rich composition of SNe II, the modified code additionally accounts for bound-free, free-free, and
collisional processes \citep{Lucy2002, Lucy2003}. The modified
\textsc{tardis} code accounts for nonlocal thermodynamic equilibrium
(NLTE) effects in the ionization and excitation balance of hydrogen
and, in selected cases, helium. Additionally, the thermal structure is
calculated from the equilibrium of the heating and cooling of the shells.

To synthesize the spectra, the properties of the propagating
energy packets are tracked and finally binned. The resulting
spectra are affected by Monte Carlo noise. To improve the
signal-to-noise ratio (S/N), multiple virtual Monte Carlo packets
(v-packets) are emitted for each emitted or interacting packet. These
additional packets propagate along randomly selected rays and get
attenuated by the integrated optical depth. When the v-packet exits
the computational region through the outer boundary, the emission
radius $r$, the propagation direction cos($\theta$), the frequency
$\nu$, and the energy $\epsilon$ are saved. These are used to calculate
the specific intensity and the impact parameter $p$ to construct the
specific intensity profiles in Sect. \ref{sec: Specific intensity
  profiles}.

We assumed power-law density profiles and a homogeneous composition to
calculate the \textsc{tardis} models for this study. The abundances of
H, He, C, N, and O were set to their CNO cycle equilibrium values
(taken from \citealt{Prantzos1986}). All other elements follow the solar
chemical composition \citep{Asplund2009}. Due to computational
expense, the code produces only parameters at a specific epoch and no
temporal development. In order to provide a time-dependent evolution
of spectra for the analysis of microlensing effects, we computed
several spectra at different epochs and used these snapshots as the
temporal evolution.

\subsection{Observed and model spectra of SN 1999em}
\label{sec: Observed and modeled spectra of SN 1999em}

For the analysis of the effect of microlensing, we used the
\textsc{tardis} models for SN 1999em. We selected SN 1999em as it is a
prototypical SN II-P \citep[e.g.,][]{Dessart2006}. Additionally, it is one of
the most well-observed and well-studied SNe II. Its distance has been
determined several times, and it has been observed over a long period
of time, resulting in many spectra \citep{Leonard2002, Baron2004,
  Hamuy2001}.  Given the availability of a huge amount of data, it is a
good candidate for SN II modeling \citep[e.g.,][]{Utrobin2007}.

The observed spectra of SN 1999em \citep{Leonard2002, Hamuy2001} were used as a reference for the
simulated spectra as \textsc{tardis} only creates snapshots at
specific epochs and does not perform fully time-dependent radiative
transfer computations. To achieve a sensible temporal evolution of the
snapshots, they were fit to the observed spectra of a specific SN.

\begin{figure*}[hbt!]
\centering
\subfigure[\hbox{Spectrum at day 11 after the explosion}]{\label{spectra_0}\includegraphics[width=0.48\textwidth]{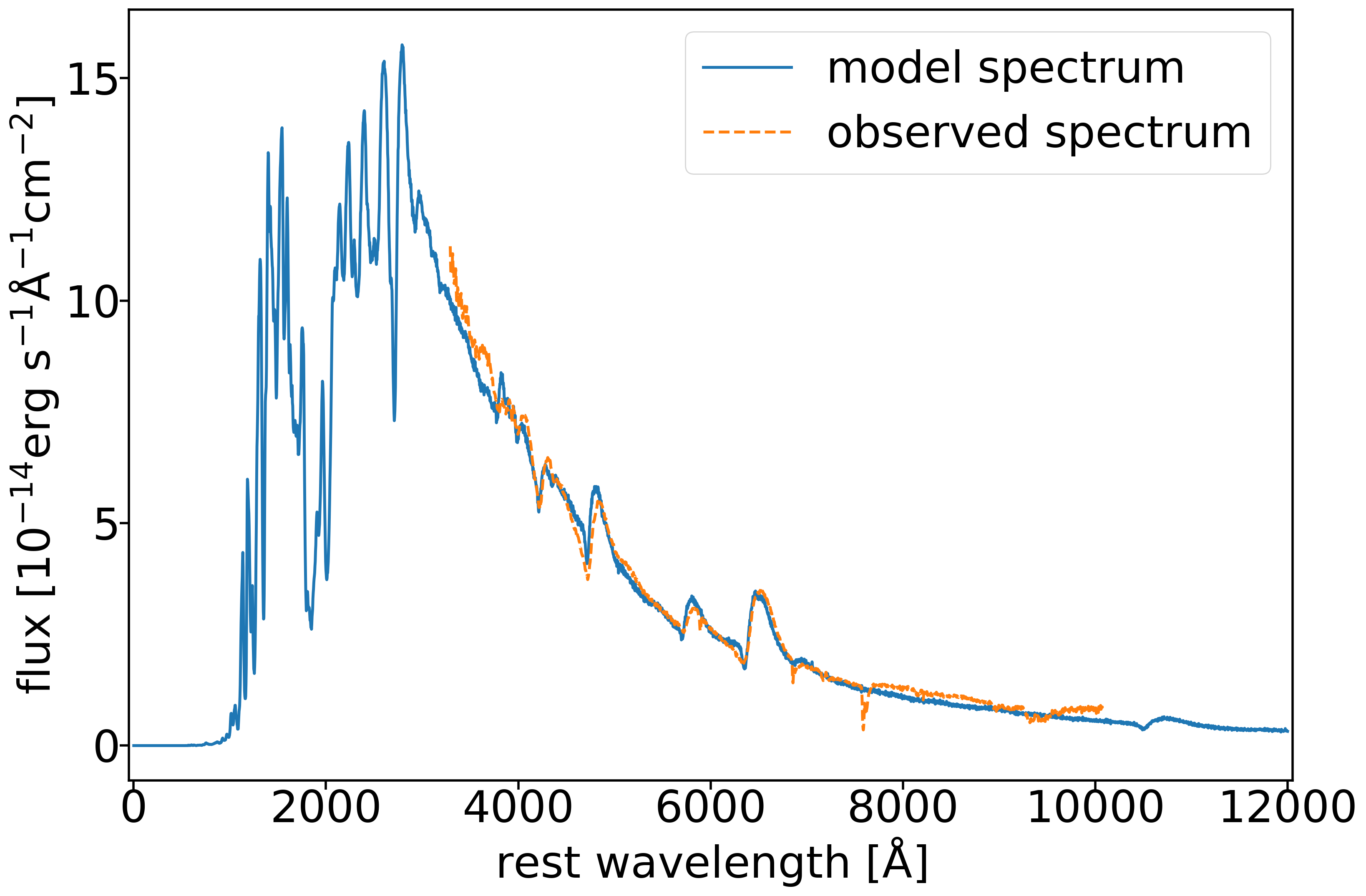}}
\subfigure[\hbox{Spectrum at day 13 after the explosion}]{\label{spectra_1}\includegraphics[width=0.48\textwidth]{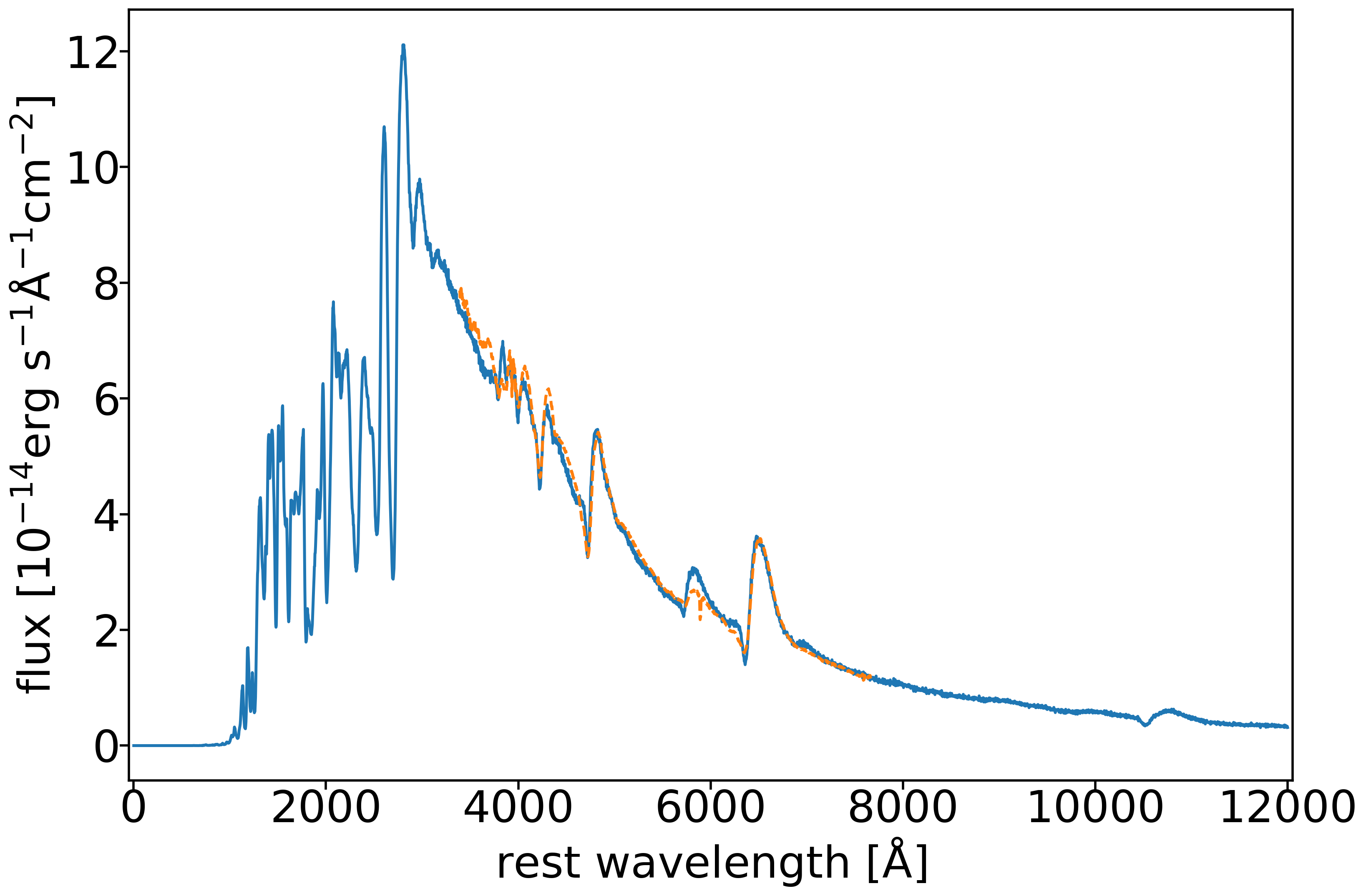}}\vskip5mm
\subfigure[\hbox{Spectrum at day 17 after the explosion}]{\label{spectra_2}\includegraphics[width=0.48\textwidth]{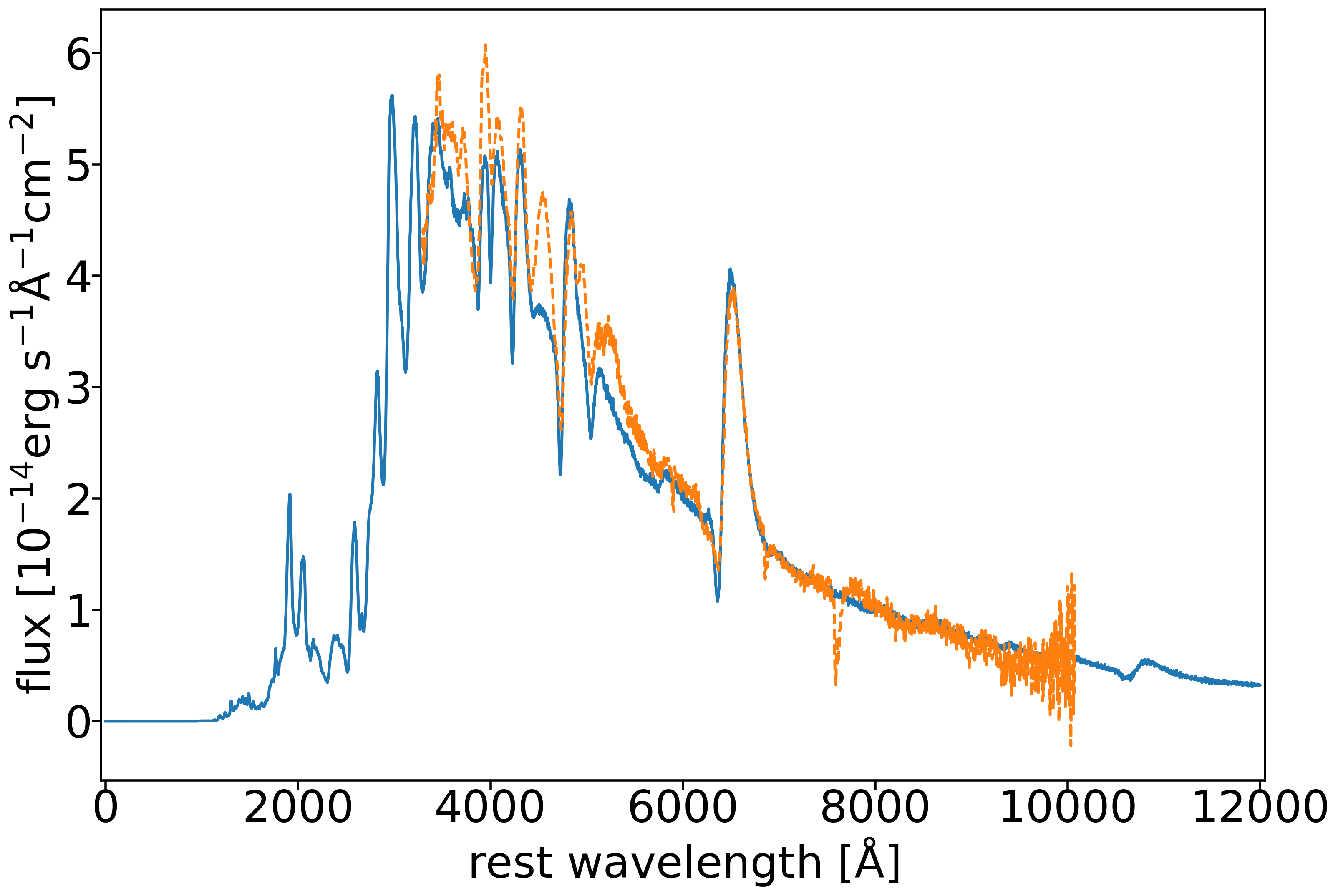}}
\subfigure[\hbox{Spectrum at day 22 after the explosion}]{\label{spectra_3}\includegraphics[width=0.48\textwidth]{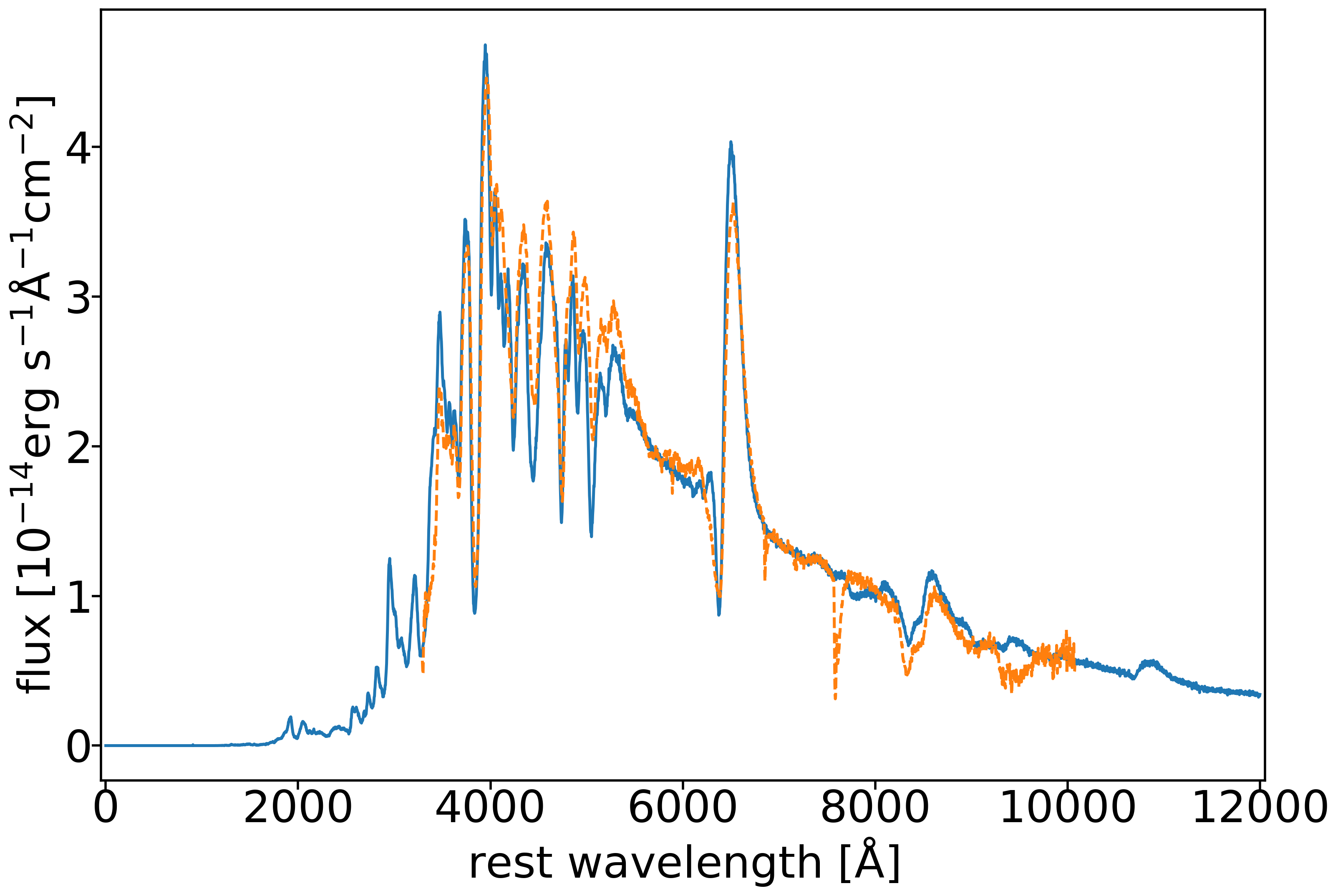}}\vskip5mm
\subfigure[\hbox{Spectrum at day 27 after the explosion}]{\label{spectra_4}\includegraphics[width=0.48\textwidth]{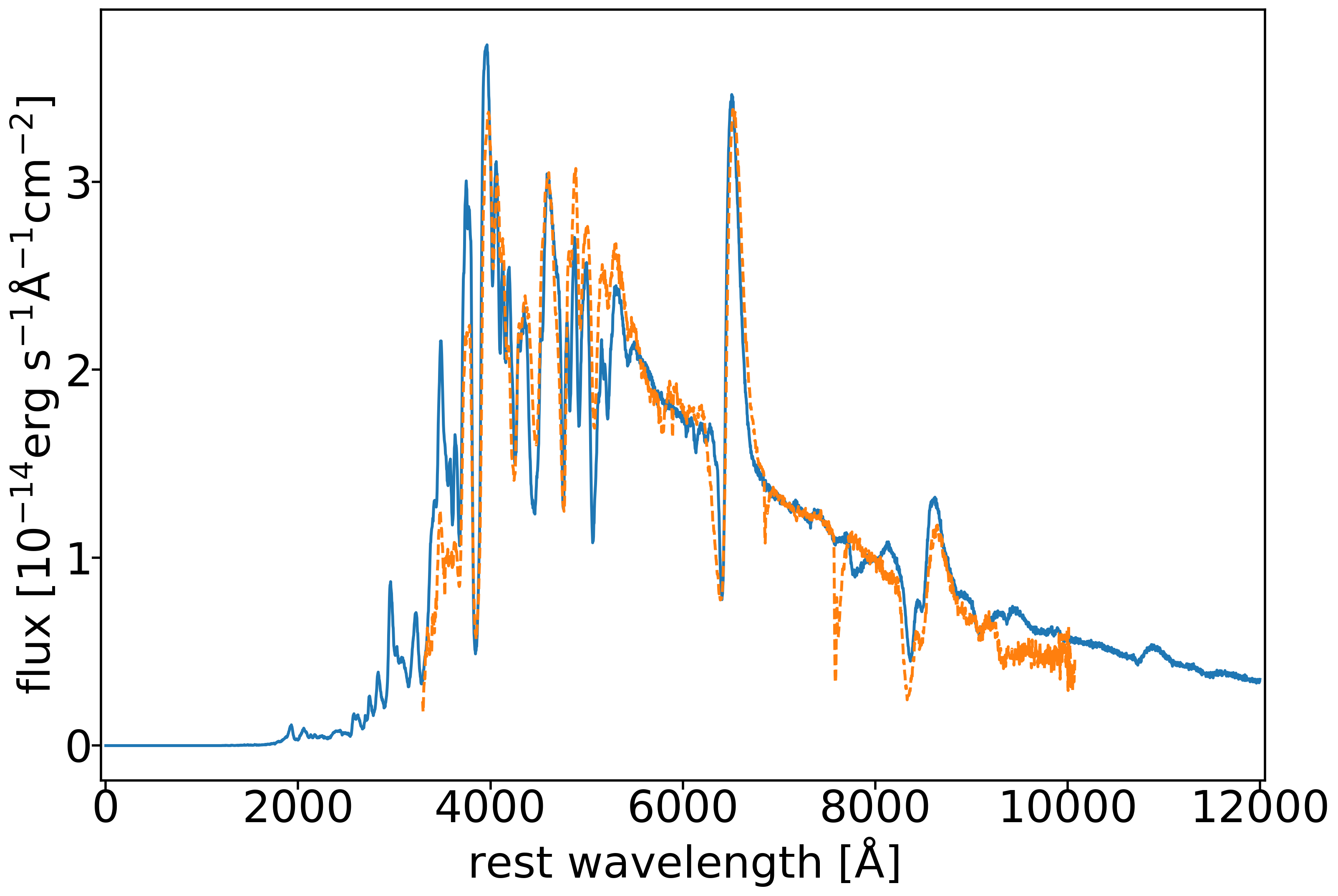}}\
\caption{\label{spectra} Model (blue) and observed (orange) spectra of SN 1999em at various epochs.  The model spectra are obtained from \textsc{tardis}.}
\end{figure*}

The observed spectra are shown in Fig. \ref{spectra} in comparison
to the model spectra selected from extensive model grids, including the
one from \cite{Vogl2020}. The observed spectra have been de-reddened
under the assumption of the \cite{Cardelli1989} extinction law. The color
excess is set to $E(\mathrm{B} - \mathrm{V}) = 0.08$ and the total-to-selective
extinction ratio to $R_{\mathrm{V}} = 3.1,$ as in \cite{Vogl2019}. The flux
of the model spectra was calculated for the SN at a distance of 10 pc
and shifted by eye to match the observed spectra. The wavelength resolution of the spectra was chosen to be about 3 $\text{\AA}$ per bin. 

At early times, around day 11 after the explosion, the spectrum is
characterized by a broad hydrogen Balmer series and a He\,\textsc{i} line at
5876 $\text{\AA}$. Roughly two weeks after the explosion, Fe\,\textsc{ii} absorption
lines become visible and stronger with time. All five spectra
were taken during the plateau phase of SN 1999em
\citep{Leonard2002, Hamuy2001}.


\subsection{Specific intensity profiles}
\label{sec: Specific intensity profiles}

We calculated the specific intensity profiles from the \textsc{tardis}
simulations using a similar approach as \cite{Huber2019}. In contrast
to these authors, we used v-packets instead of regular Monte Carlo
packets. We further neglected flight-time differences between packets
that are emitted at different distances to the observer. The parameters
of the v-packets used for the calculation are the emission radius $r$,
the energy $\epsilon$, the frequency $\nu$, the time elapsed since explosion $t,$ and
cos($\theta$), where $\theta$ is the angle between the position vector
of the observer and the position vector of the v-packet leaving the
computational domain.

\begin{figure}[hbt!]
\centering
\subfigure[\hbox{Day 11 after explosion}]{\label{specific_i_p_SN1_0}\includegraphics[width=0.241\textwidth]{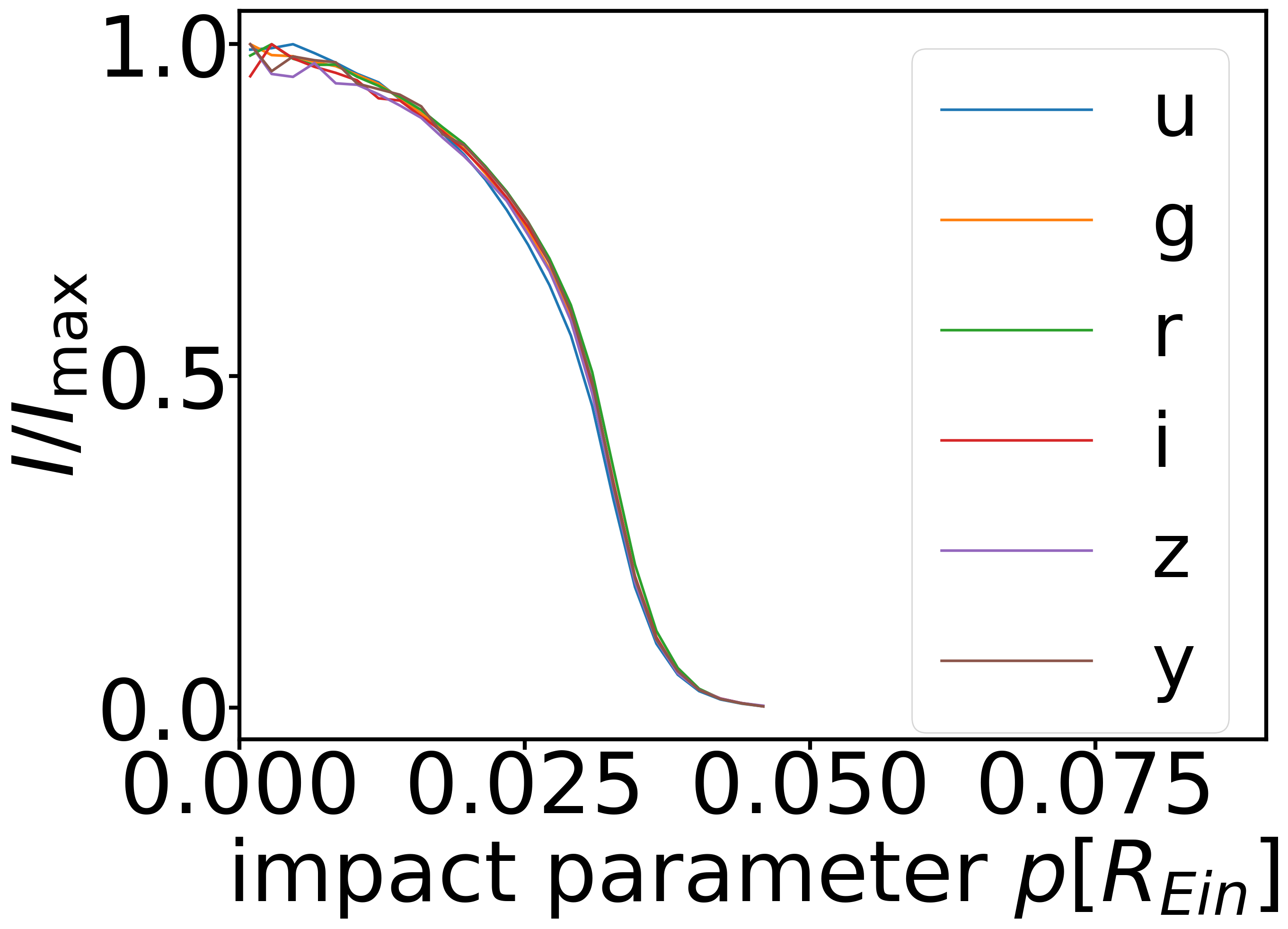}}
\subfigure[\hbox{Day 13 after explosion}]{\label{specific_i_p_SN1_1}\includegraphics[width=0.241\textwidth]{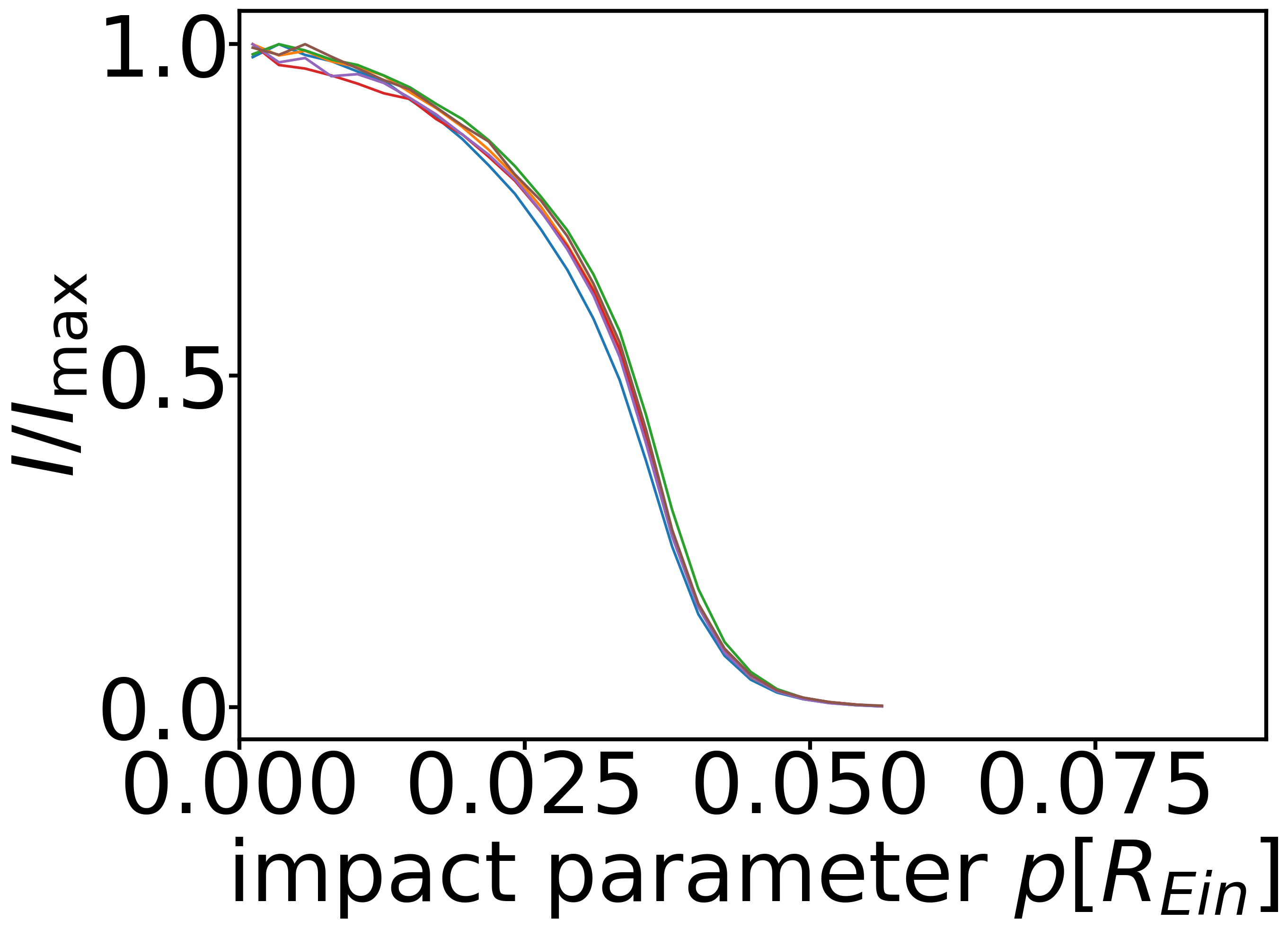}}\vskip5mm
\subfigure[\hbox{Day 17 after explosion}]{\label{specific_i_p_SN1_2}\includegraphics[width=0.241\textwidth]{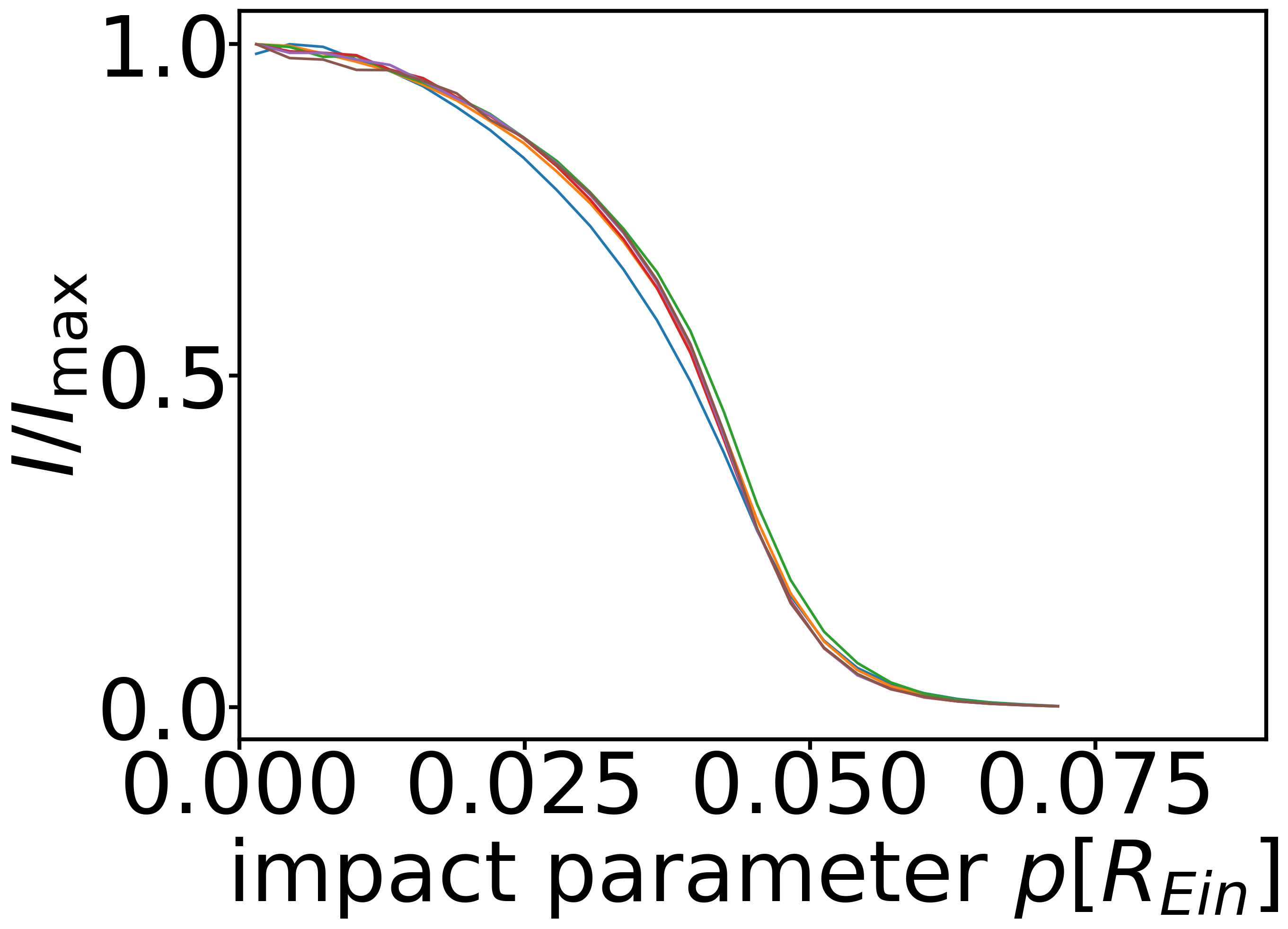}}
\subfigure[\hbox{Day 22 after explosion}]{\label{specific_i_p_SN1_3}\includegraphics[width=0.241\textwidth]{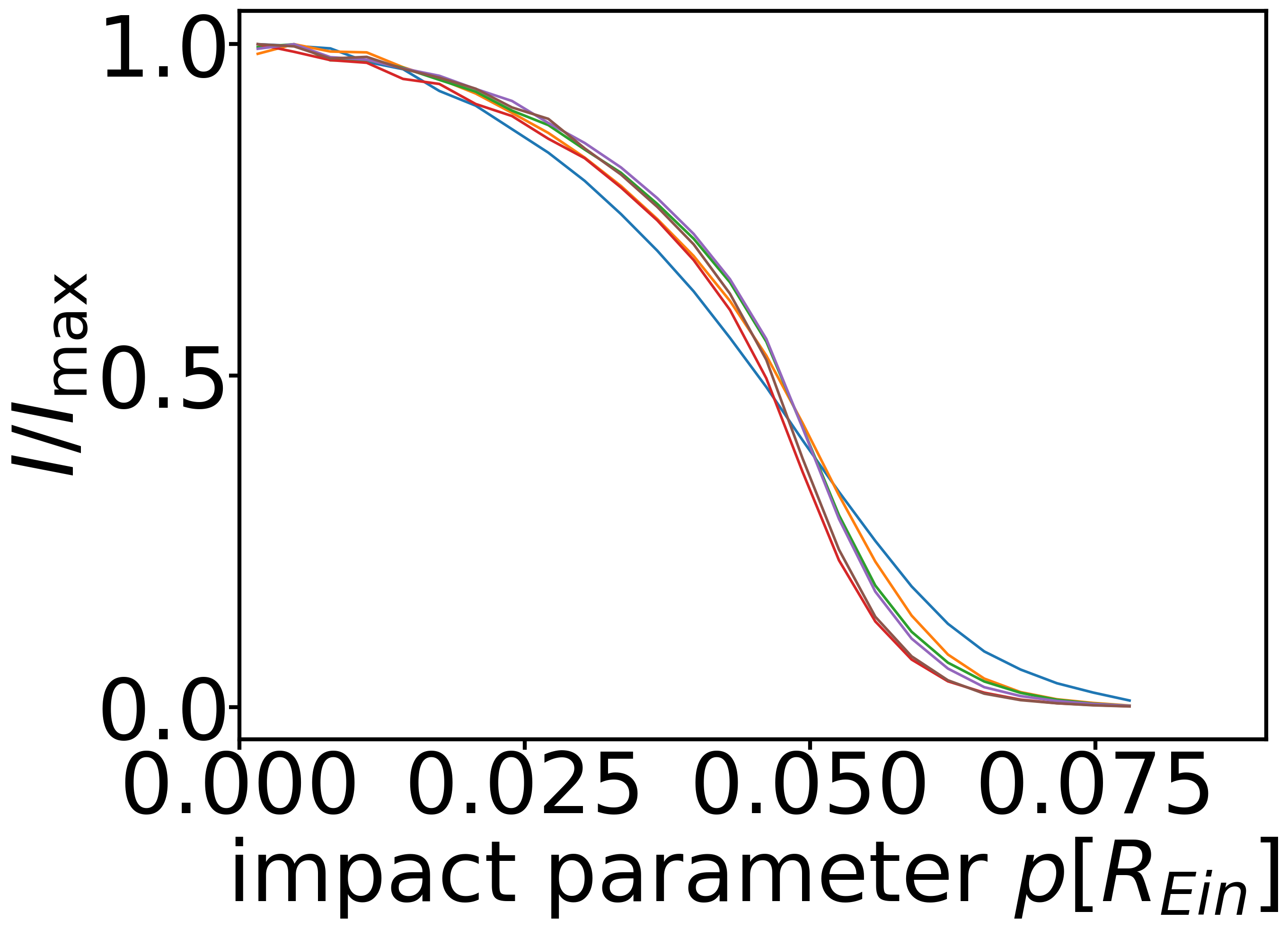}}\vskip5mm
\subfigure[\hbox{Day 27 after explosion}]{\label{specific_i_p_SN1_4}\includegraphics[width=0.241\textwidth]{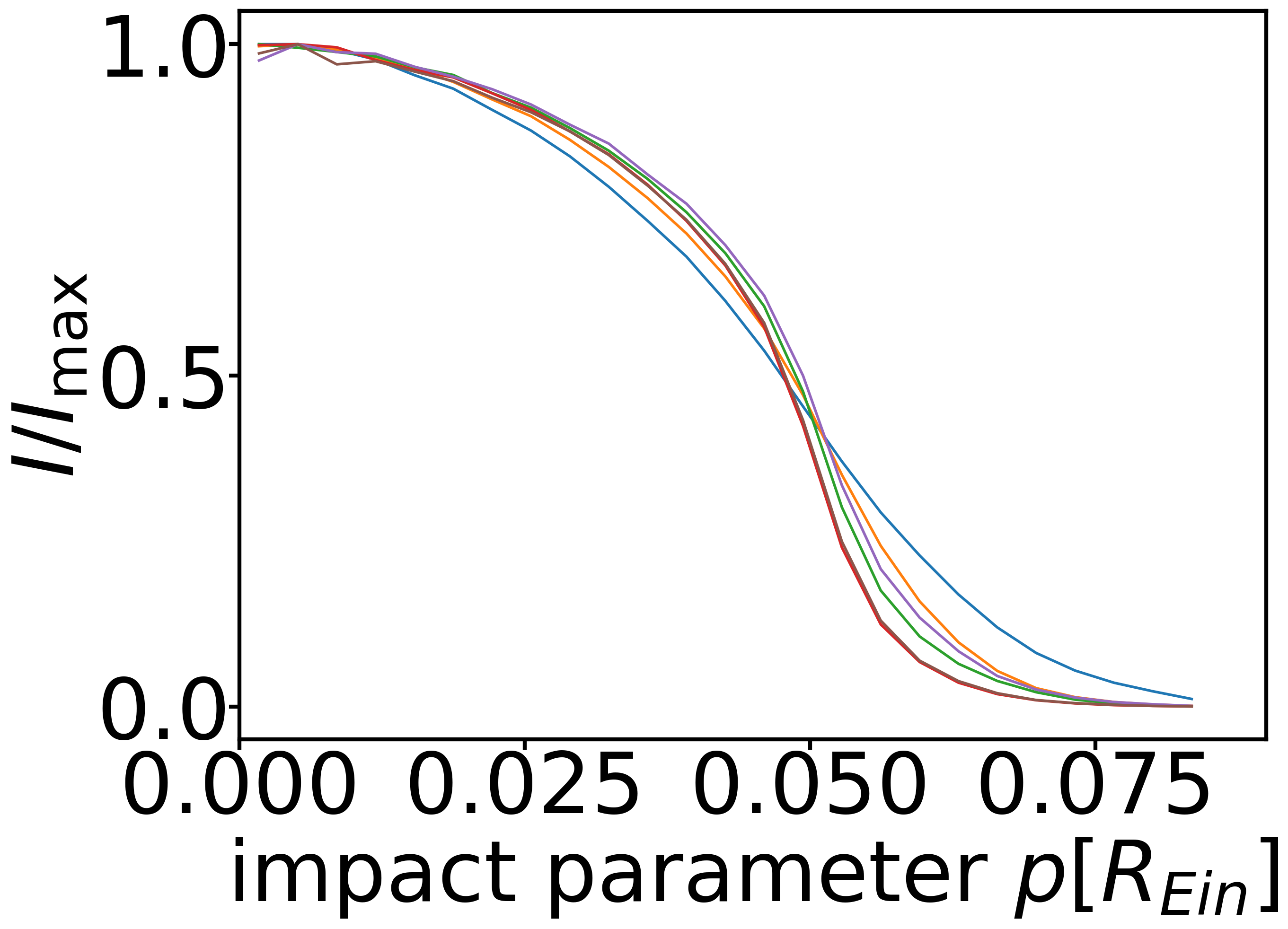}}\
\caption{\label{specific_i_p_SN1} Normalized specific intensity profiles as a function of the impact parameter, $p$, in units of $R_{\mathrm{Ein}}$ (with $R_{\mathrm{Ein}}$ = 2.9 $\times 10^{16}$ cm defined in Eq. (\ref{R_Ein}), setting the SN to a redshift of z$_{\mathrm{s}}$ = 0.77 and the lens to a redshift of z$_{\mathrm{l}}$ = 0.32).}
\end{figure}

In Fig. \ref{specific_i_p_SN1} the spatial distributions of the
normalized intensity profiles for the six LSST filters, $u$, $g$, $r$,
$i$, $z$, and $y$ \citep{LSSTScienceCollaboration2009}, are shown. The
profiles for the different filters are calculated by convolving the
transmission profile of each filter with the specific intensity.
Based on the plot, we chose 25 bins of the impact parameter $p$ as
an appropriate binning resolution for each epoch. 
The resulting intensity was normalized by the maximum of the
specific intensity for each filter.
The normalized specific intensity profiles are
very achromatic, especially at the first three investigated epochs, 
which show barely any difference between the profiles for the
different filters.

\section{Microlensing of type II supernovae}
\label{sec: Microlensing on Type II Supernovae}

With the upcoming LSST, thousands of LSNe will
be detected, including SNe Ia and SNe II. The possible use of SNe Ia
for time-delay measurements, and the influence of microlensing on it,
has already been investigated by \cite{Huber2019} and
\cite{Goldstein2018} using characteristic structures within color
curves to infer a time delay between two lensing images. However,
microlensing on SNe II is still an open question from a theoretical
viewpoint and is the motivation for this work.

The microlensed flux was calculated according to \cite{Huber2019} using specific intensities, retrieved with the \textsc{tardis} simulation at the source plane, which were microlensed using magnification
maps. To investigate the effect of microlensing on time-delay
measurements of LSNe II, we inserted 10000 random SN positions
into a microlensing map.

\subsection{Microlensing magnification maps}
\label{sec: Microlensing magnification maps}

We used the magnification maps from \textsc{gerlumph} \citep{Vernardos2015} to calculate the microlensed SN spectra.
These maps, which were calculated using the inverse ray-shooting technique \citep{Kayser1986,
  Wambsganss1992, Vernardos2013}, contain the magnification
$\mu(x,y)$ in the source plane as a function of the Cartesian coordinates
$x$ and $y$ in units of the Einstein radius $R_{\mathrm{Ein}}$.
The characteristic scale of the map, $R_{\mathrm{Ein}}$, is defined as:
\begin{equation} \label{R_Ein}
R_{\mathrm{Ein}} = \sqrt{\frac{4G \langle M \rangle}{c^{2}}\frac{D_{\mathrm{s}}D_{\mathrm{ds}}}{D_{\mathrm{d}}}},
\end{equation}
where $D_{\mathrm{s}}$ is the angular-diameter distance from the observer
to the source, $D_{\mathrm{ds}}$ the angular-diameter distance from the
lens to the source, $D_{\mathrm{d}}$ the angular-diameter distance from
the observer to the lens, and $c$ the speed of light. We further assumed a Salpeter
initial mass function with a mean mass of the point mass microlenses
of $\langle M \rangle$ = 0.35 $\mathrm{M_{\odot}}$ as described by
\cite{Huber2020}.

\begin{figure}[hbt!]
\centering
\subfigure[\hbox{Type I microlensing map}]{\label{micromap1}\includegraphics[width=0.21\textwidth]{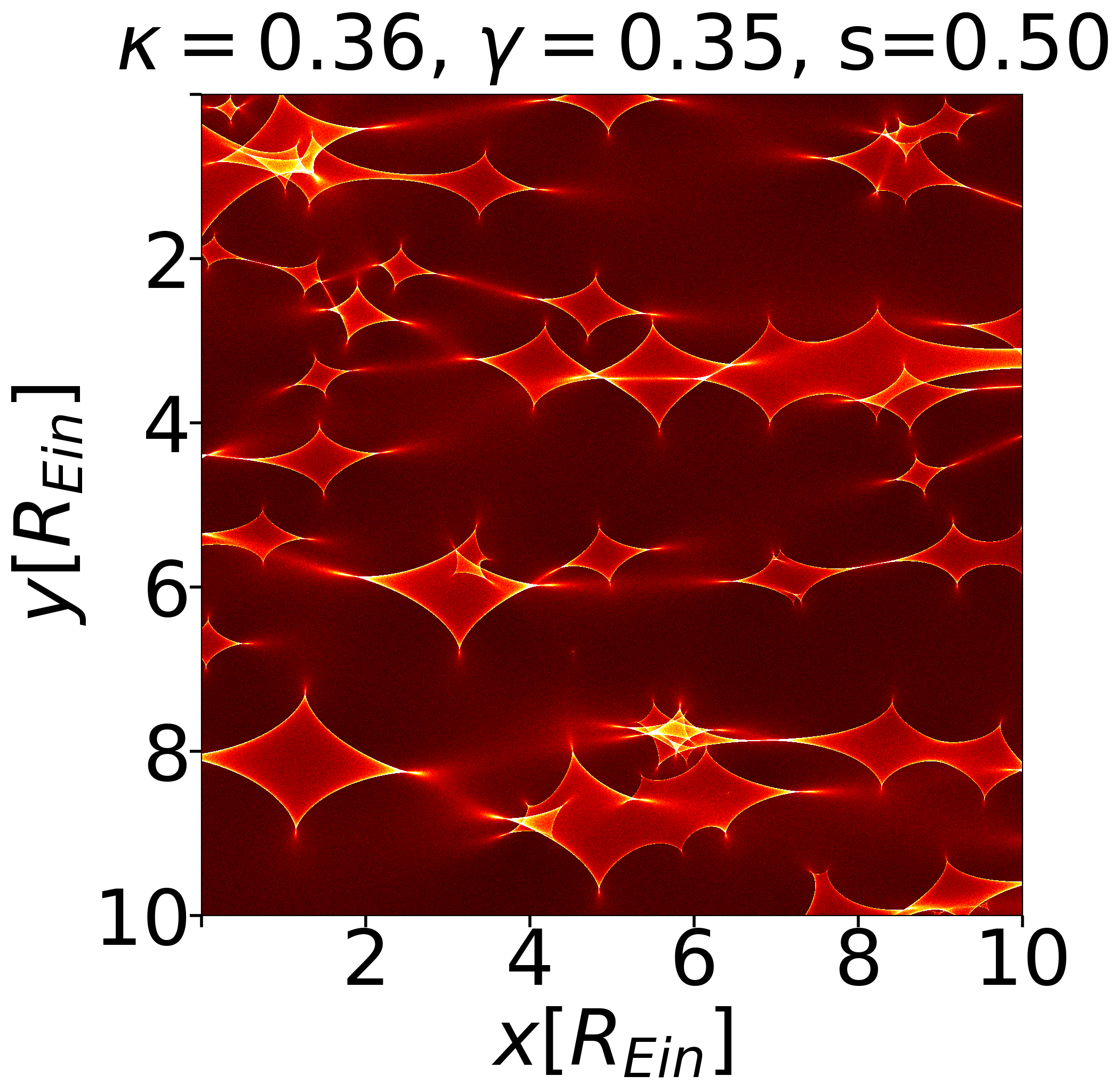}}
\subfigure[\hbox{Type II microlensing map}]{\label{micromap2}\includegraphics[width=0.21\textwidth]{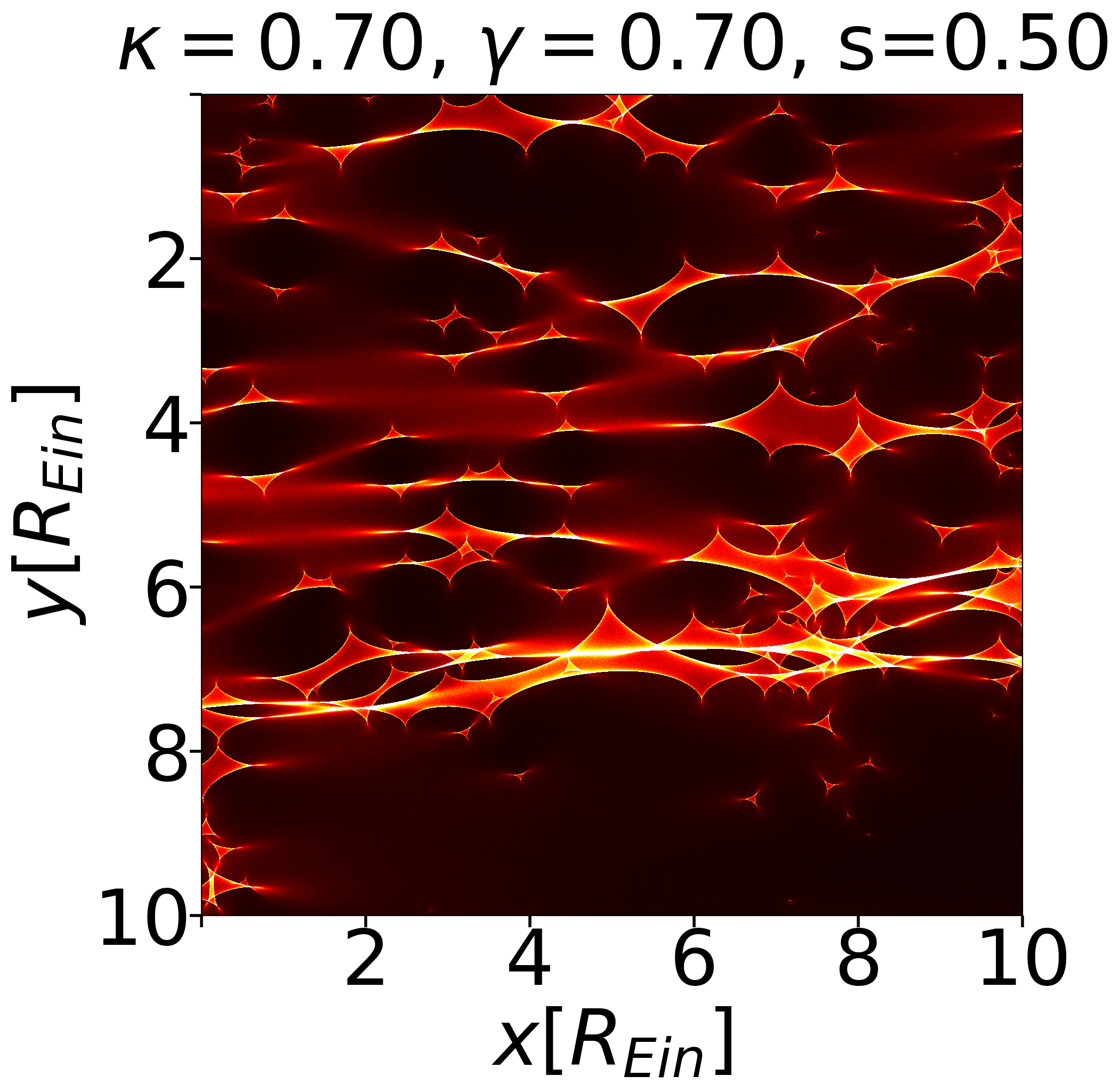}}
\subfigure{\label{micromap3}\includegraphics[width=0.059\textwidth]{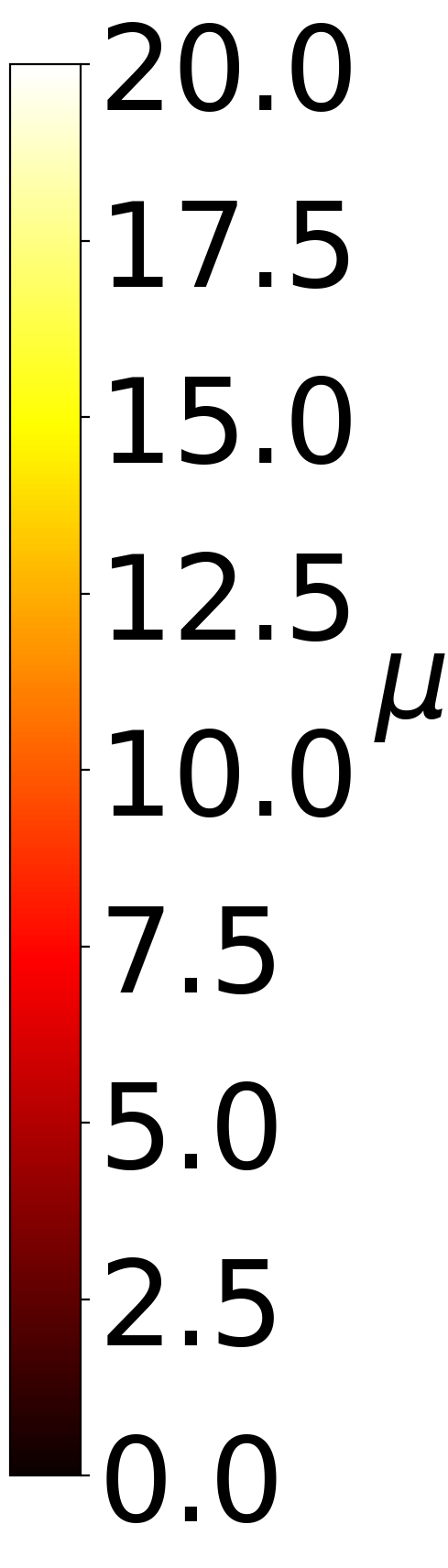}}\
\caption{\label{micromaps} Magnification maps for two different $\kappa$ and $\gamma$ values with smooth matter fraction $s$ = 0.50. The magnification $\mu(x, y)$ is indicated by the color scale on the right-hand side.}
\end{figure}

The magnification maps are primarily defined by three parameters: the
convergence $\kappa$, the shear $\gamma$, and the smooth matter
fraction $s = 1-\frac{\kappa_{*}}{\kappa}$, where $\kappa_{*}$ is the
convergence of the stellar component. The maps used in this work have a
resolution of 20000 $\times$ 20000 pixels, corresponding to a size of
10 $R_{\mathrm{Ein}} \times  10 R_{\mathrm{Ein}}$.

In this work we consider two SN images of a lens system with different microlensing maps. Galaxy-scale strong lens systems typically have
two or four images of the background source.  For a two-image
lensed system, the first and second images that appear are time-delay
minimum (type I) and saddle (type II) images\footnote{The type I and II
refer to the types of lensing images, not SN types.}, respectively. For a
four-image lensed system, the first two images are of type
I and the next two images are of type II.  For our current study, we consider two SN images from either a double or a quad system, with
the first SN image as type I and the second one as type II.
Therefore, we simulated microlensing for these two SN images using two
different microlensing magnification maps, which are shown in Fig.
\ref{micromaps}.

For the first image, we chose $\kappa$ = 0.36 and $\gamma$ = 0.35,
which are the median values of type I lensing images of the OM10
catalog \citep{Oguri2010}. For the second image, we chose $\kappa$ =
0.7 and $\gamma$ = 0.7, which are the median values of
type II lensing images of the OM10 catalog \citep{Huber2020, Suyu2020}. We adopted $s$ = 0.5 for both cases.

\begin{figure}[hbt!]
\centering
{\includegraphics[width=0.489\textwidth]{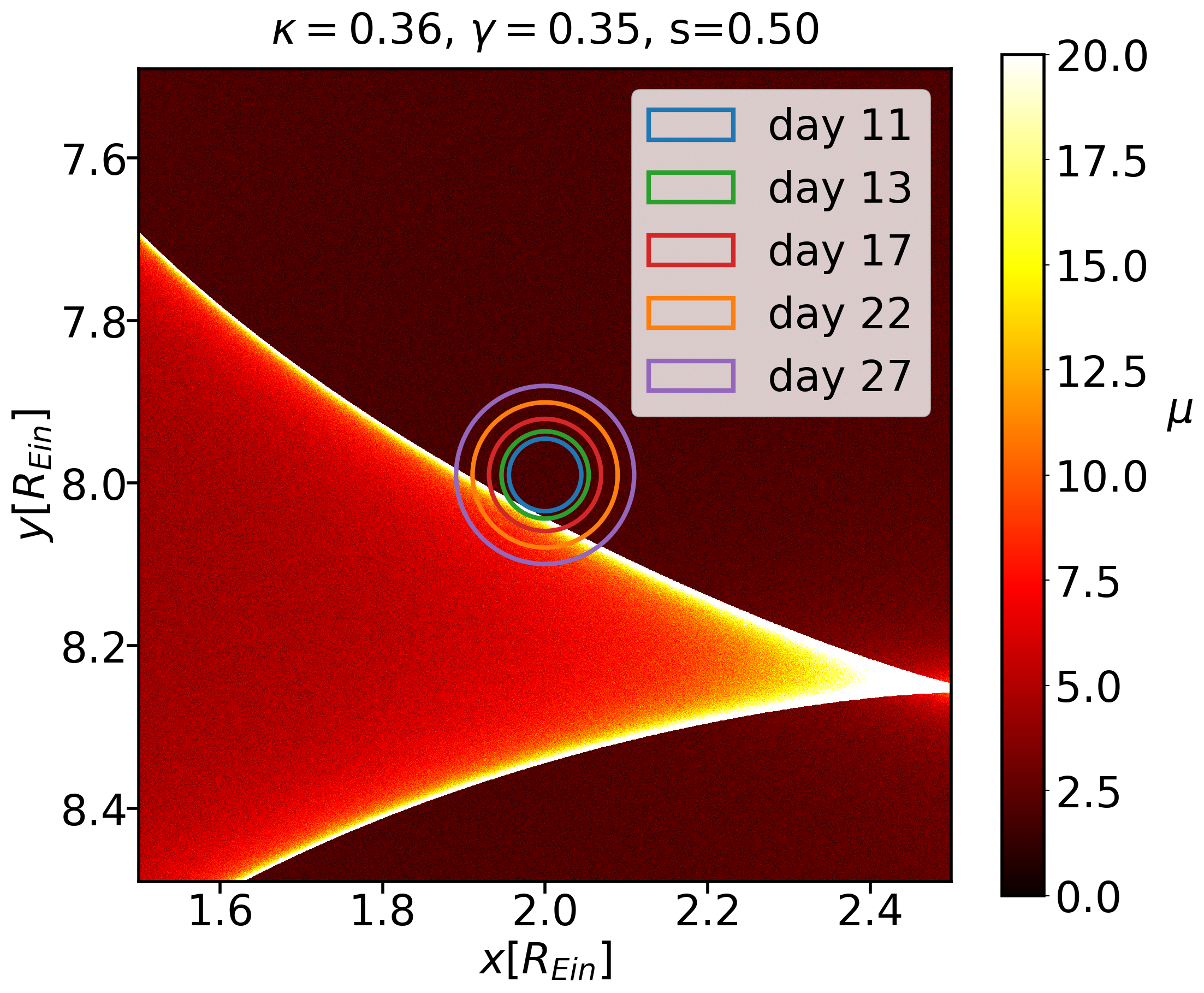}}\
\caption{\label{lightcurve_map} Magnification map for type I
  lensing with the position of the SN for the microlensed light curve. The different circles represent the sizes of the SN at the five
  different epochs we investigate.}
\end{figure}

The observed microlensed flux $F_{\lambda, \mathrm{o}}(t)$ was computed
by placing the SN into the magnification map and solving:
\begin{equation} \label{flux_eq}
F_{\lambda, \mathrm{o}}(t) = \frac{1}{D_{\mathrm{lum}}^{2}(1+z_{s})} \int \mathrm{d}x \int \mathrm{d}y \ I_{\lambda, \mathrm{e}}(t, x, y) \ \mu(x, y),
\end{equation}
where $D_{\mathrm{lum}}$ is the luminosity distance to the source and
$I_{\lambda, \mathrm{e}}(t, x, y)$ is the specific intensity in the
source plane.  For the calculation we interpolated the specific
intensity onto a 2D Cartesian grid of the same spatial resolution as the magnification map and integrated over the SN \citep{Huber2019}.
The AB magnitudes were then calculated by:
\begin{equation} \label{mag_AB}
m_{\mathrm{AB,X}}(t) = -2.5 \log_{10}\left( \frac{\int \mathrm{d}\lambda \ \lambda S_{\mathrm{X}}(\lambda) \ F_{\lambda \mathrm{o}}(t)}{\int \mathrm{d}\lambda \ S_{\mathrm{X}}(\lambda) \ c / \lambda} \times \mathrm{\frac{cm^{2}}{erg}} \right) - 48.6
\end{equation}
\citep{Bessell2012}, where $S_{\mathrm{X}}(\lambda)$ is the transmission function of the LSST filter X \citep{LSSTScienceCollaboration2009}.

\subsection{Type II supernova light curves and color curves}
\label{sec: SNe II light curves and color curves}

\begin{figure}[hbt!]
\centering
\subfigure{\label{lightcurve_u}\includegraphics[width=0.24\textwidth]{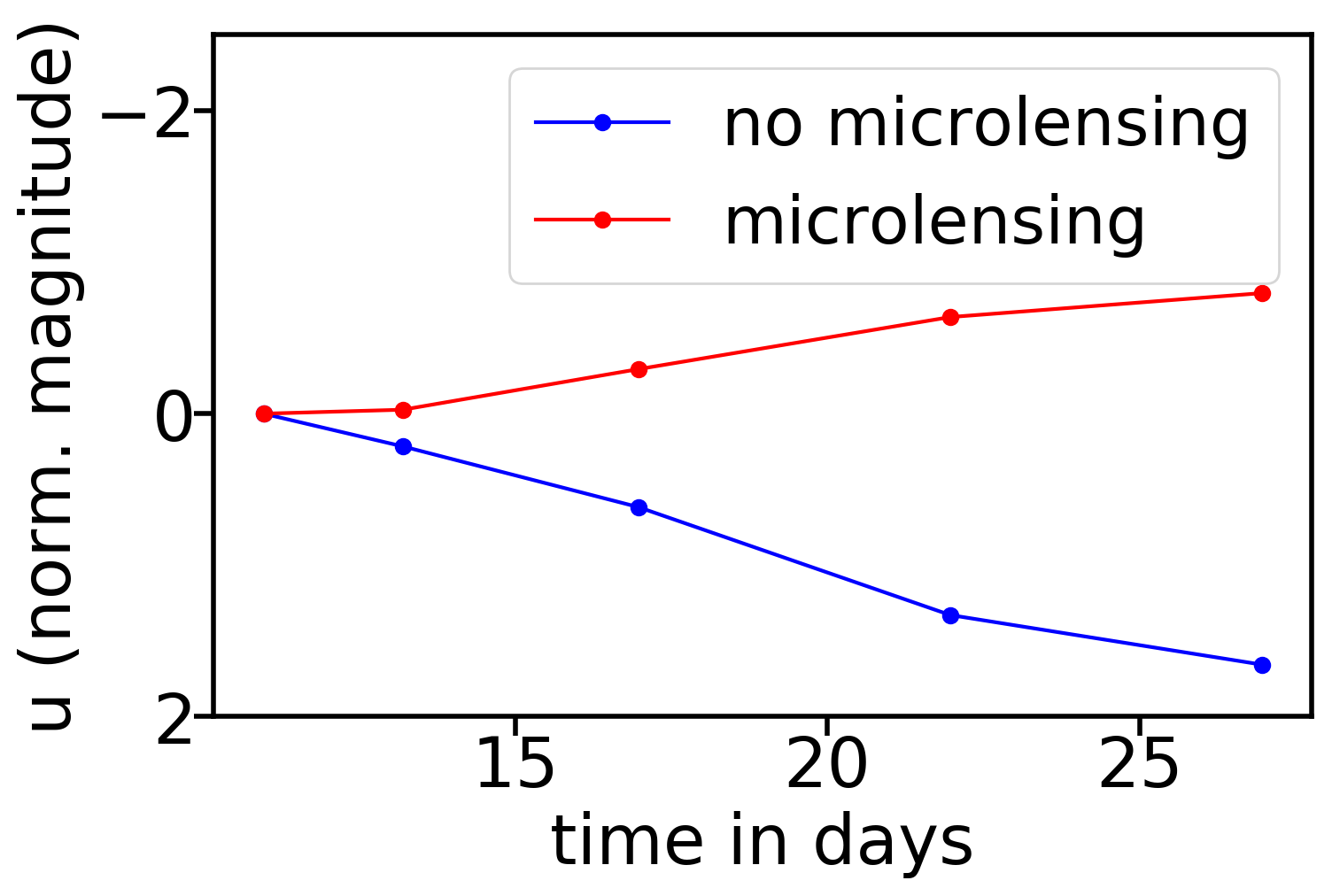}}\
\subfigure{\label{lightcurve_g}\includegraphics[width=0.24\textwidth]{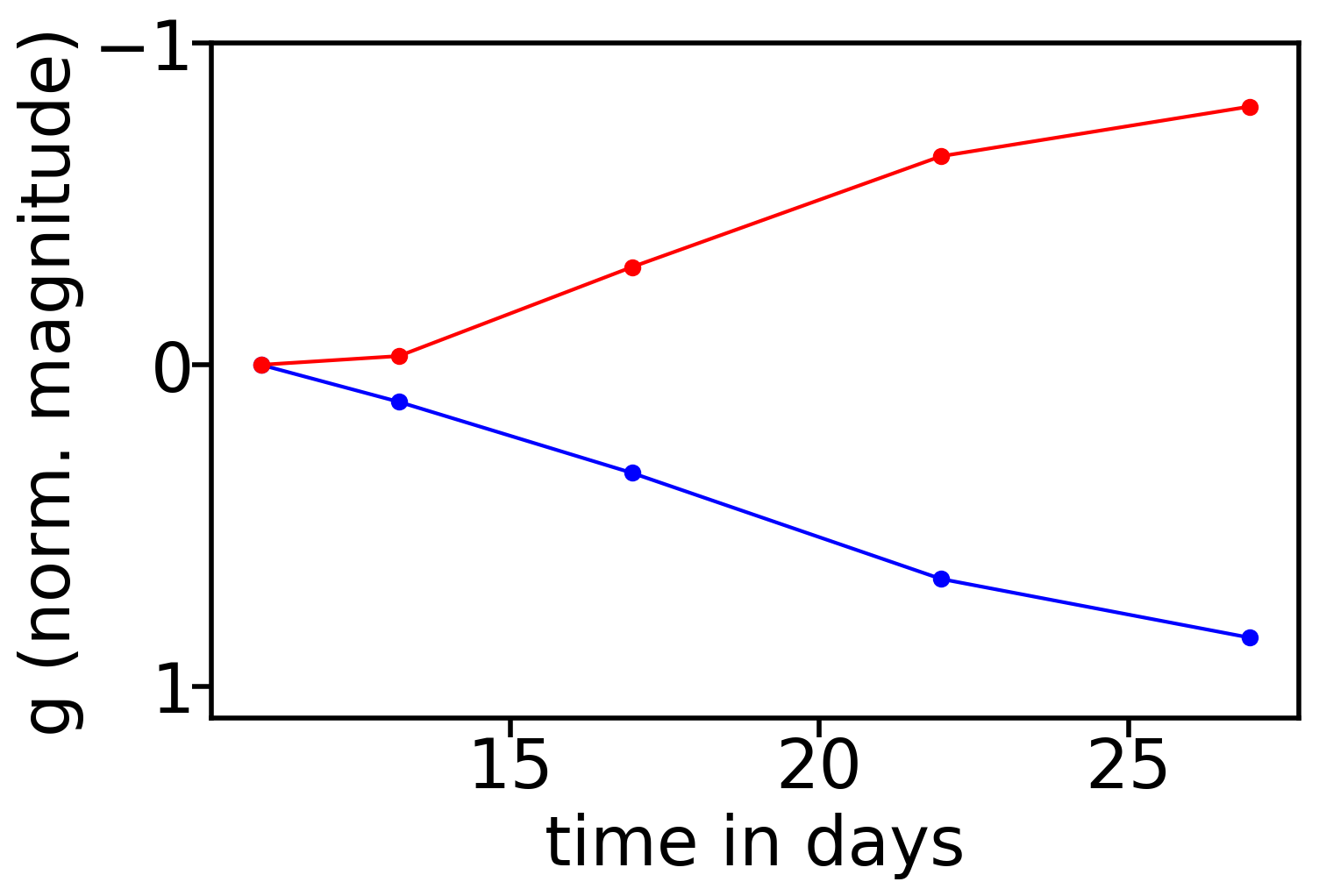}}\\
\subfigure{\label{lightcurve_r}\includegraphics[width=0.24\textwidth]{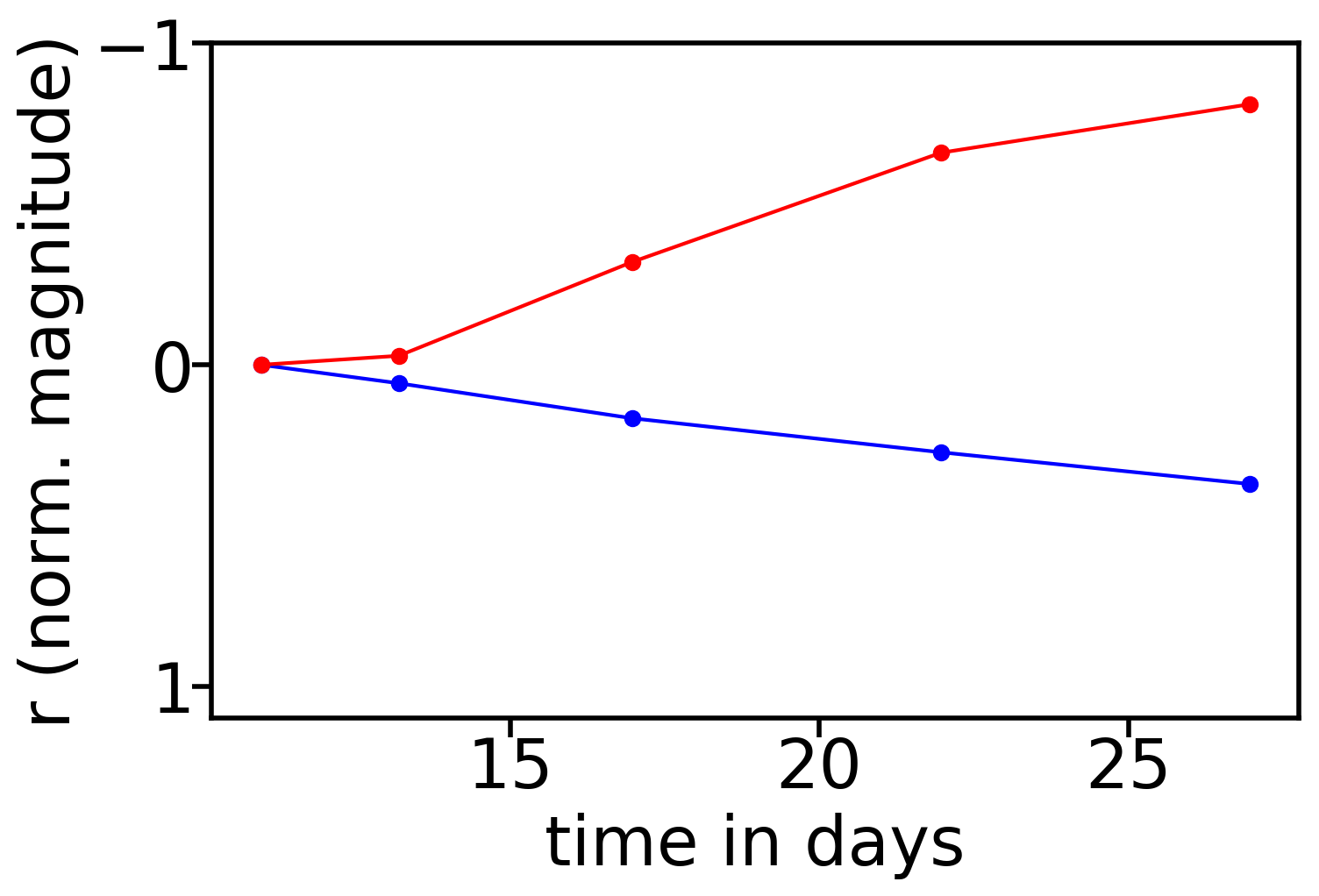}}\
\subfigure{\label{lightcurve_i}\includegraphics[width=0.24\textwidth]{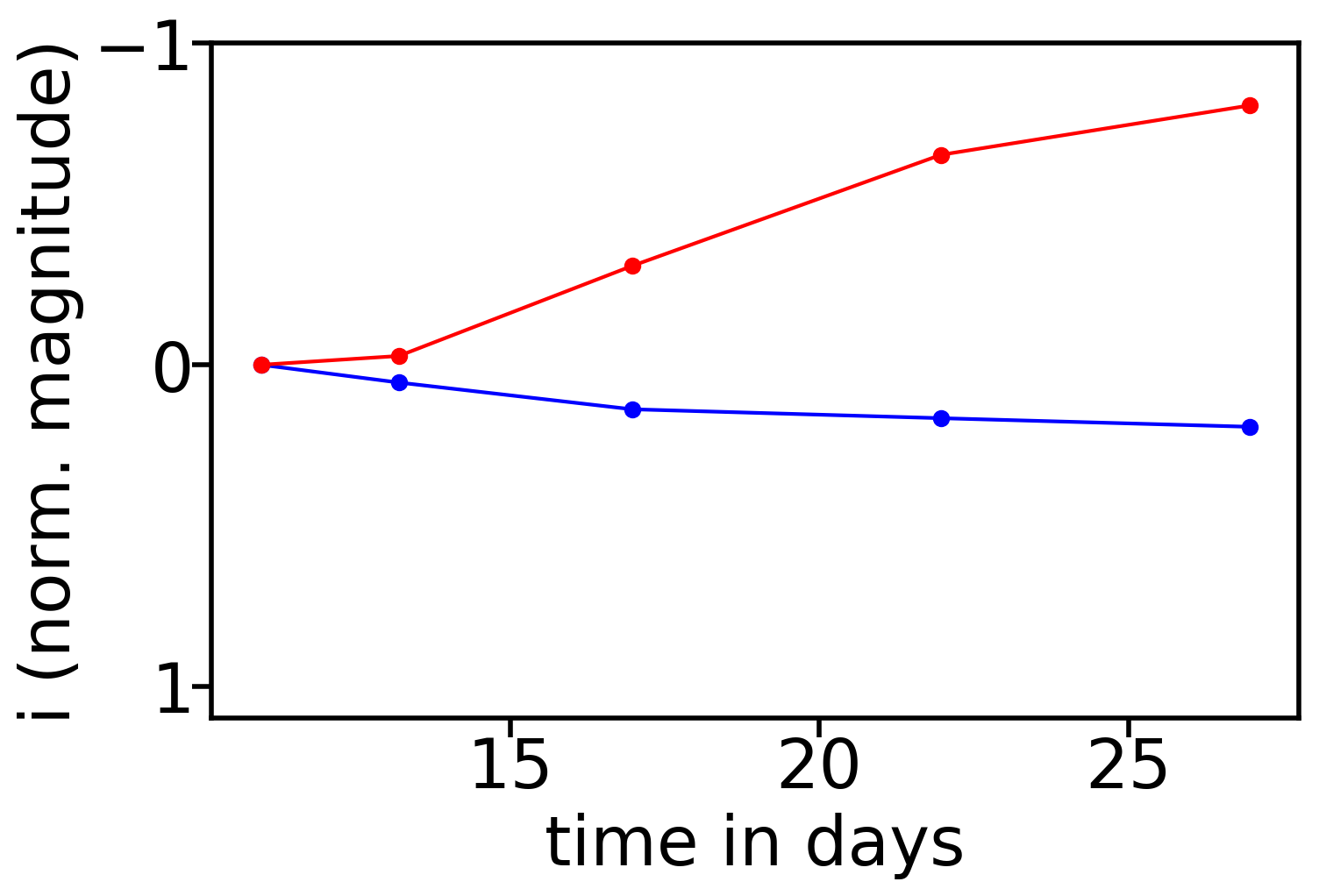}}\\
\subfigure{\label{lightcurve_z}\includegraphics[width=0.24\textwidth]{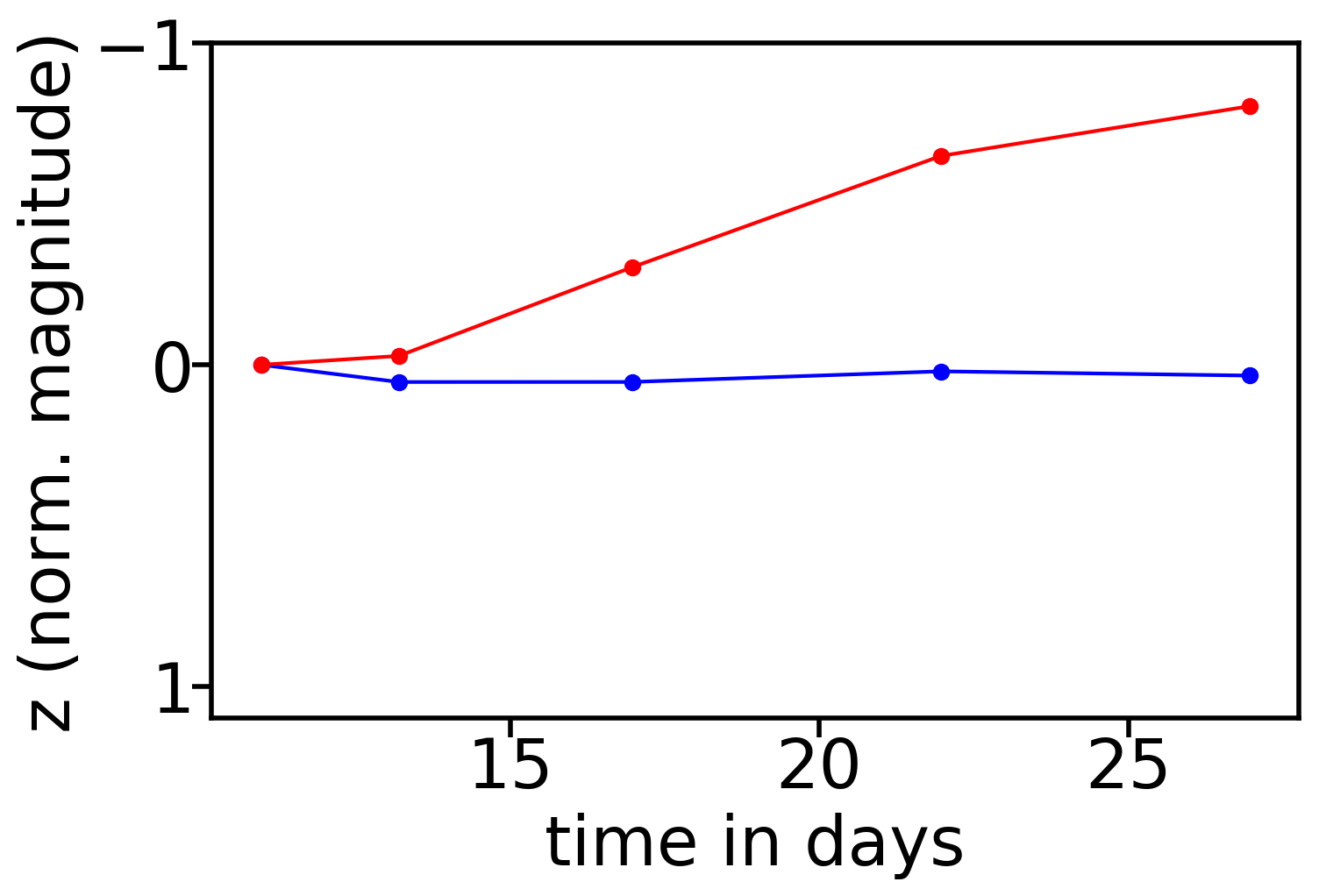}}\
\subfigure{\label{lightcurve_y}\includegraphics[width=0.24\textwidth]{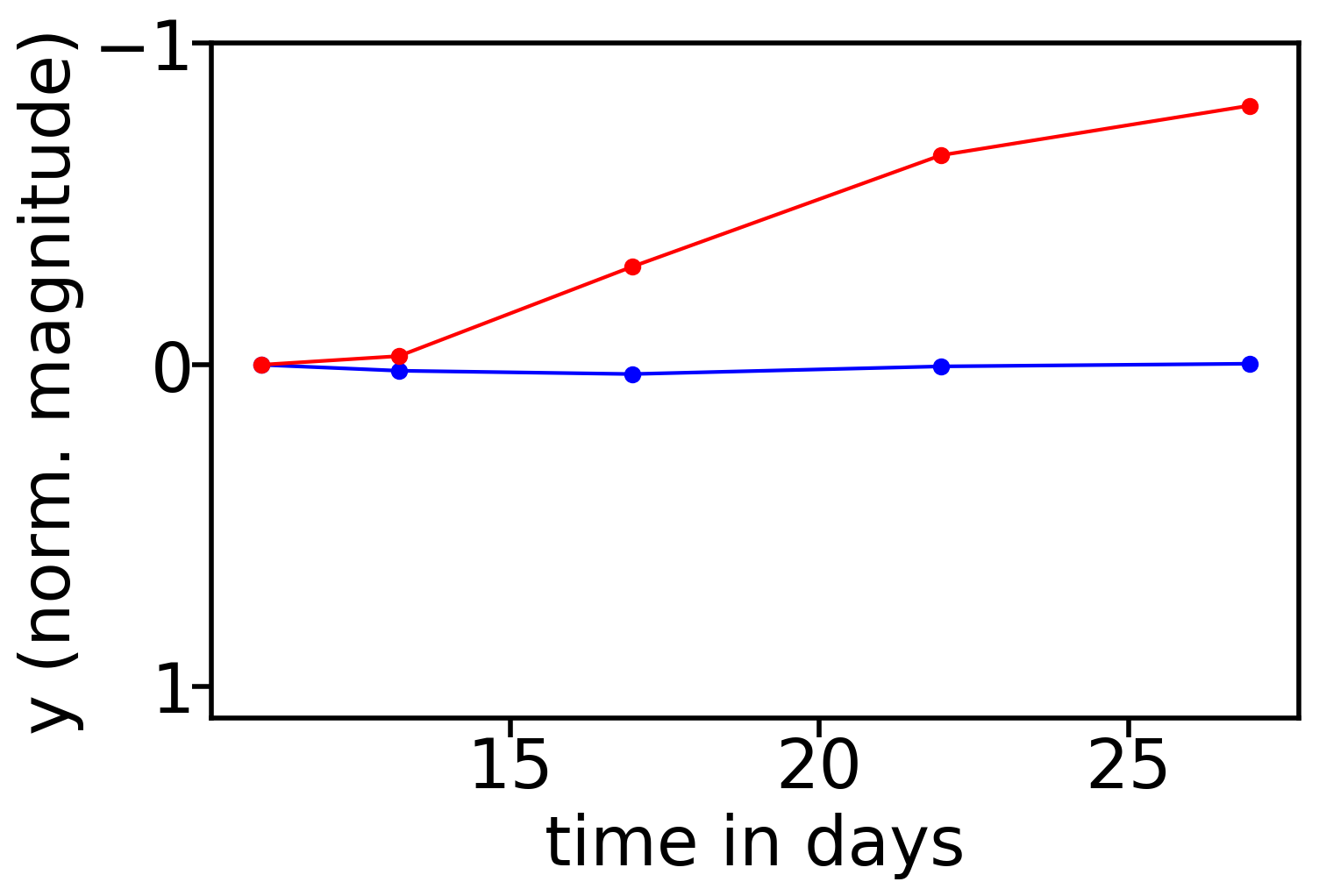}}\
\caption{\label{lightcurves} Microlensed and non-microlensed
  normalized light curves in the LSST filters $u, g, r, i, z$, and $y$
  for SN 1999em at the position marked in the microlensing map in
  Fig. \ref{lightcurve_map}.}
\end{figure}

In this subsection we study the effect of microlensing on the light
curves and color curves retrieved from the \textsc{tardis} simulations and
investigate if they can be used for time-delay
measurements. The impact on the color curves is of
particular interest as the achromatic behavior of the specific intensity profiles suggests
that microlensing has little influence on the color curves.

\begin{figure*}[hbt!]
\centering
\subfigure[]{\label{colorcurve_u_g}\includegraphics[width=0.475\textwidth]{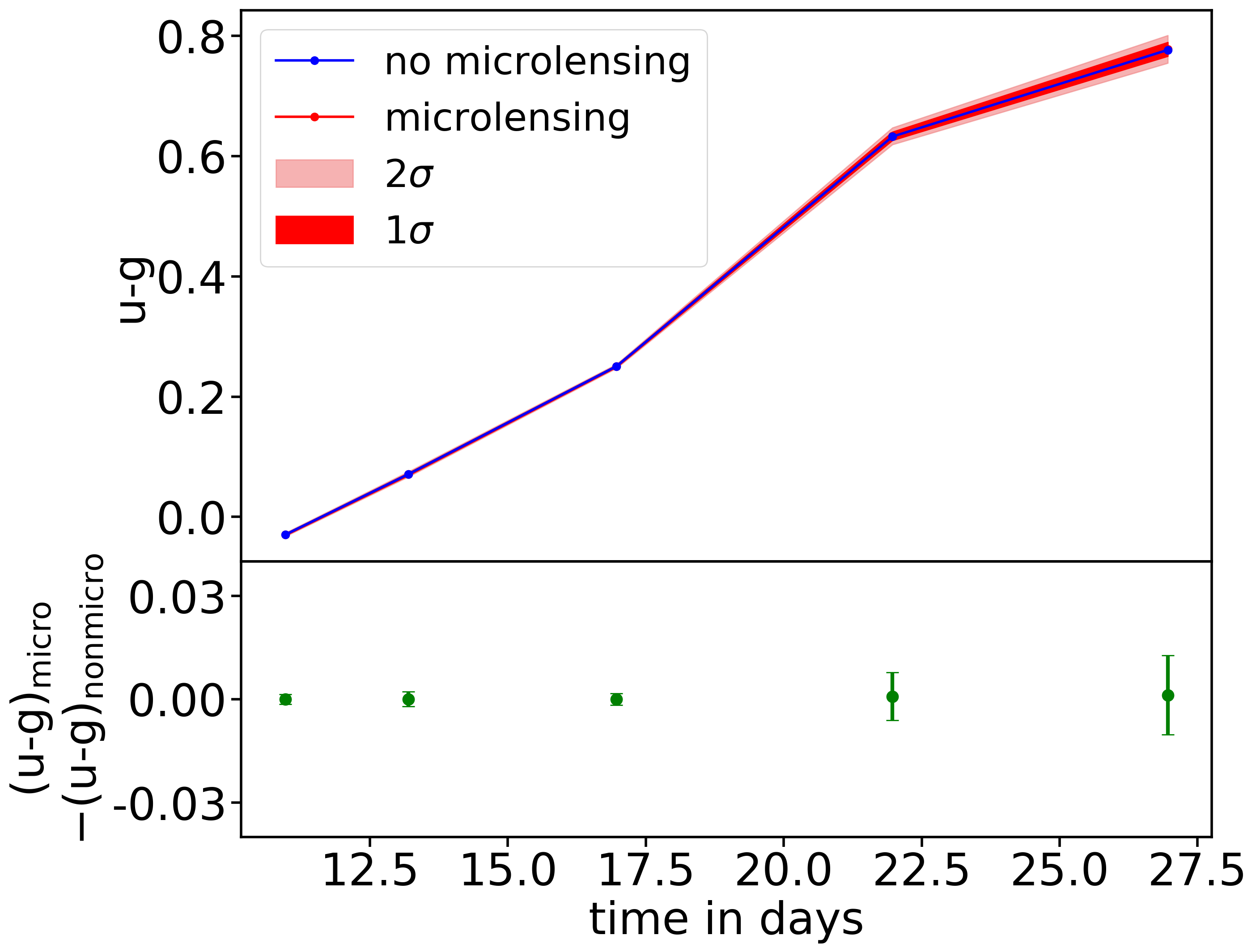}}
\subfigure[]{\label{colorcurve_r_i}\includegraphics[width=0.475\textwidth]{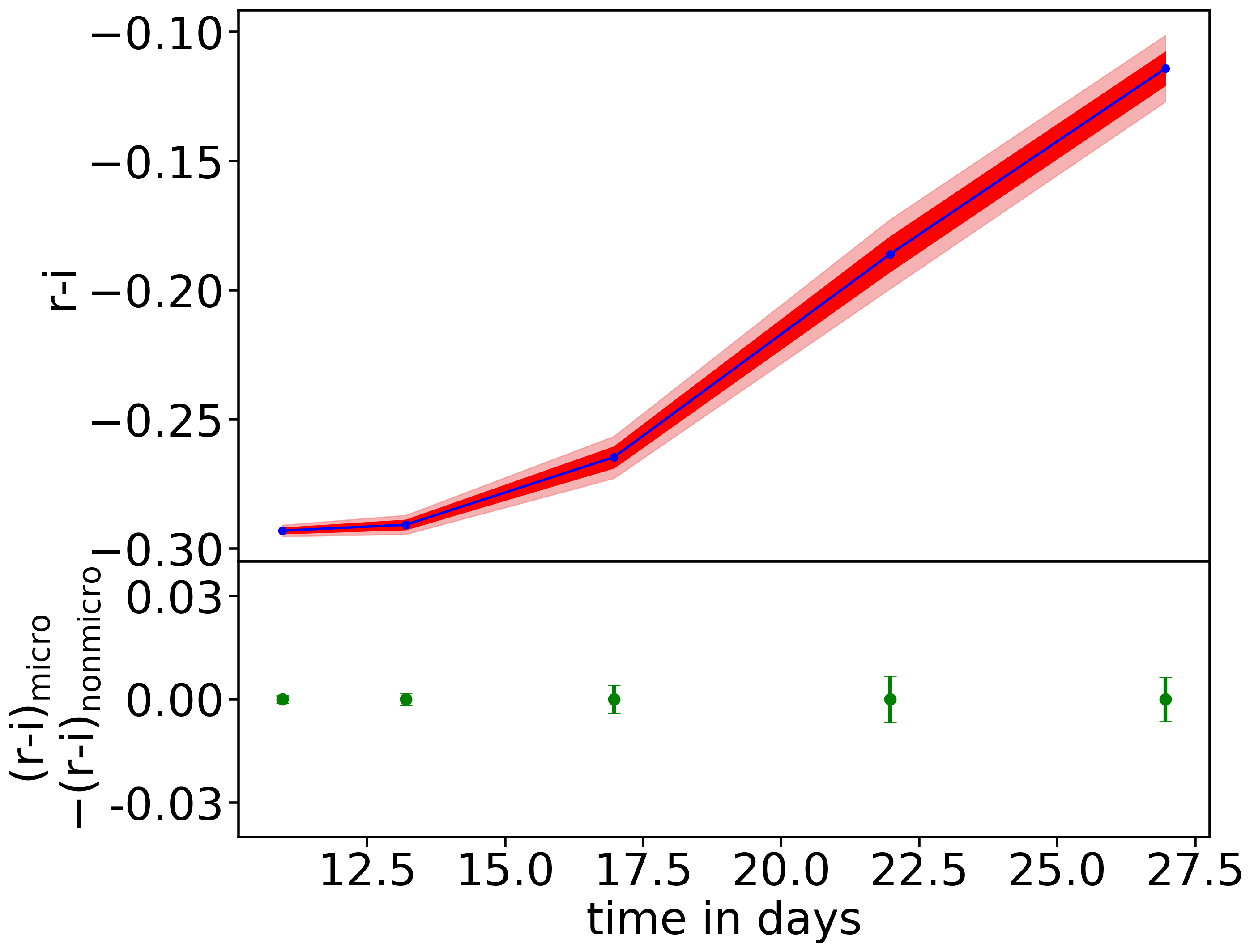}}\\
\subfigure[]{\label{colorcurve_i_z}\includegraphics[width=0.475\textwidth]{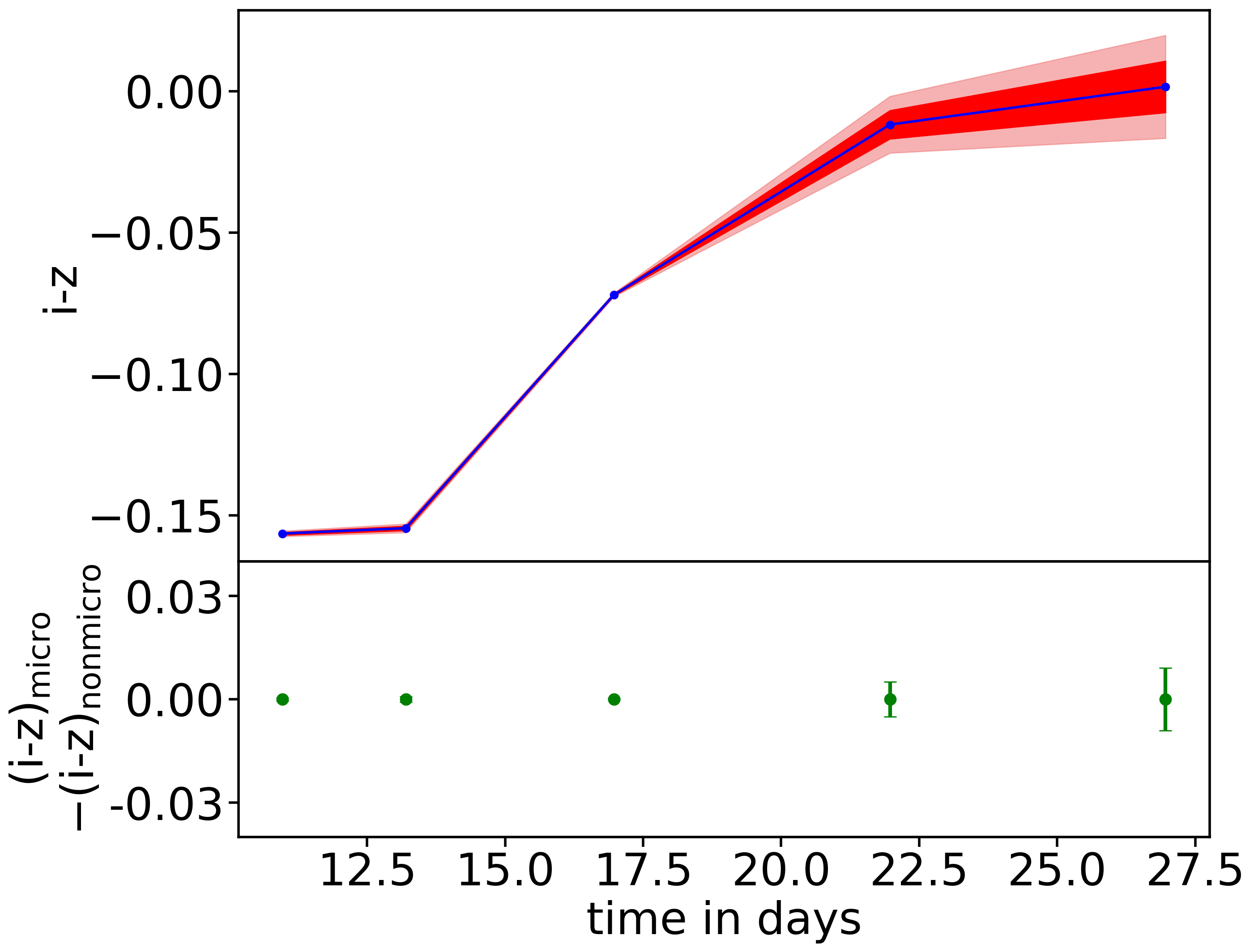}}
\subfigure[]{\label{colorcurve_z_y}\includegraphics[width=0.475\textwidth]{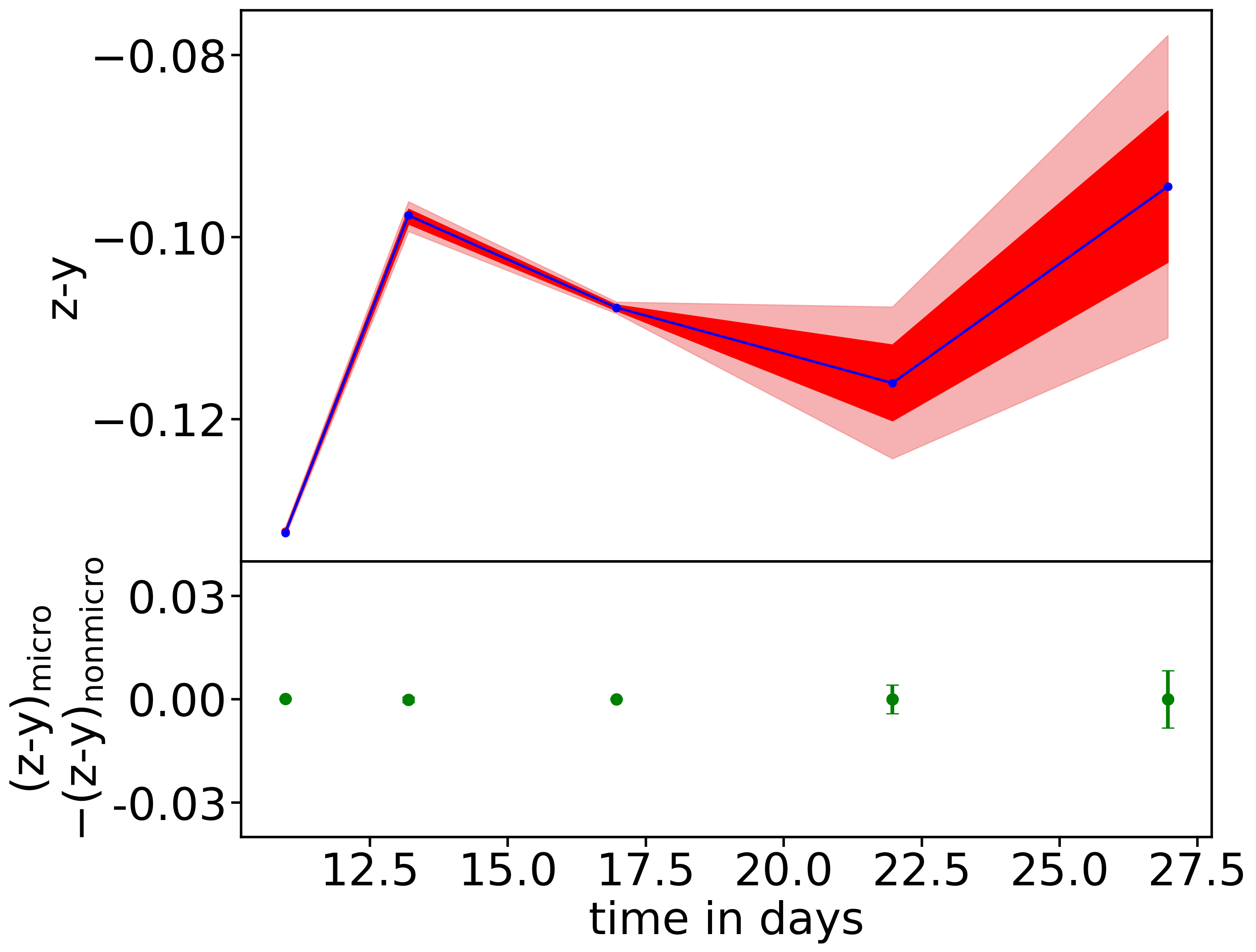}}\
\caption{\label{colorcurves} Color curves of the non-microlensed and
  microlensed spectra with median, 1$\sigma$, and 2$\sigma$ ranges obtained
  from 10000 random positions in the type I lensing magnification map. In addition, the bottom part of each panel (\ref{colorcurve_u_g} to \ref{colorcurve_z_y}) shows the deviation of the microlensed color from the non-microlensed color with 1$\sigma$ uncertainties.}
\end{figure*}
\begin{figure}[hbt!]
\centering
{\includegraphics[width=0.48\textwidth]{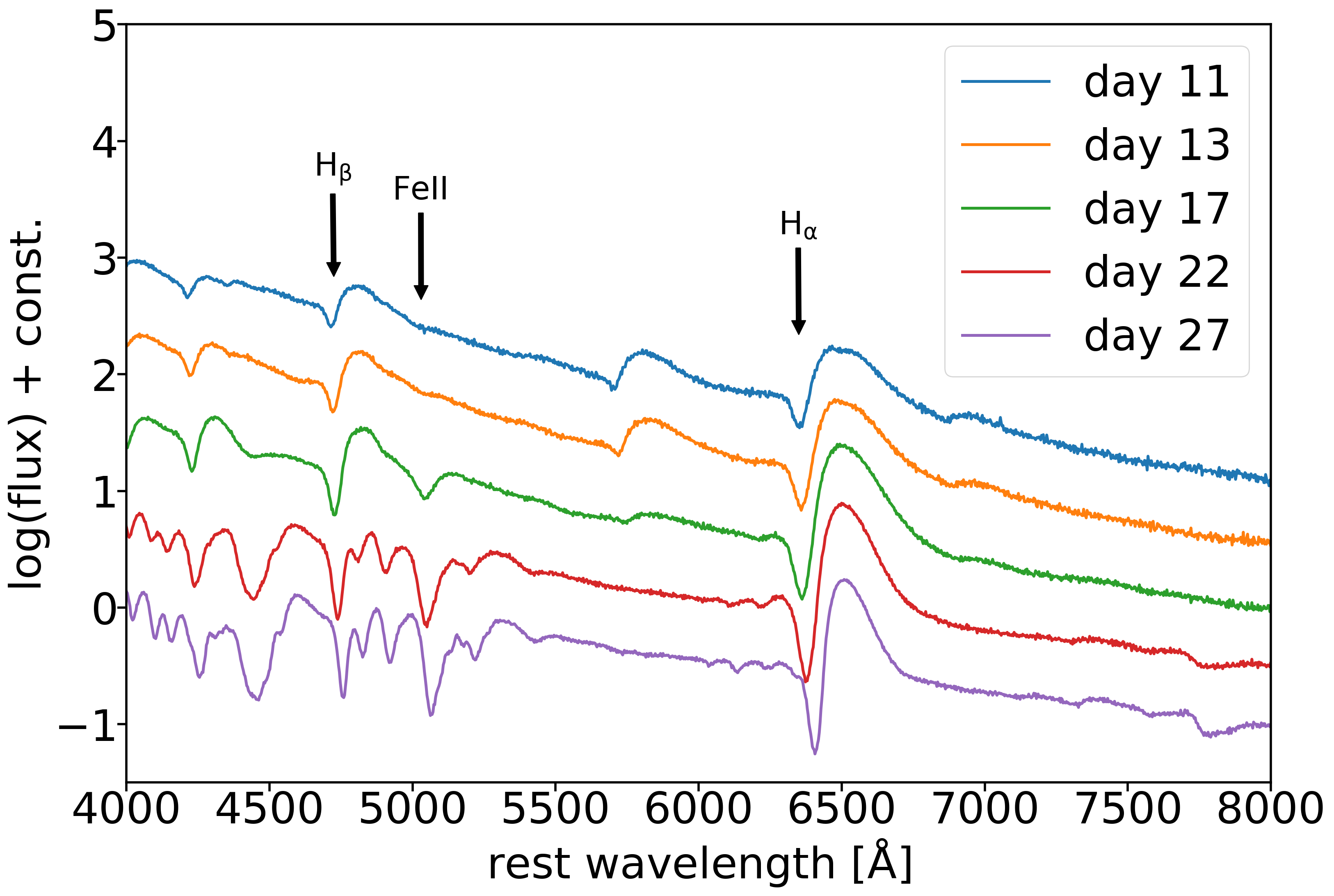}}\
\caption{\label{absorption_lines} Model spectra for the five epochs with the
  three absorption lines of interest, labeled H$\alpha$,
  H$\beta$, and Fe\,\textsc{ii}.}
\end{figure}

The curves were created using the filters $u, g, r, i, z$, and $y$ of
LSST \citep{LSSTScienceCollaboration2009}. The transmission of
each filter was convolved with the supernova flux to
obtain the light curve for that filter. The microlensed flux was
calculated according to Eq. (\ref{flux_eq}) with
$D_{\mathrm{lum}} = 10$ pc to get the absolute magnitudes when calculating the light curves. For the non-microlensed flux, the
magnification was set to $\mu(x, y)$ = 1. The luminosity distance was chosen such that further computations could be made with the
flux instead of the luminosity density. The microlensed and
non-microlensed AB magnitudes for the light curves were then calculated
via Eq. (\ref{mag_AB}).
As an example for the microlensed light curve, we
chose a position in the magnification map of type I
lensing where the expanding layers cross a caustic with
time. The position is shown in Fig. \ref{lightcurve_map}, with the
circles representing the radius of the SN containing 99.9 \% of the
luminosity for the five different epochs. We see that the caustic is
crossed from day 13 onward.

The caustic crossing can also be seen within the light
curves shown in Fig. \ref{lightcurves} for the
non-microlensed and microlensed case. We normalized the light curves by shifting the magnitude of the first epoch to zero. The
light curves show that after day 11 the difference in the absolute
magnitude between the non-microlensed and microlensed case is
increased. This means that the SN is magnified by the caustic crossing
after day 11. The intrinsic light curves do not have
characteristic structures, such as peaks, as they include only epochs
of the plateau phase of SN 1999em. The plateau phase is the part
in the light curve of a SN II-P where the magnitude barely changes
over a certain period of time. The light curves within the plateau phase of our investigation can be strongly affected by microlensing, which adds large uncertainties to time-delay measurements.

Even though the non-microlensed light curves are almost linear, which
leads to nearly linear color curves, we wanted to show the influence of
microlensing on these color curves, following \cite{Goldstein2018},
\cite{Huber2019}, and \cite{Huber2020}. Due to the achromatic specific intensity profiles, the influence is expected to be low. We calculated the LSN
signal at 10000 different random positions in the microlensing
map. For each position, we calculated the magnitude difference,
$m_{\mathrm{AB,X}}(t_{i}) - m_{\mathrm{AB,Y}}(t_{i})$, for two filters,
X and Y. For each filter combination, we plot the median and the 1$\sigma$
and 2$\sigma$ deviations retrieved from the 10000 individual color
curves.
Four comparisons of the microlensed color curves to the
non-microlensed color curves are shown in Fig. \ref{colorcurves}. Additional color curves are shown in Appendix \ref{sec: Appendix A}. It can been seen that the influence of microlensing on the color curves is small overall. Yet, these color curves cannot be used in time-delay measurements in a straightforward way because they do not show large characteristic structures that would simplify the time-delay measurements,
given the possibility of differential dust extinction between the multiple SN images \citep{Eliasdottir2006}.
Therefore, to use color curves, a longer time span, covering more than the first part of the plateau phase that we consider in this work, would be helpful for inferring the time delays.  We defer this to future studies.

\subsection{Type II supernova spectra and absorption features}
\label{sec: SNe II spectra and absorption features}

As the light curves and color curves lack significant features within
the plateau phase, we investigate another
approach to perform time-delay measurements on SNe II. We used the
temporal evolution of the minima of the absorption lines of
H$\alpha$, H$\beta$, and Fe\,\textsc{ii} as they are common features used for velocity measurements that can be detected in at least three epochs. The hydrogen lines can even be detected in all five of the epochs (see Fig. \ref{absorption_lines}).

\begin{figure*}[hbt!]
\centering
{\includegraphics[width=\textwidth]{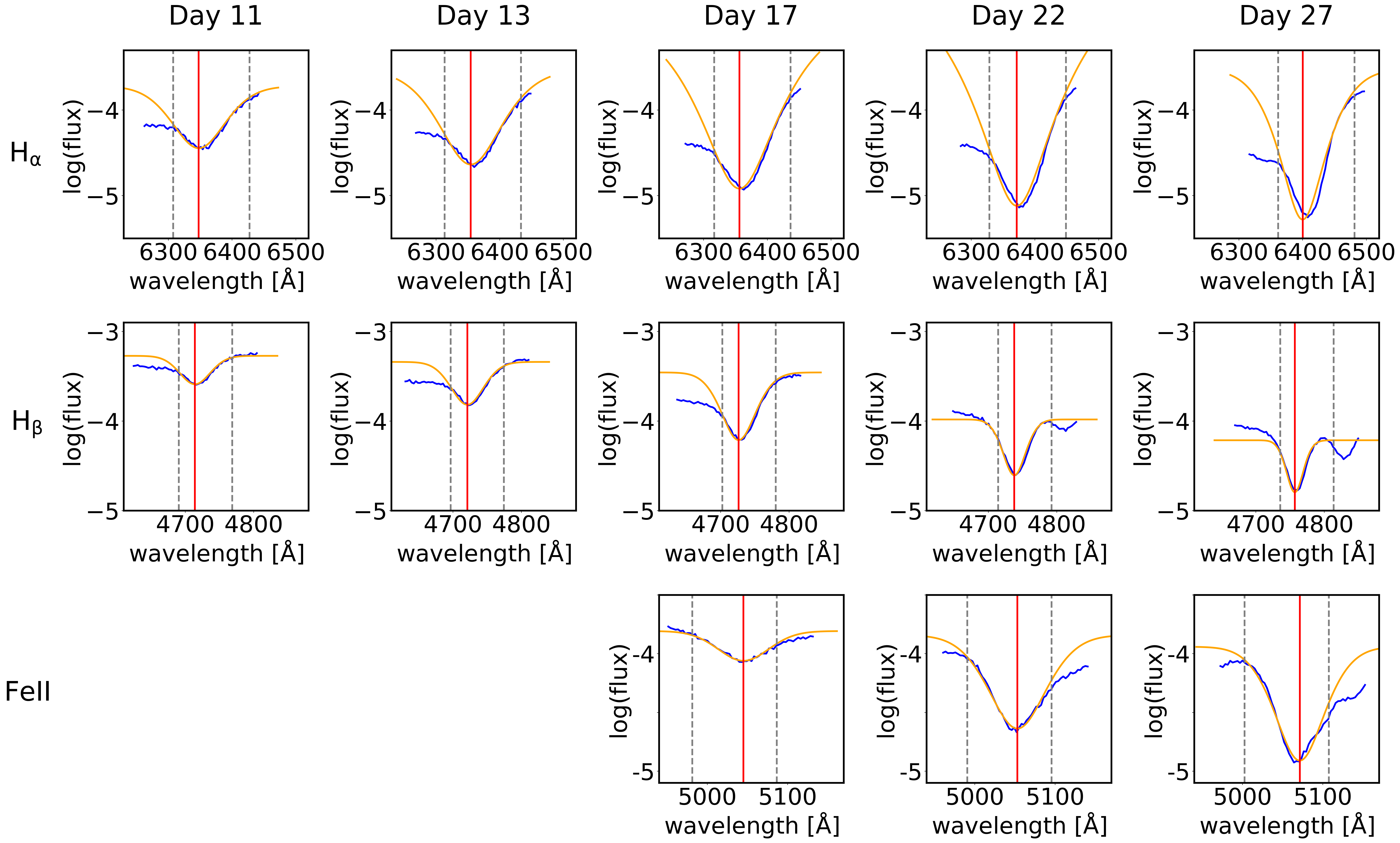}}\
\caption{\label{non_micro_fits} Noiseless absorption lines without microlensing in blue for H$\alpha$, H$\beta$, and Fe\,\textsc{ii} including the Gaussian fits in orange used to measure
the minima of the absorption lines, marked by the vertical red line. The vertical dashed lines in gray mark the borders of the fitting range.}
\end{figure*}

We applied a two-step procedure to determine the wavelength of an absorption line from noisy spectra.  In the first step,  
we visually estimated the absorption minimum and selected a range of
wavelength bins around that minimum to refine the estimate through the
function {\tt signal.find\_peaks} of the python module {\tt scipy}
\citep{2020SciPy-NMeth}.  We adopted the output of
{\tt signal.find\_peaks} as $\lambda_{\mathrm{init}}$, the initial wavelength
of the absorption minimum.  In the second step, we fit a Gaussian to
the absorption line profile to further refine the wavelength
measurement, especially in the presence of noise. As the lines have individual asymmetrical P-Cygni profiles, different fitting ranges were
selected for each feature. For H$\alpha$ we chose 15 bins toward smaller wavelengths and 25 bins toward
longer wavelengths, for H$\beta$ 8 bins toward smaller and 18 bins toward longer wavelengths, and for Fe\,\textsc{ii} 20 bins toward smaller and 15 toward longer wavelengths. The bins are counted from $\lambda_{\mathrm{init}}$, and each bin has a width of $3\AA$.
The 
success of the fitting is dependent on the fitting range. 
If the range is too small, then the fit might fail. Noisy spectra in particularly  can sometimes have
two small minima within the absorption feature. If the range is
too large, then the fit might shift to longer wavelengths as a result 
of the asymmetric P-Cygni line shape.
The optimal wavelength range to use for the Gaussian fit was
determined through simulations of noisy spectra. We applied different fit windows onto the noisy spectra and chose the window that best reproduced the known absorption minimum of the model spectrum.\footnote{By considering various fitting ranges, we estimate that different fitting ranges close to our chosen range could change the uncertainties of the inferred time delays by $\lesssim0.2$ days.}
The result of the Gaussian fit yields the wavelength of the absorption
line, as detailed below.

The Gaussian fit was performed using the {\tt optimize.curve\_fit} function of the python
module {\tt scipy}, a least-squares minimization using a Levenberg-Marquardt optimizer with a fitting function:
\begin{equation}
G(x, A, M, \sigma, h) = \left(-\frac{A^{2} }{\sigma \sqrt{2 \pi}} \exp\left(-\frac{(x - M)^{2}}{2 \sigma^{2}}\right)\right) + h 
.\end{equation}
The input $x$ is the wavelength range over which the fit is computed. The parameter $A$ is the square
root of the amplitude, $M$ the mean of the Gaussian, $\sigma$ the standard deviation, and $h$ the offset.
If the fitting failed, which is indicated by a determined minimum that
is larger than 
$\lambda_{\mathrm{init}} + 60\,\text{\AA}$ (outside the fitting range) or smaller than
$\lambda_{\mathrm{init}} - 60\,\text{\AA}$, the fitting was
repeated with a fit range shifted by five bins, which equals 15 $\text{\AA}$, to longer wavelengths. In these few cases, it is the effect of microlensing that deforms the spectral feature and shifts it to the red or blue once the SN crosses a caustic. The shift to redder wavelengths cannot be compensated for by the fitting method itself, making this additional calculation necessary. In Fig. \ref{non_micro_fits} examples of fits
to the non-microlensed absorption lines are shown. We see in these plots that some fits, especially for later times, tend to the bluer or redder part of the absorption line due
to the asymmetrical line profile and fitting range. As this shift is always in the same direction, this does not affect the following phase retrieval method.

\begin{figure*}[hbt!]
\centering
\subfigure{\label{temp_wave_legend}\includegraphics[width=\textwidth]{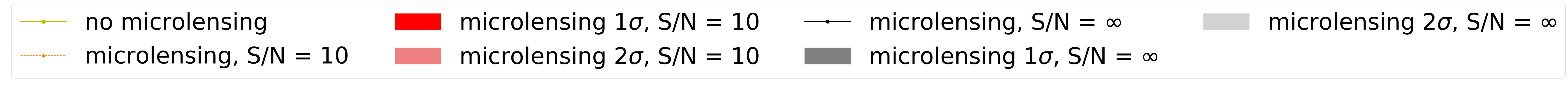}}\hfill
\subfigure{\label{temp_wave_min_H_alpha_1}\includegraphics[width=0.49\textwidth]{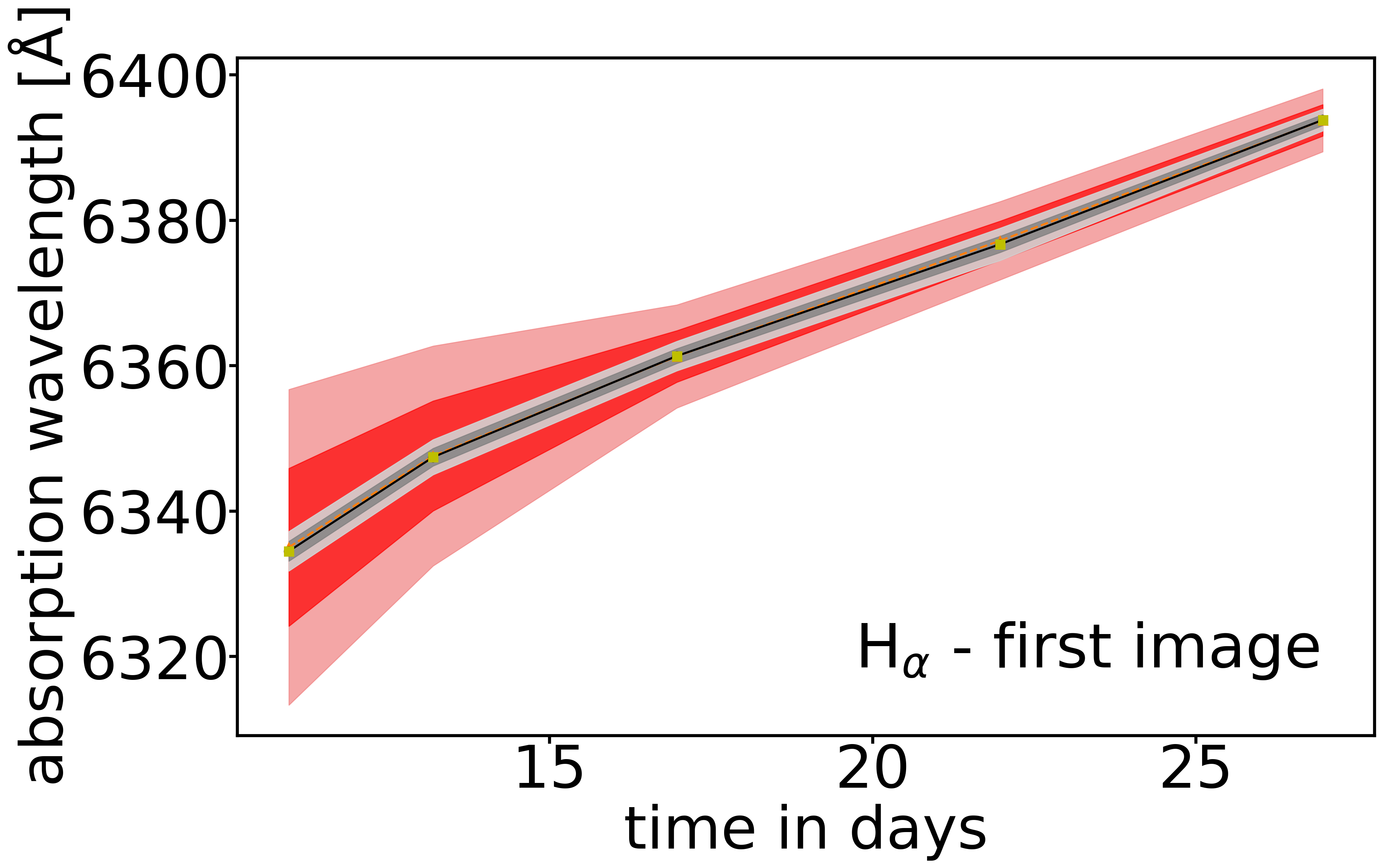}}\hfill
\subfigure{\label{temp_wave_min_H_alpha_2}\includegraphics[width=0.49\textwidth]{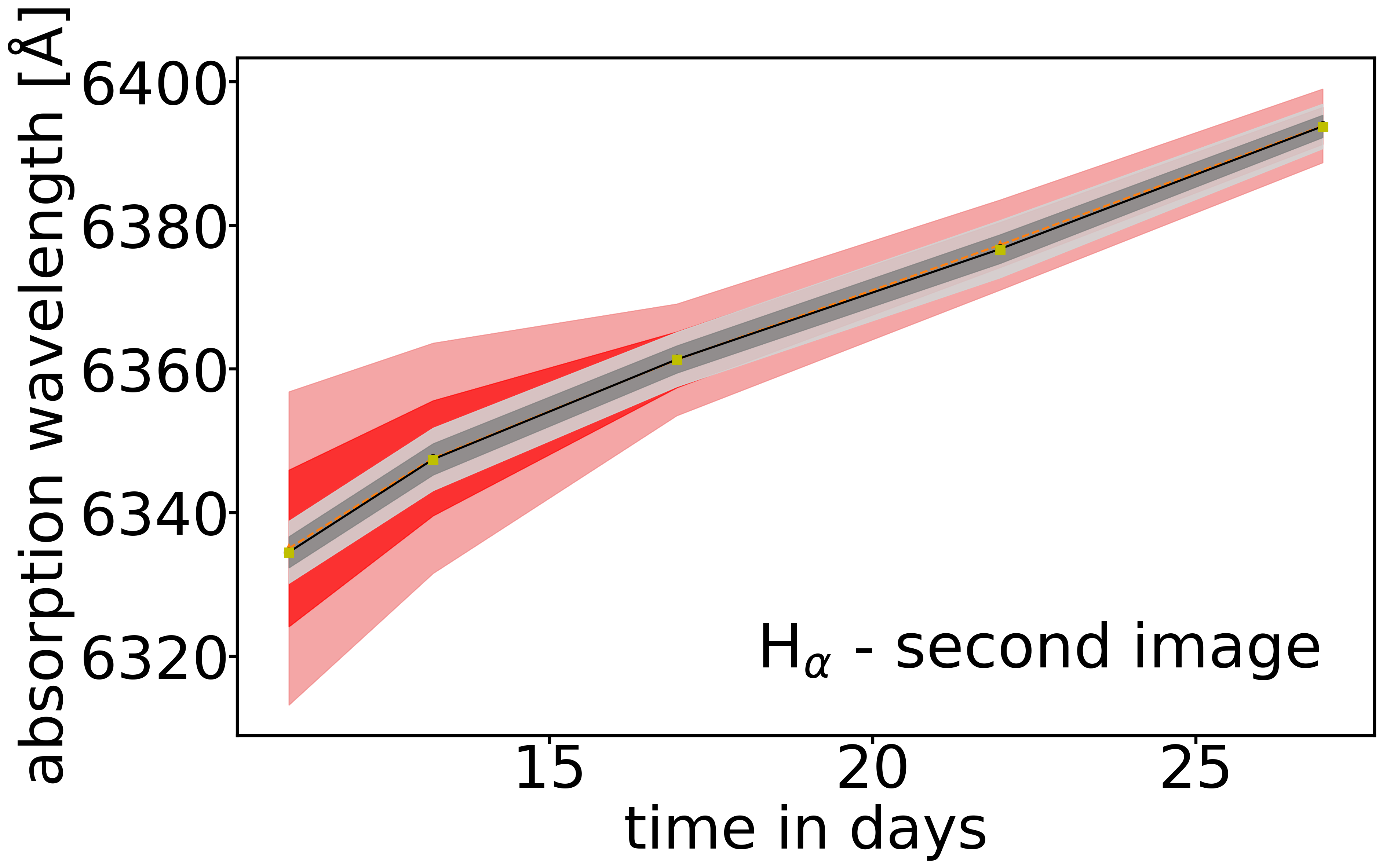}}\\
\subfigure{\label{temp_wave_min_H_beta_1}\includegraphics[width=0.49\textwidth]{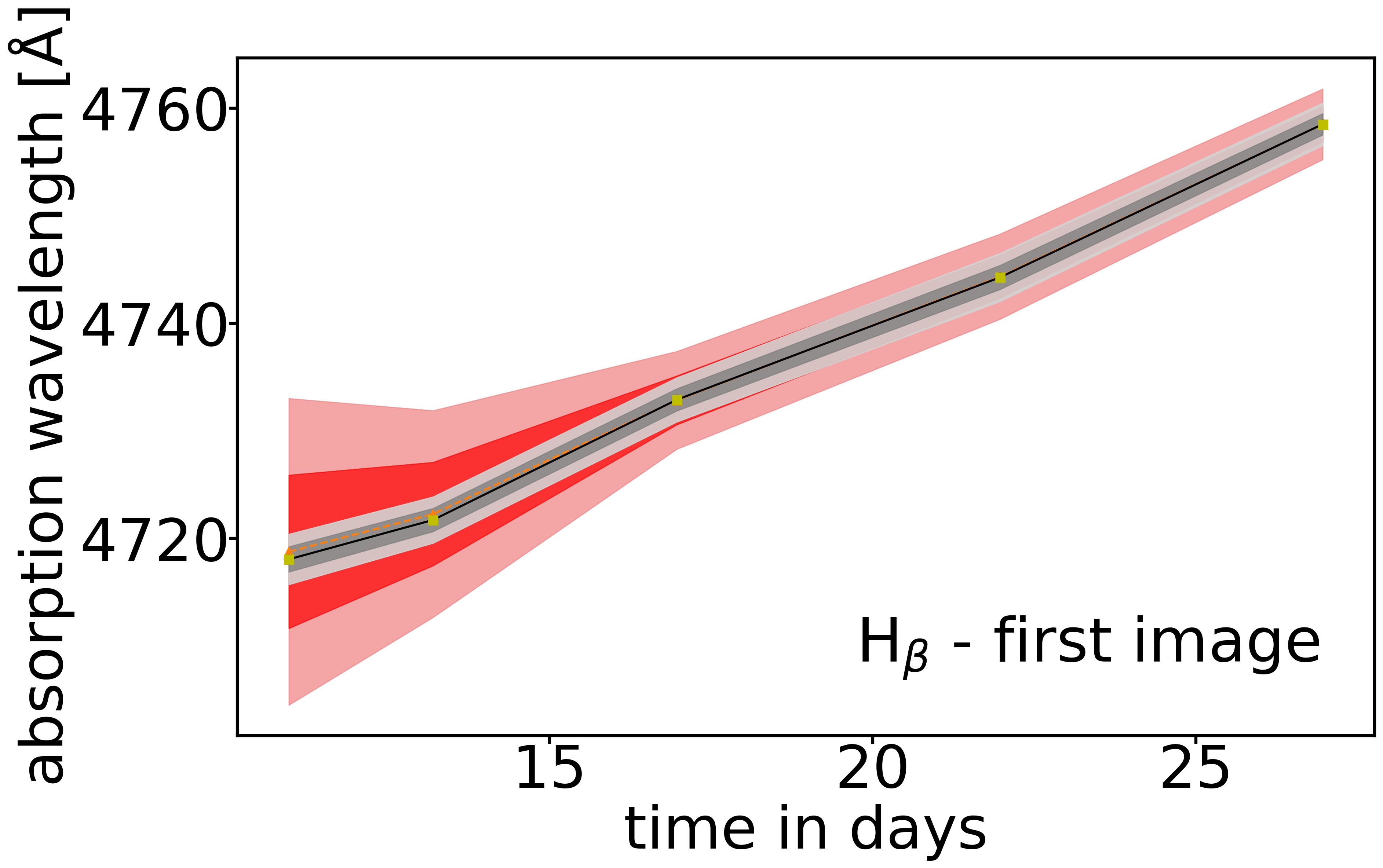}}\hfill
\subfigure{\label{temp_wave_min_H_beta_2}\includegraphics[width=0.49\textwidth]{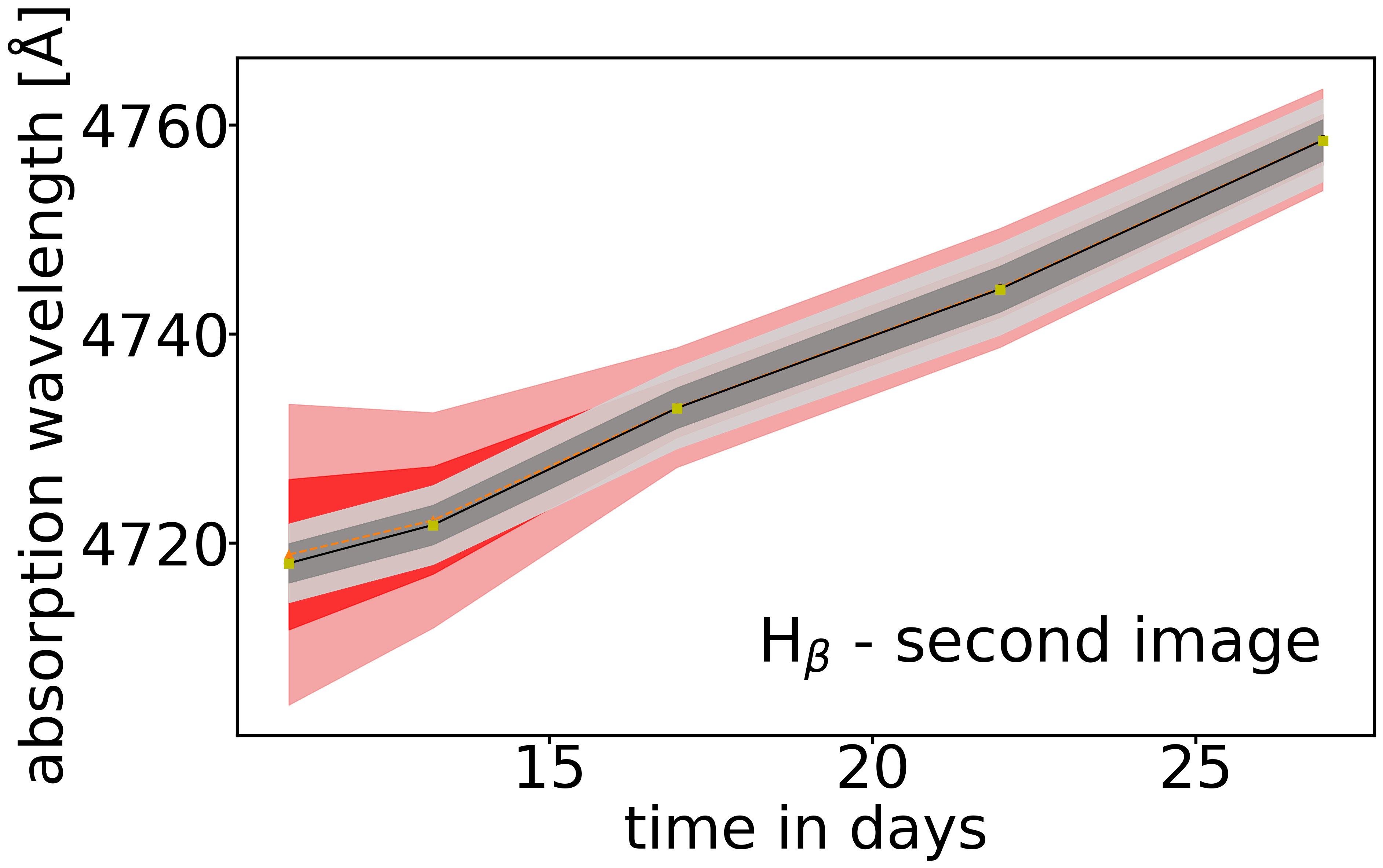}}\\
\subfigure{\label{temp_wave_min_FeII_1}\includegraphics[width=0.49\textwidth]{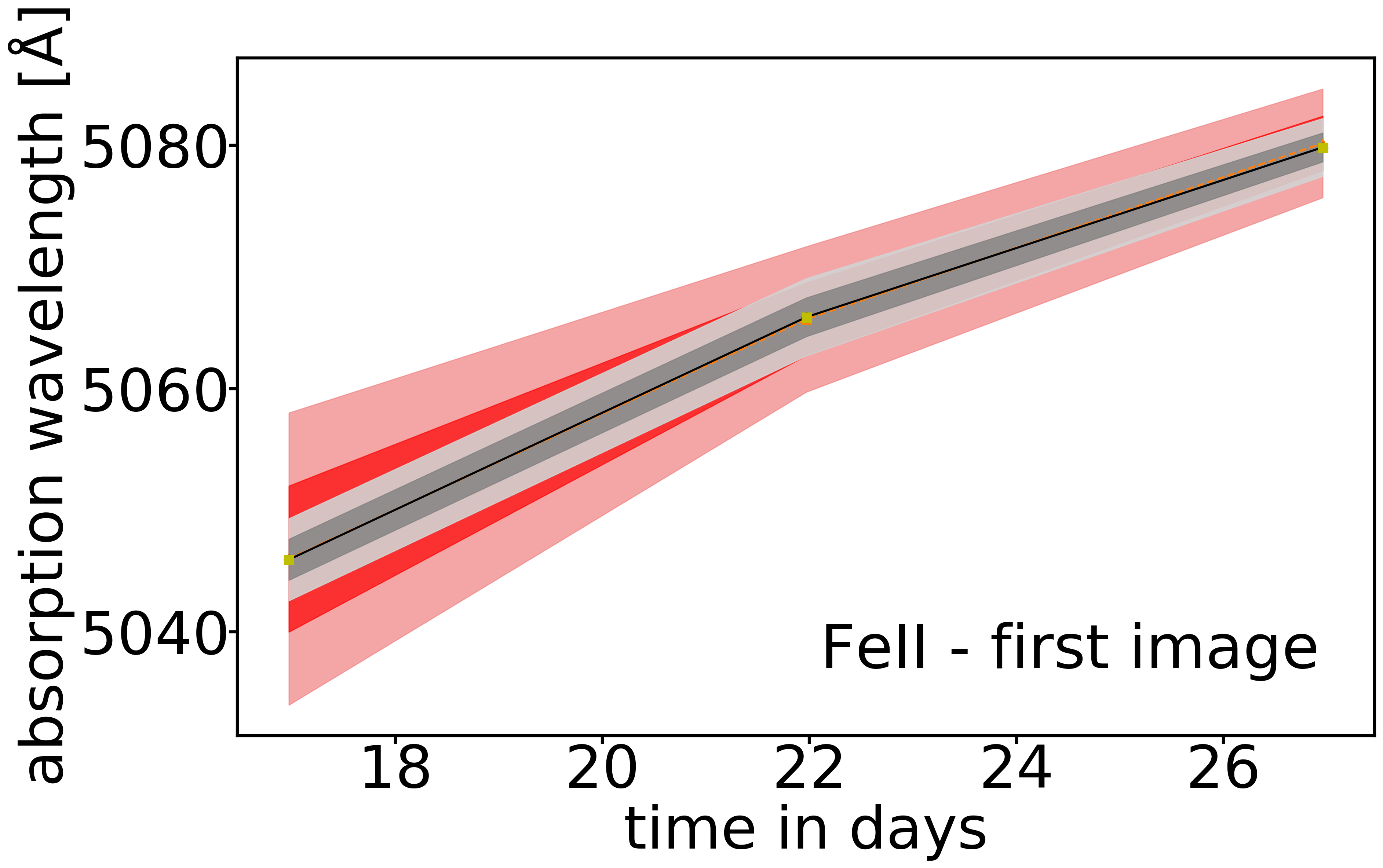}}\hfill
\subfigure{\label{temp_wave_min_FeII_2}\includegraphics[width=0.49\textwidth]{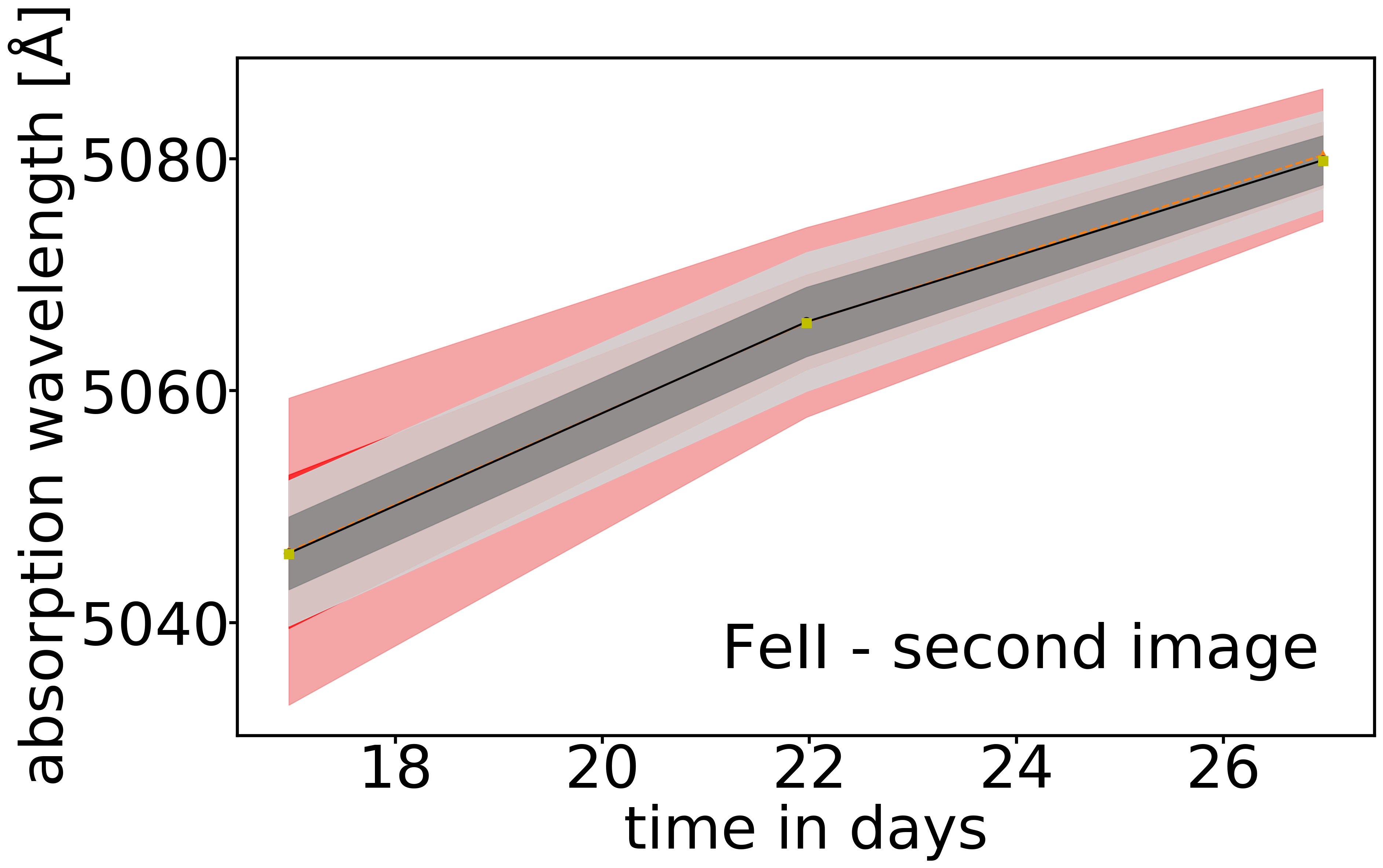}}\
\caption{\label{temp_wave_min} Temporal evolution of the absorption
  minima of the H$\alpha$, H$\beta$, and Fe\,\textsc{ii}
  lines for non-microlensed and microlensed spectra using the two
  different magnification maps of Fig. \ref{micromaps} (left: Fig.
  \ref{micromap1}; right: Fig. \ref{micromap2}).
  For the microlensed
  spectra without noise (labeled S/N = $\infty$), the median of the 10000 positions
  and the 1$\sigma$ and 2$\sigma$ ranges
  are shown in black and gray, respectively. The temporal evolution of the absorption lines
  with added noise of $S/N = 10$ is shown in orange and red.}
\end{figure*}

\begin{figure*}[hbt!]
\centering
\subfigure[\hbox{$S/N = 10$}]{\label{corner_10}\includegraphics[width=0.48\textwidth]{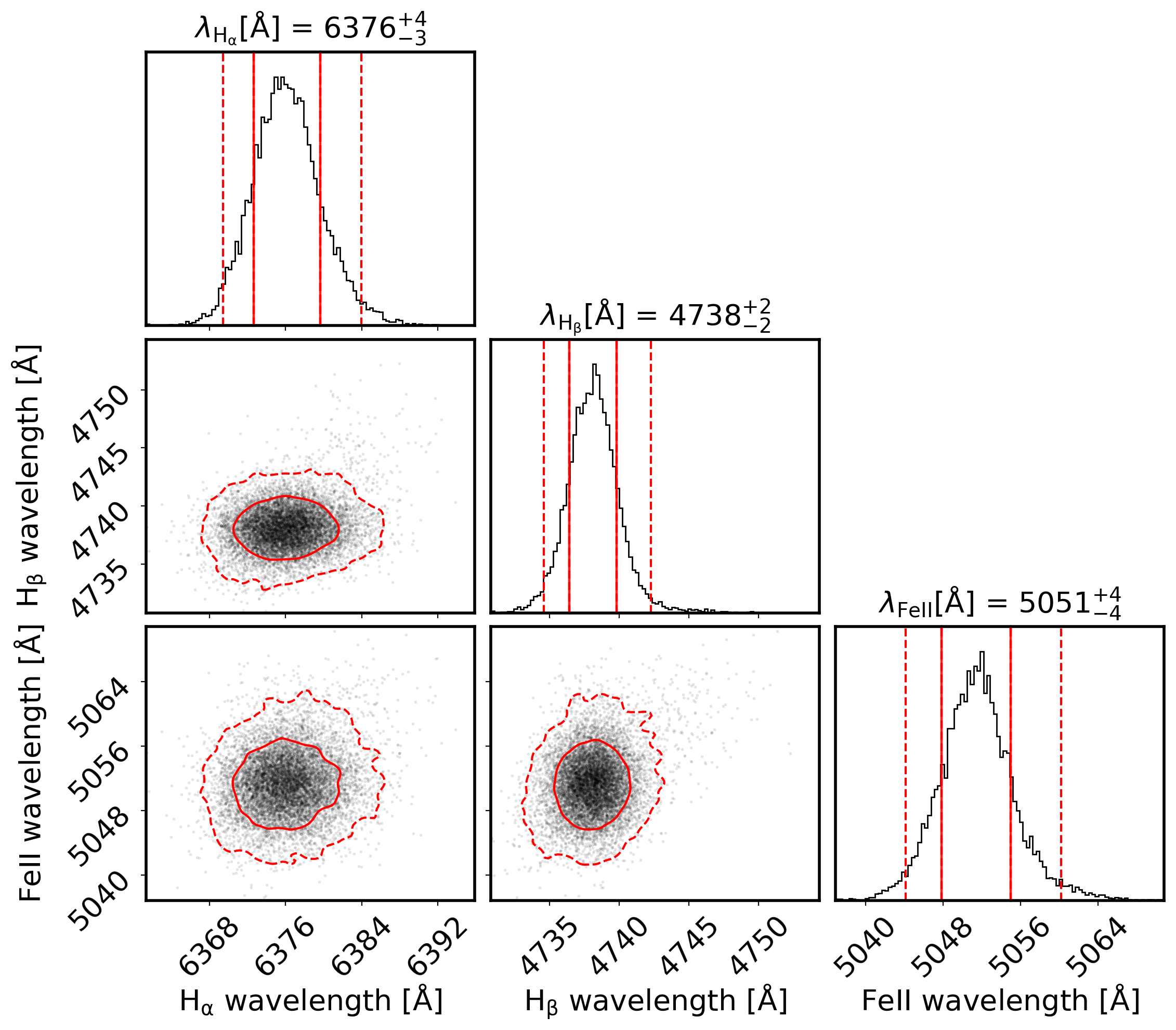}}
\subfigure[\hbox{$S/N = 20$}]{\label{corner_20}\includegraphics[width=0.48\textwidth]{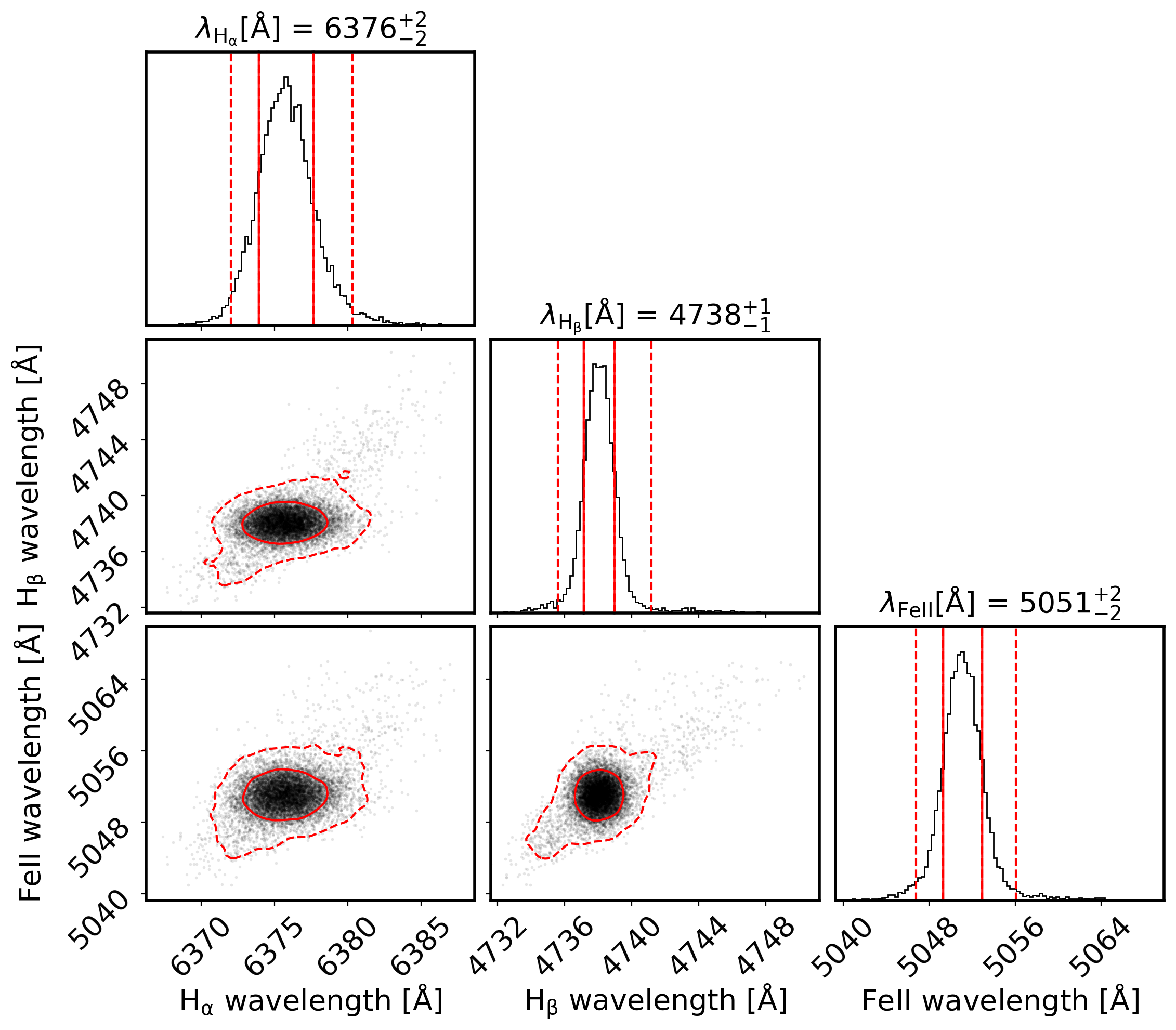}}\\
\subfigure[\hbox{$S/N = 30$}]{\label{corner_30}\includegraphics[width=0.48\textwidth]{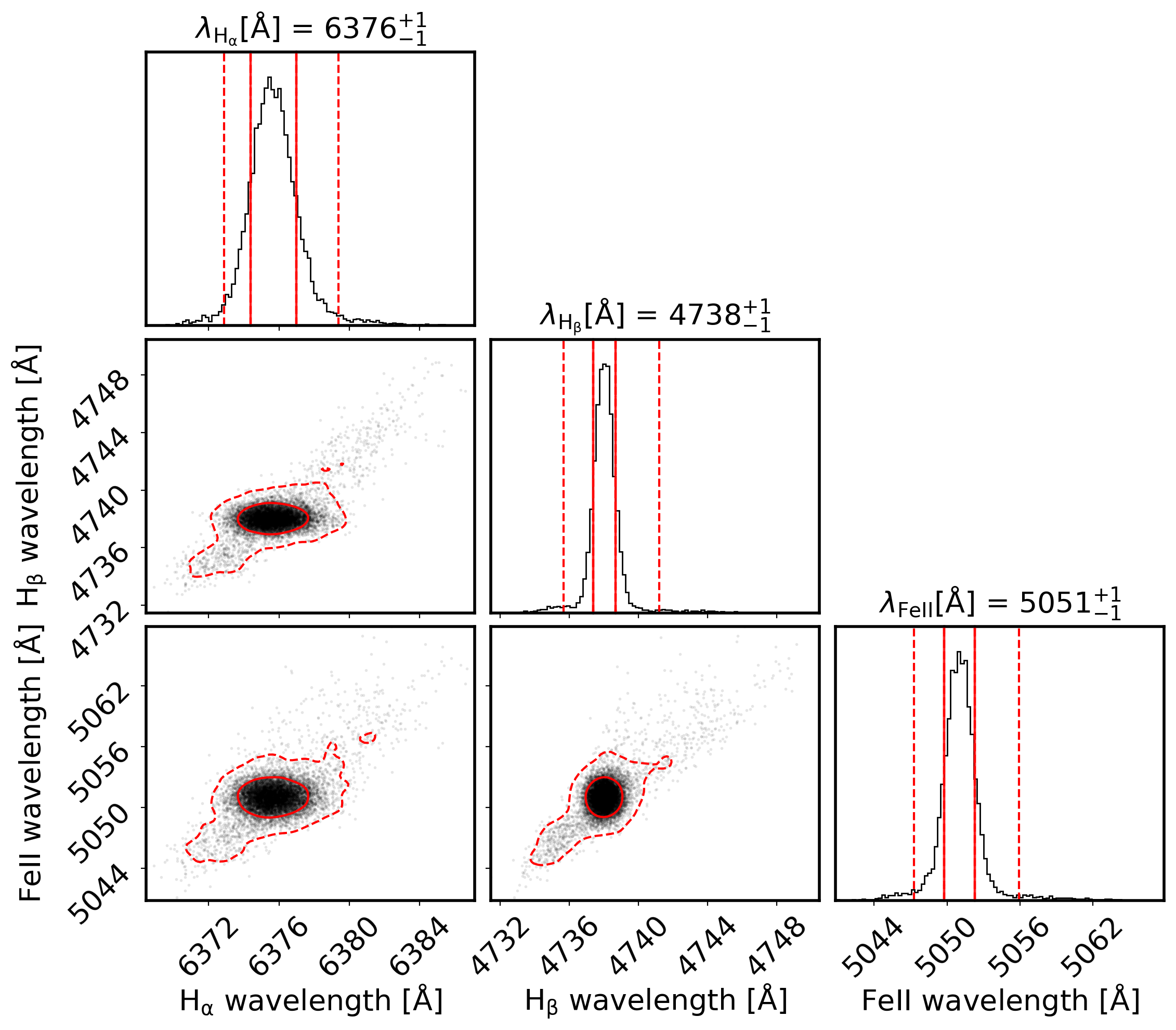}}
\subfigure[\hbox{$S/N = \infty$}]{\label{corner_100}\includegraphics[width=0.48\textwidth]{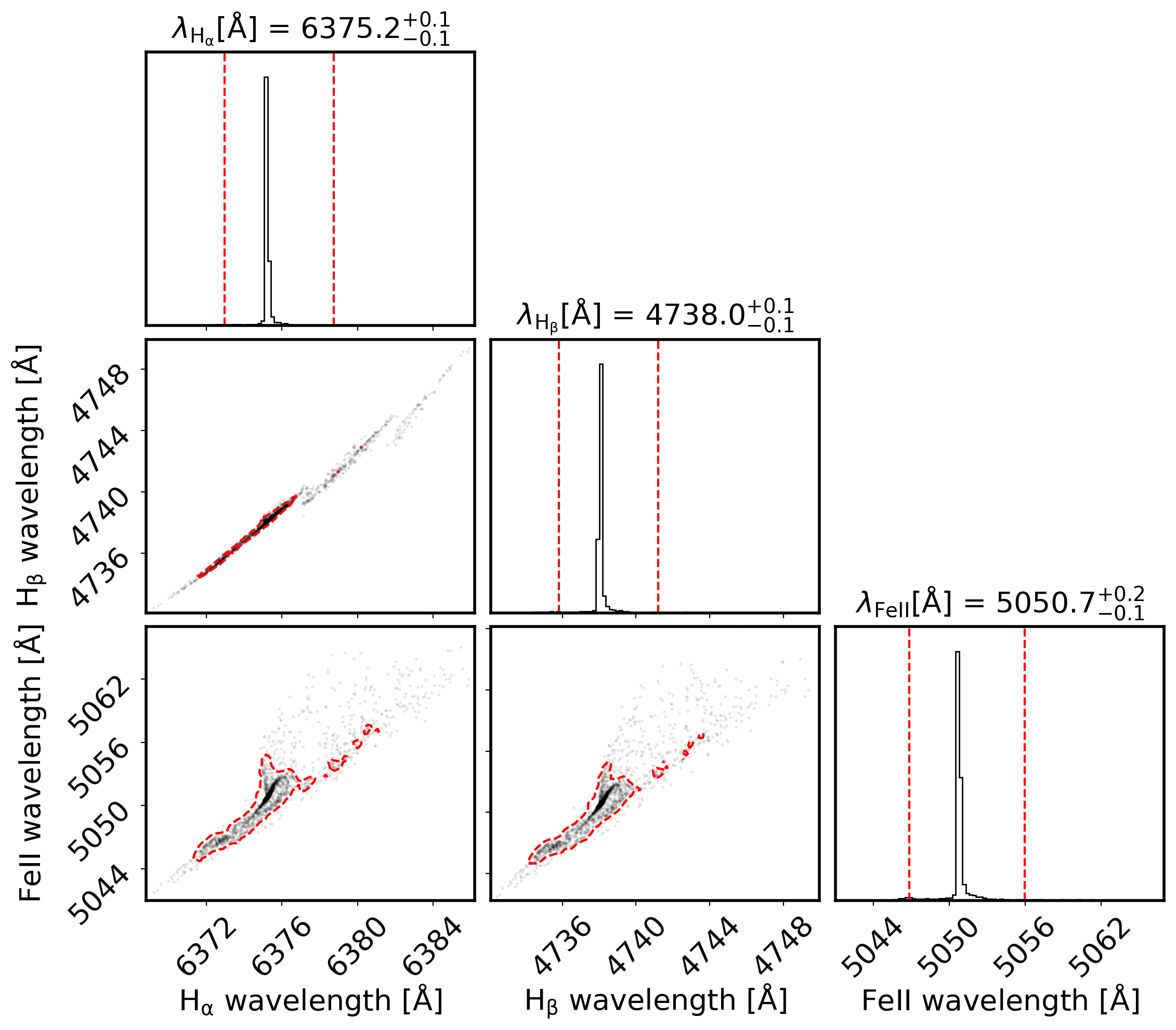}}\
\caption{\label{corner_plots}Correlations between the absorption minima of Fe\,\textsc{ii}, H$\alpha$, and H$\beta$ at day 22 for the four investigated $S/N$: 10, 20, 30, and $\infty$.  The solid 1$\sigma$ contours in the 2D histograms contain 68\% of the data and the dashed 2$\sigma$ contours 95\%. The 1D histograms contain the 1$\sigma$ (vertical solid lines) and the 2$\sigma$ (vertical dashed lines) ranges. For $S/N = \infty$ we only show the contour containing 95\% of the data and the 2$\sigma$ range. Above the 1D histograms the mean wavelength of the absorption minimum with 1$\sigma$ uncertainties is indicated.
  While there are significant correlations in the absorption minima of the different lines in the case of noiseless spectra, the correlations become weak or negligible with noisy spectra.}
\end{figure*}

The fitting was performed for every position in the microlensing map, for
each of the five epochs and for each absorption line in the
microlensed spectra without noise. The temporal evolution of the
absorption minima of H$\alpha$, H$\beta$, and
Fe\,\textsc{ii} is shown in Fig. \ref{temp_wave_min} for the two magnification
maps of Fig. \ref{micromaps}.
The plots include the median of all
determined minima of the positions in the microlensing map and the 1$\sigma$ and
2$\sigma$ uncertainties.  The temporal evolution of the
wavelengths of the absorption minima of the spectra without
microlensing is shown in yellow for reference.  The evolution is similar for the two images, but
microlensing tends to introduce larger uncertainties for the second
lensing image.
In addition to the noise-free case, we repeated the calculations
assuming a S/N per pixel of 10 for the spectra.
To simulate a noisy spectrum, we added Gaussian noise to the flux of each
wavelength bin of the microlensed spectrum; in other words, we drew a random
Gaussian number with standard deviation given by $\sigma_{\mathrm{G}}$ =
flux/(S/N) and added it to the flux. We simulated a noisy microlensed
spectrum for each position in the microlensing map and applied a Savitzky-Golay filter to make the fitting procedure easier. We obtained the
distribution of the absorption minima shown in red in Fig.
\ref{temp_wave_min}.  In all panels of Fig.
\ref{temp_wave_min} (particularly the top panels for H$\alpha$ images), the uncertainties due to microlensing (gray bands) are
smaller than the uncertainties due to noise in the spectra (red bands)
when $S/N=10$.\\
\noindent\hspace*{4mm} Furthermore, we investigated the correlation between the absorption minima of the three lines for different S/N in Fig. \ref{corner_plots}. We show the plots for $S/N = 10$, 20, 30, and $\infty$ at day 22. We consider these three additional S/N values besides the noiseless case as feasible for future observations. We also use these S/N and day 22 for the phase retrieval in the following section. 
In contrast to the noisy cases, the noiseless case has a very tight data distribution, and therefore we only show the 2$\sigma$ contour.
For $S/N = \infty$, the absorption minima of H$\alpha$, H$\beta$, and Fe\,\textsc{ii} are strongly correlated within the 1$\sigma$ and 2$\sigma$ contours. For $S/N = 30$, the 1$\sigma$ contour already shows no correlation, whereas in the 2$\sigma$ contour slight correlations are still present. As the S/N gets smaller, the 1$\sigma$ stays uncorrelated and the 2$\sigma$ contour become increasingly more uncorrelated.

\section{Supernova phase inference from spectra and time-delay measurement}
\label{sec: SN phase inference from spectra}

\begin{figure*}[hbt!]
\centering
\subfigure[\hbox{Retrieved phase using the absorption line of Fe\,\textsc{ii}}]{\label{phase_FeII}\includegraphics[width=0.48\textwidth]{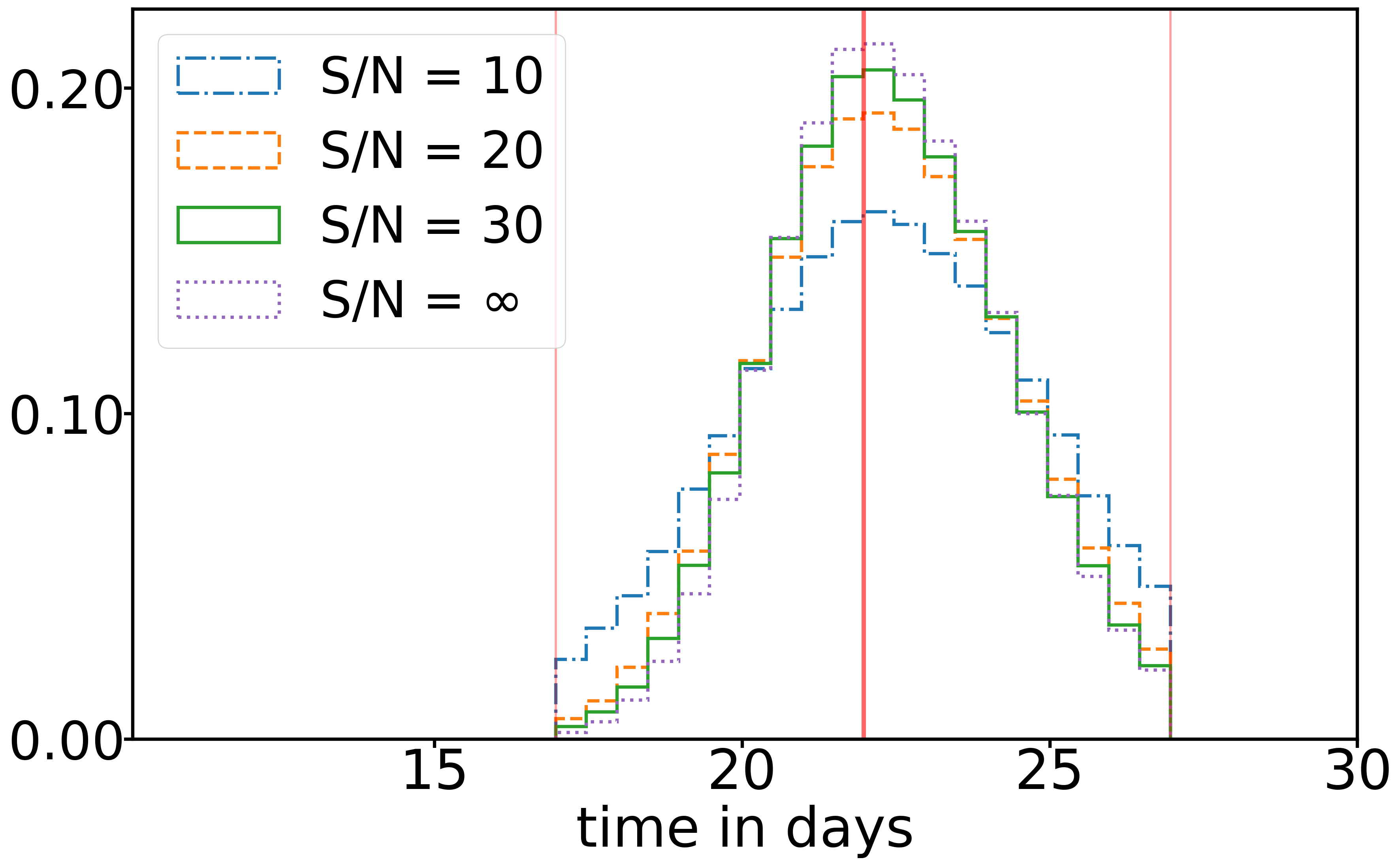}}
\subfigure[\hbox{Retrieved phase using the absorption line of H$\alpha$}]{\label{phase_H_alpha}\includegraphics[width=0.48\textwidth]{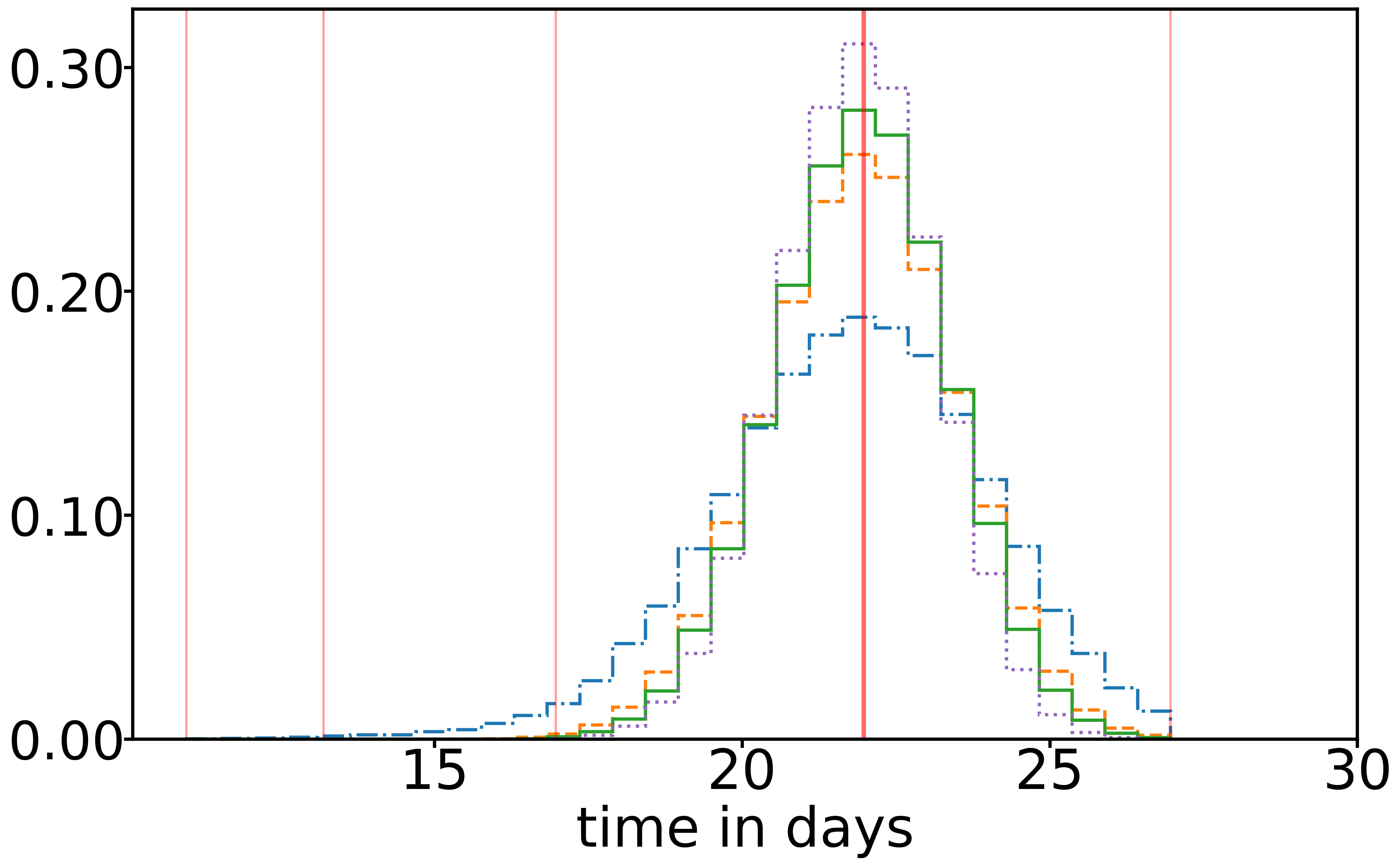}}\\
\subfigure[\hbox{Retrieved phase using the absorption line of H$\beta$}]{\label{phase_H_beta}\includegraphics[width=0.48\textwidth]{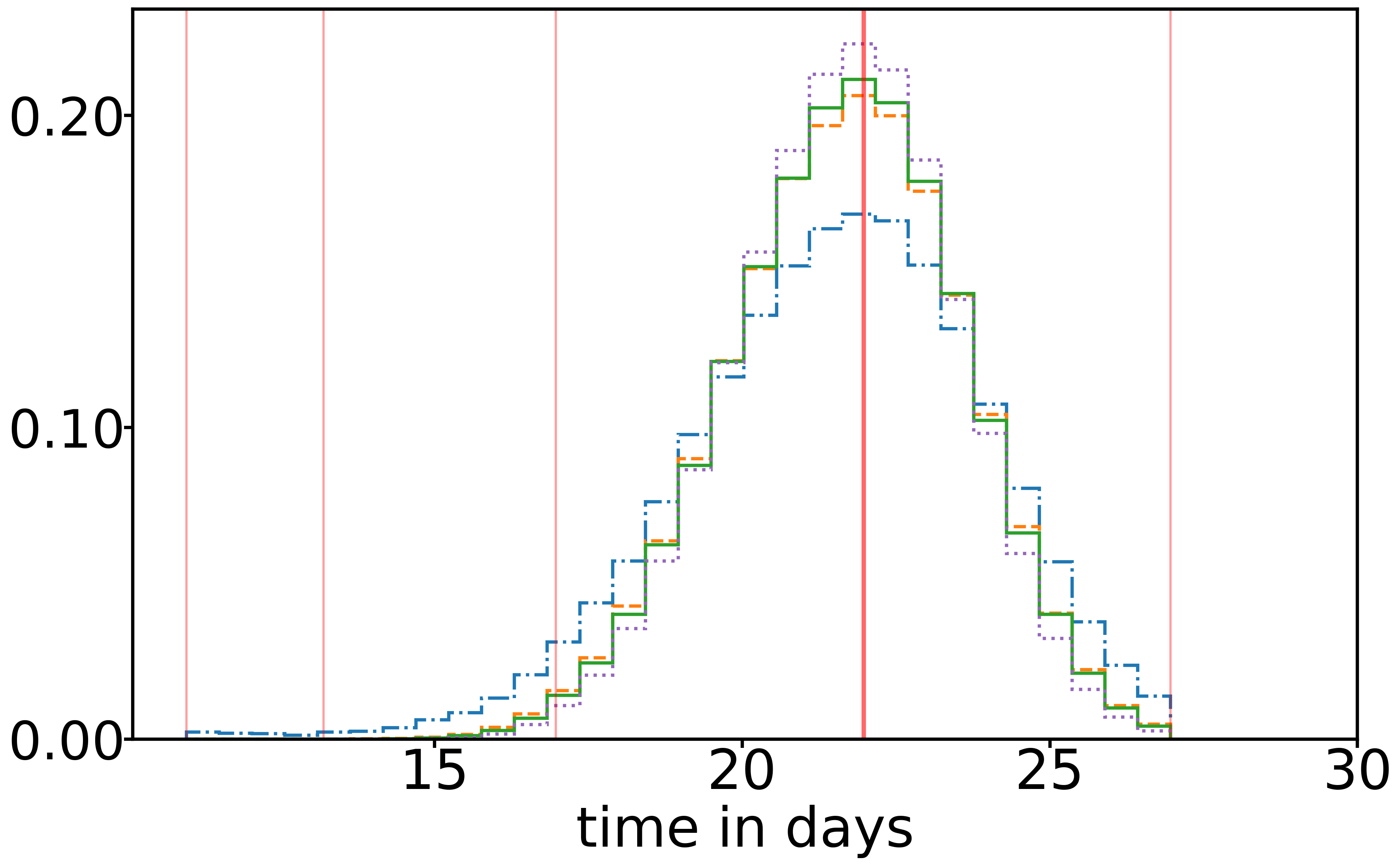}}
\subfigure[\hbox{Retrieved phase combining all three absorption lines}]{\label{phase_combined}\includegraphics[width=0.48\textwidth]{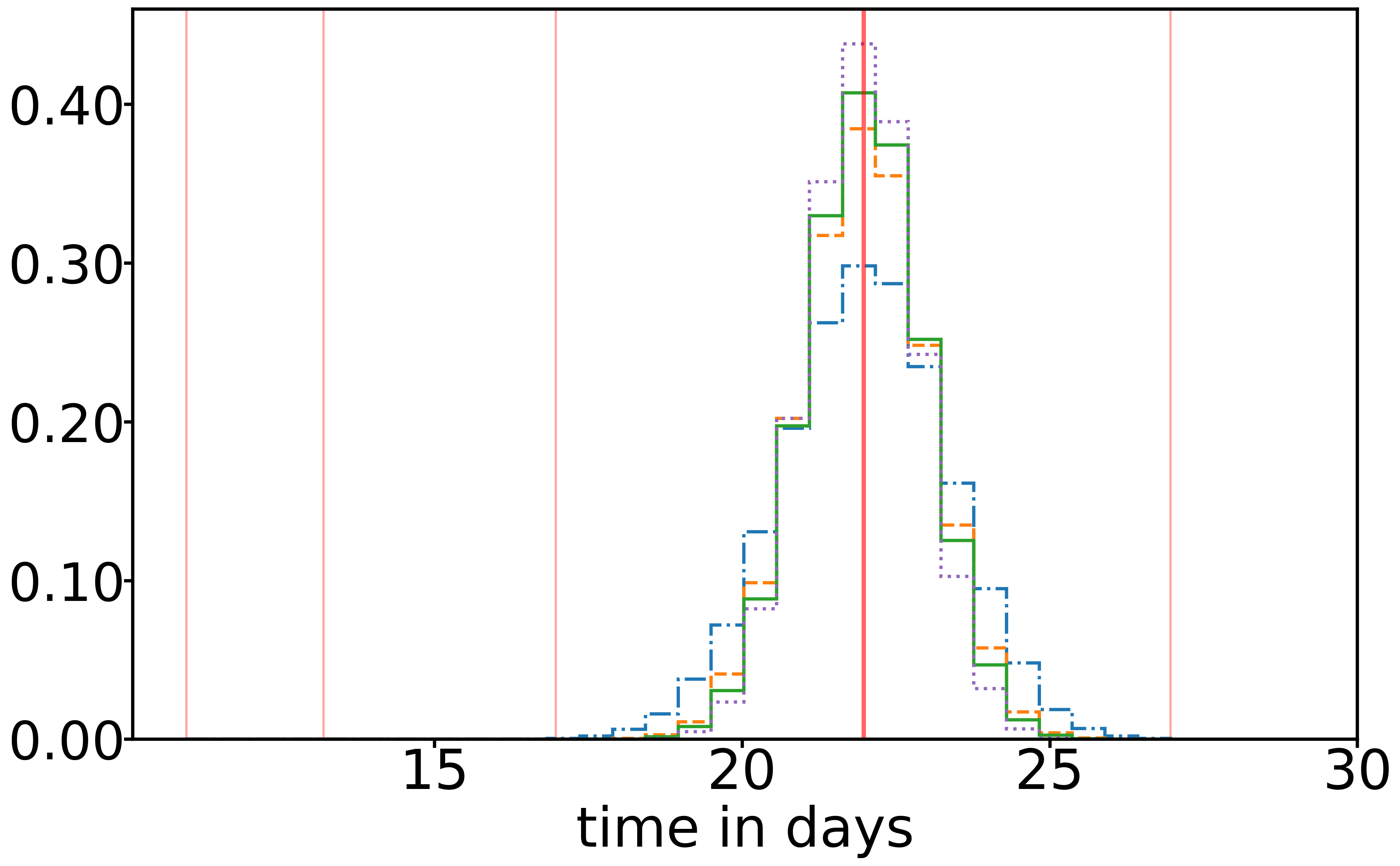}}\
\caption{\label{phases} Histograms of the retrieved phases of the
  second SN image using the absorption lines Fe\,\textsc{ii},
  H$\alpha$, and H$\beta$. The phase inference
  is done for $S/N = 10$, 20, and 30 and for noiseless spectra. The red
  vertical lines in the histograms indicate the epochs of the five
  available spectra of the first SN image. The thick vertical line
  marks the input value. By using the wavelengths of absorption
  features, the phase of the second SN image is correctly recovered.}
\end{figure*}

In this section we investigate the possibility of inferring phases from
the wavelengths of the absorption minima of the spectral lines.
The basic idea is based on Fig. \ref{temp_wave_min}, which shows
that the wavelengths of the absorption minima evolve over time.  When
we detect a LSN system based on its first appearing image,
which we denote as SN-A, we can obtain spectra of this SN-A
image at different epochs to determine the evolution of the absorption
features (as shown in Fig. \ref{temp_wave_min}).  A subsequent
spectrum taken at time $t_{\mathrm{B}}$ of the next appearing SN image,
which we denote as SN-B, can then be used to infer the phase of SN-B based on the measured wavelength (i.e., the time $t_{\mathrm{A}}$ at which
image SN-A had the same absorption wavelength value as that measured
in the SN-B spectrum).  The difference between these two times, $t_{\mathrm{B}}-t_{\mathrm{A}}$, then gives the time delay between the SN-B and
SN-A images.  Below, we describe in more detail how we quantitatively
inferred the phase and time delays, taking microlensing
and noise in the spectrum into account.

Here we consider the scenario where we have obtained spectra of SN-A at five
epochs (as in the previous sections), denoted by $t_i$, where
$i=1,2,...,N_{\mathrm{epoch}}$ and $N_{\mathrm{epoch}} = 5$. The spectra have
the same S/N. Through the microlensing simulations
described in the last two sections, we obtained the distribution of
absorption line wavelengths, accounting for microlensing and noise as shown in
the left panels of Fig. \ref{temp_wave_min}.  In particular, for the
microlensing of the SN-A image, we adopted the magnification map with
$\kappa$ = 0.36 and $\gamma$ = 0.35 since the first
appearing SN-A image must be a time-delay minimum image.
By sampling
$10000$ positions on the microlensing map and thus simulating 10000
noisy microlensed spectra for the given S/N at each $t_i$, we obtained the distribution of
absorption line wavelengths $\lambda(t_i)$ following the absorption
line fitting method described in Sect. \ref{sec: SNe II spectra and
  absorption features}.

We approximated the distribution of $\lambda(t_i)$ at each epoch as a
Gaussian with mean $\lambda_{{\mathrm{d}},i}$ and standard deviation
$\sigma_{{\mathrm{d}},i}$.  To obtain a continuous distribution of
$\lambda_{\mathrm{d}}(t)$ and $\sigma_{\mathrm{d}}(t)$ for any time $t$, we
linearly interpolated between the discrete $\lambda_{{\mathrm{d}},i}$ and
$\sigma_{{\mathrm{d}},i}$ values.  This allows us to describe
the likelihood of the data $d_{\mathrm{A}}$ (i.e., the spectra and microlensing
maps of SN-A) for a given time $t$ and absorption wavelength $\lambda$ as
\begin{equation}
\label{prob}
P(d_{\mathrm{A}} | \lambda, t) = C_{\mathrm{A}} \mathrm{exp} \left( \frac{-(\lambda - \lambda_{\mathrm{d}}(t))^{2}}{2 \sigma_{\mathrm{d}}(t)^{2}} \right),
\end{equation}
where $C_{\mathrm{A}}$ is the normalization constant. 

Next we consider a spectrum taken of the SN-B image at $t_{\mathrm{B}}$.
Similar to the procedure described above for the SN-A image, we repeated the microlensing simulations to produce 10000
microlensed spectra of SN-B for a given S/N to account for
uncertainties associated with microlensing and noise in the spectrum.
The main difference from the procedure of SN-A is that the microlensing
map for SN-B is for the time-delay saddle image with $\kappa$ = 0.70,
$\gamma$ = 0.70.  This is appropriate since at least
one of the next appearing SN images is a time-delay saddle. Furthermore,
considering the saddle image allows us to probe uncertainties due to
both types of lensing images (time-delay minimum and time-delay saddle
images) given that the microlensing maps are sensitive to the image
types.  From the distribution of 10000 microlensed noisy realizations
of the SN-B single epoch spectrum, we can measure the mean wavelength
$\lambda_{\mathrm{B}}$ and standard deviation $\sigma_{\mathrm{B}}$.  We used the wavelength measurement $\lambda_{\mathrm{B}}\pm\sigma_{\mathrm{B}}$ to
infer the times in SN-A that are consistent with these values.

Quantitatively, we are interested in obtaining the probability distribution
of time $t$ of SN-A given the data $d_{\mathrm{A}}$ and measurement $d_{\mathrm{B}}$
(of $\lambda_{\mathrm{B}}\pm\sigma_{\mathrm{B}}$) -- that is, $P(t | d_{\mathrm{A}}, d_{\mathrm{B}})$ --
which can be expressed by
\be
\label{eq:Pt}
P(t | d_{\mathrm{A}},d_{\mathrm{B}}) = \int P(t, \lambda | d_{\mathrm{A}}, d_{\mathrm{B}}) {\mathrm{d}}\lambda,
\ee
where $\lambda$ are the absorption line wavelengths associated with SN-A.

Using Bayes' rule, we can rewrite the integrand as:
\be
\label{eq:bayes}
P(t, \lambda | d_{\mathrm{A}}, d_{\mathrm{B}}) = \frac{P(d_{\mathrm{A}} | \lambda, t) P(\lambda | d_{\mathrm{B}}) P(t) }{P(d_{\mathrm{A}})},
\ee
where we have separated dependences ($d_{\mathrm{A}}$ and $t$ do not depend on
$d_{\mathrm{B}}$).  The expression for $P(d_{\mathrm{A}} | \lambda, t) $ is given by 
Eq. (\ref{prob}). Through our Gaussian approximation of the measurement 
$d_{\mathrm{B}}$, we can write
\be
\label{eq:Plambda}
P(\lambda | d_{\mathrm{B}}) = \frac{1}{\sqrt{2 \pi} \sigma_{\mathrm{B}}} \cdot \mathrm{exp}\left( -\frac{(\lambda - \lambda_{\mathrm{B}})^{2}}{2 \sigma_{\mathrm{B}}^{2}} \right).
\ee
Given that our data $d_{\mathrm{A}}$ is fixed, $P(d_{\mathrm{A}})$ is a
constant.  Assuming a uniform prior on $t$ (i.e., $P(t)$ is constant), we
can simplify Eq. (\ref{eq:Pt}) to
\be
\label{eq:Pt_final}
P(t | d_{\mathrm{A}}, d_{\mathrm{B}}) = C_{\mathrm{A}\mathrm{B}} \int P(d_{\mathrm{A}}|\lambda,t) P (\lambda | d_{\mathrm{B}}) {\mathrm{d}}\lambda,
\ee
where $C_{\mathrm{A}\mathrm{B}}$ is a constant.

\begin{table*}[hbt!]
\centering
\begin{tabular}{|c|c|c|c|c|}
\hline 
$S/N$ & 10 & 20 & 30 & $\infty$ \\ 
\hline 
Fe\,\textsc{ii} & 22.3 $\pm$ 2.3 days & 22.4 $\pm$ 1.9 days & 22.4 $\pm$ 1.9 days & 22.4 $\pm$ 1.8 days \\ 
\hline 
H$\alpha$ & 21.8 $\pm$ 2.2 days & 21.9 $\pm$ 1.5 days & 21.9 $\pm$ 1.4 days & 21.9 $\pm$ 1.3 days \\ 
\hline 
H$\beta$ & 21.5 $\pm$ 2.4 days & 21.7 $\pm$ 1.9 days & 21.7 $\pm$ 1.9 days & 21.7 $\pm$ 1.8 days \\ 
\hline 
Fe\,\textsc{ii} + H$\alpha$ + H$\beta$ & 22.0 $\pm$ 1.3 days & 22.0 $\pm$ 1.0 days & 22.0 $\pm$ 1.0 days & 22.0 $\pm$ 0.9 days \\ 
\hline 
\end{tabular}
\caption{\label{phase_uncertainties}Weighted average and the 1$\sigma$ uncertainties of the
  phase retrievals considering all five epochs of the first image. We list the results for the three absorptions lines (Fe\,\textsc{ii},
  H$\alpha$, and H$\beta$) for the three
  different S/N values (10, 20, and 30) and for the noiseless case, indicated by
  $S/N = \infty$. The last row shows the uncertainty on the phase when
  using all three lines, assuming the noise and the microlensing in these lines are
  uncorrelated.}
\end{table*}

We can compute $P(t | d_{\mathrm{A}}, d_{\mathrm{B}})$ through importance sampling \citep[e.g.,][]{Lewis2002}. 
We first generated samples of
$P(d_{\mathrm{A}}|\lambda,t)$ by (1) randomly drawing a value of $t$, assuming they follow a Gaussian distribution, which
we denote as $t_{\mathrm{samp}}$, (2) computing $\lambda_{\mathrm{d}}(t_{\mathrm{samp}})$ and $\sigma_{\mathrm{d}}(t_{\mathrm{samp}})$, and (3) using these sampled values to draw a
random Gaussian deviate, $\lambda_{\mathrm{samp}}$. We repeated steps 
(1)-(3) to obtain 31981 samples of $(t_{\mathrm{samp}}, \lambda_{\mathrm{samp}})$ (or 19980 samples for the specific case of the Fe\,\textsc{ii} absorption line). The number of samples was the result of setting the amount between the third and fourth epoch to 10000 samples. The number of samples within the other epochs resulted from choosing the same density of samples along the wavelength axis.
Each of these samples was then weighted by $P (\lambda_{\mathrm{samp}} | d_{\mathrm{B}})$ as given by Eq. (\ref{eq:Plambda}).  The
weighted distribution of $t$ finally yields $P(t | d_{\mathrm{A}}, d_{\mathrm{B}})$.

As an example, we applied this procedure with $t_B$ set equal to the fourth epoch at day 22.0 of the SN-A 
image modeled with \textsc{tardis}. We chose this epoch as it is the central 
epoch of the temporal evolution of the Fe\,\textsc{ii} absorption line and enabled us to evaluate all three absorption lines of interest, 
H$\alpha$, H$\beta$, and Fe\,\textsc{ii}. For the noise 
addition, we again selected the three different S/N values of 10, 20, and 30.

The retrieved phases of all three absorption lines are shown in Fig.
\ref{phases} as histograms indicating the distribution of the weighted
samples. 
The input epoch (thick vertical red line) 
is recovered very well by our phase retrieval method. In particular,
assuming a $S/N = 20$ or higher, the histogram shows a sharp peak
centered around the input phase value. We additionally combined the
three separate phase retrievals in Fig.~\ref{phase_combined}. This
assumes the best case scenario with negligible correlation
in the noise and microlensing, which means that we can simply multiply the three
distributions with one another. The assumption of uncorrelated data is justified based on the correlation plots in Fig.~\ref{corner_plots}, which show that in the case of noisy spectra the absorption minima are only very weakly correlated. Only the noiseless spectra show correlations, which we can safely disregard since $S/N = \infty$ is a limiting case, only included in this study to give an upper limit for the accuracy. Additionally, its median and width are very similar to the case of $S/N = 30$, suggesting that the correlation does not matter much for the present work. The peak around the epoch that
should be retrieved is sharper than for the single line
retrievals. 
Another possibility for reducing the uncertainties would be to combine several spectra of the SN-B image, if they are available, but taking possible correlations introduced by microlensing into account.

In Table \ref{phase_uncertainties} we list the weighted average and the corresponding 1$\sigma$ deviations of
the phases retrieved for the different
absorption lines and their combination. For Fe\,\textsc{ii} and H$\beta$ the averages deviate more than the average of H$\alpha$ from the input value of 22.0 days. For H$\alpha$ the input value is recovered within its 1$\sigma$ uncertainties. For all considered S/N, we recover the input value when combining the phase retrieval results of the three absorption lines.

Even though a S/N of 10 already yields a good result for the phase retrieval with uncertainties smaller than 2.5 days, the uncertainties of
$S/N = 20$ and 30 are both smaller than 2 days. The difference in 1$\sigma$ of $S/N = 20$ and 30 is marginal or even negligible, with $S/N = 20$ also being very close
(within 0.2 days in uncertainty) to the noiseless
1$\sigma$ deviation, suggesting
that a S/N of 20 would be sufficient to apply our phase retrieval method.

Our method also does not require very high resolution spectra. Well-calibrated spectra with resolution R $\gtrsim$ 500 are sufficient for determining the relative phase of SN images to fractions of a day since the line profiles of SNe are broad.

To further explore the effect of microlensing on the phase retrieval,
we investigate a scenario with extremely high micro-magnification
using two additional magnification maps in Appendix \ref{sec: Appendix B}.  We find the phase retrieval to be only negligibly affected by microlensing even in this high microlensing scenario since we obtain the same retrieved phases within $0.2\,$days for the original and the high-magnification microlensing scenarios.

\begin{figure*}[hbt!]
\centering
\subfigure[\hbox{Retrieved phase using the absorption line of Fe\,\textsc{ii}}]{\label{phase_FeII_short}\includegraphics[width=0.48\textwidth]{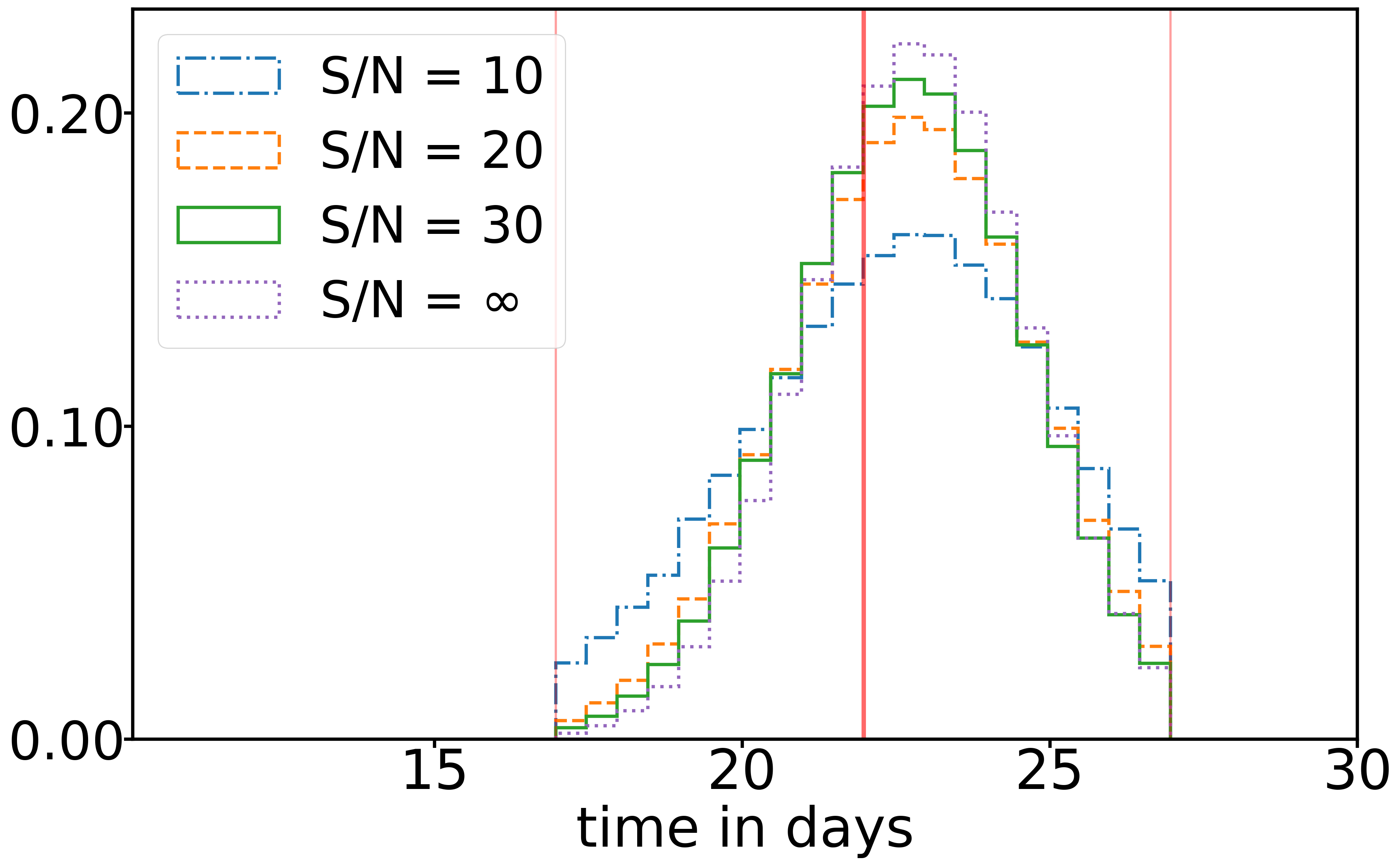}}
\subfigure[\hbox{Retrieved phase using the absorption line of H$\alpha$}]{\label{phase_H_alpha_short}\includegraphics[width=0.48\textwidth]{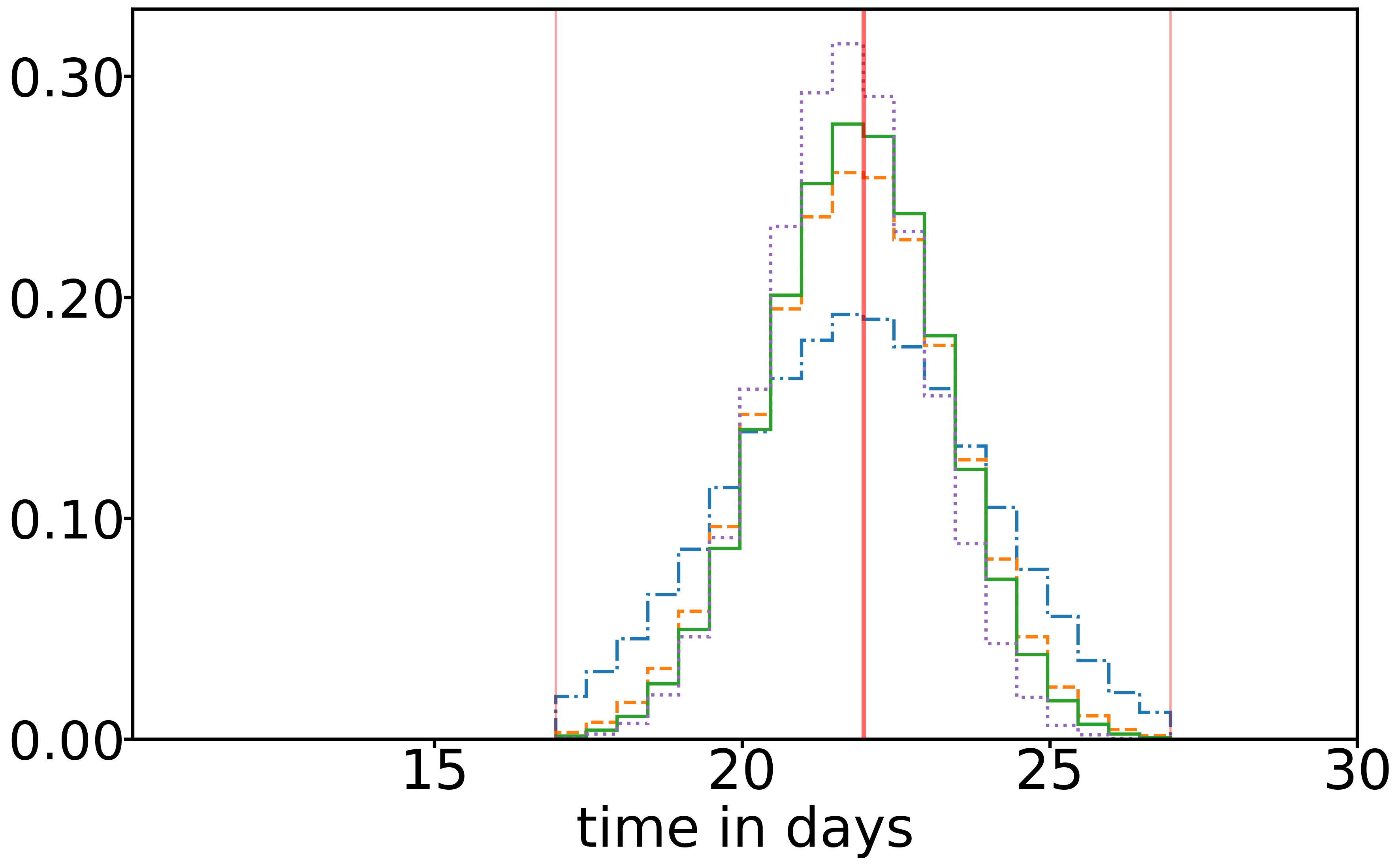}}\\
\subfigure[\hbox{Retrieved phase using the absorption line of H$\beta$}]{\label{phase_H_beta_short}\includegraphics[width=0.48\textwidth]{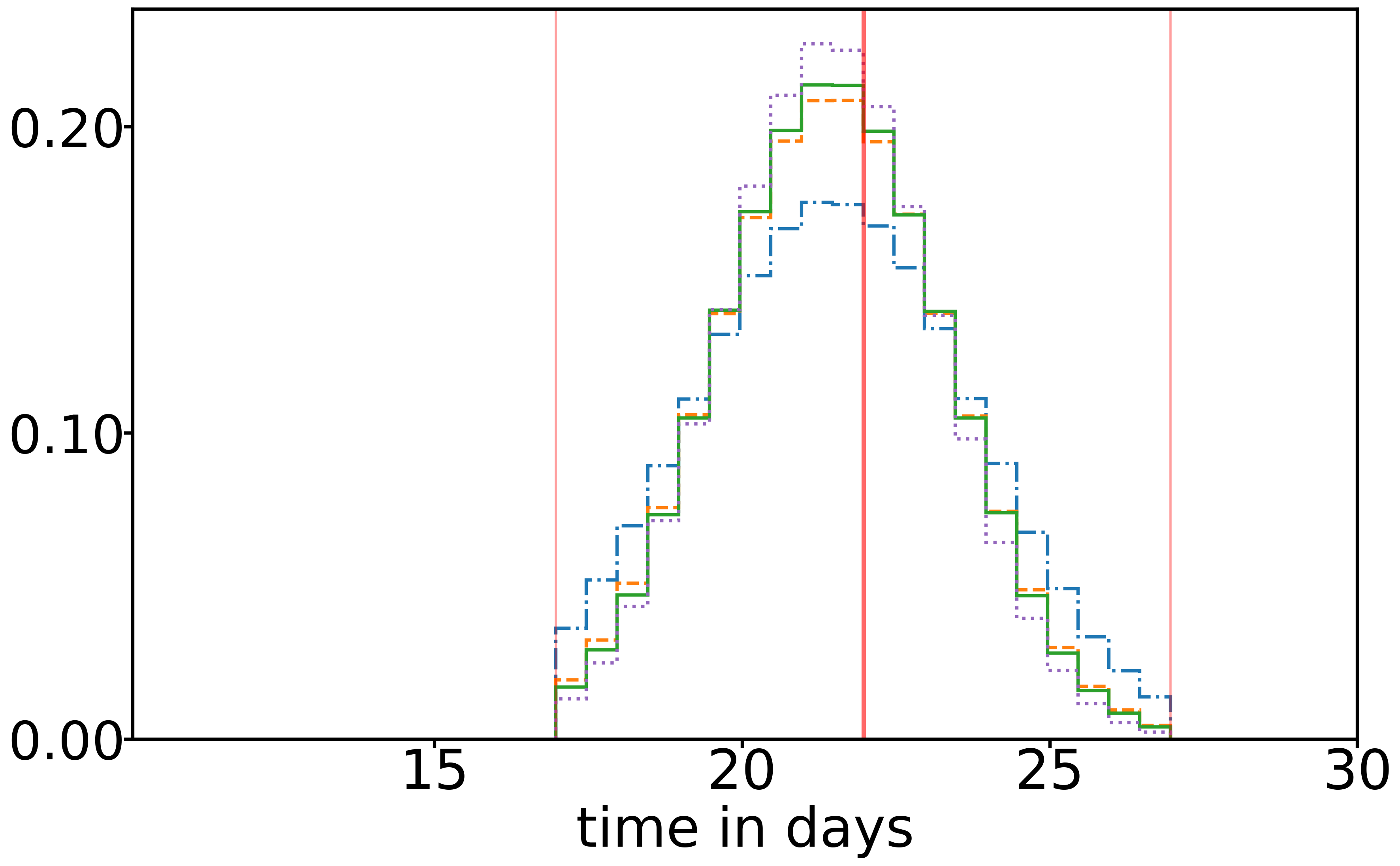}}
\subfigure[\hbox{Retrieved phase combining all three absorption lines}]{\label{phase_combined_short}\includegraphics[width=0.48\textwidth]{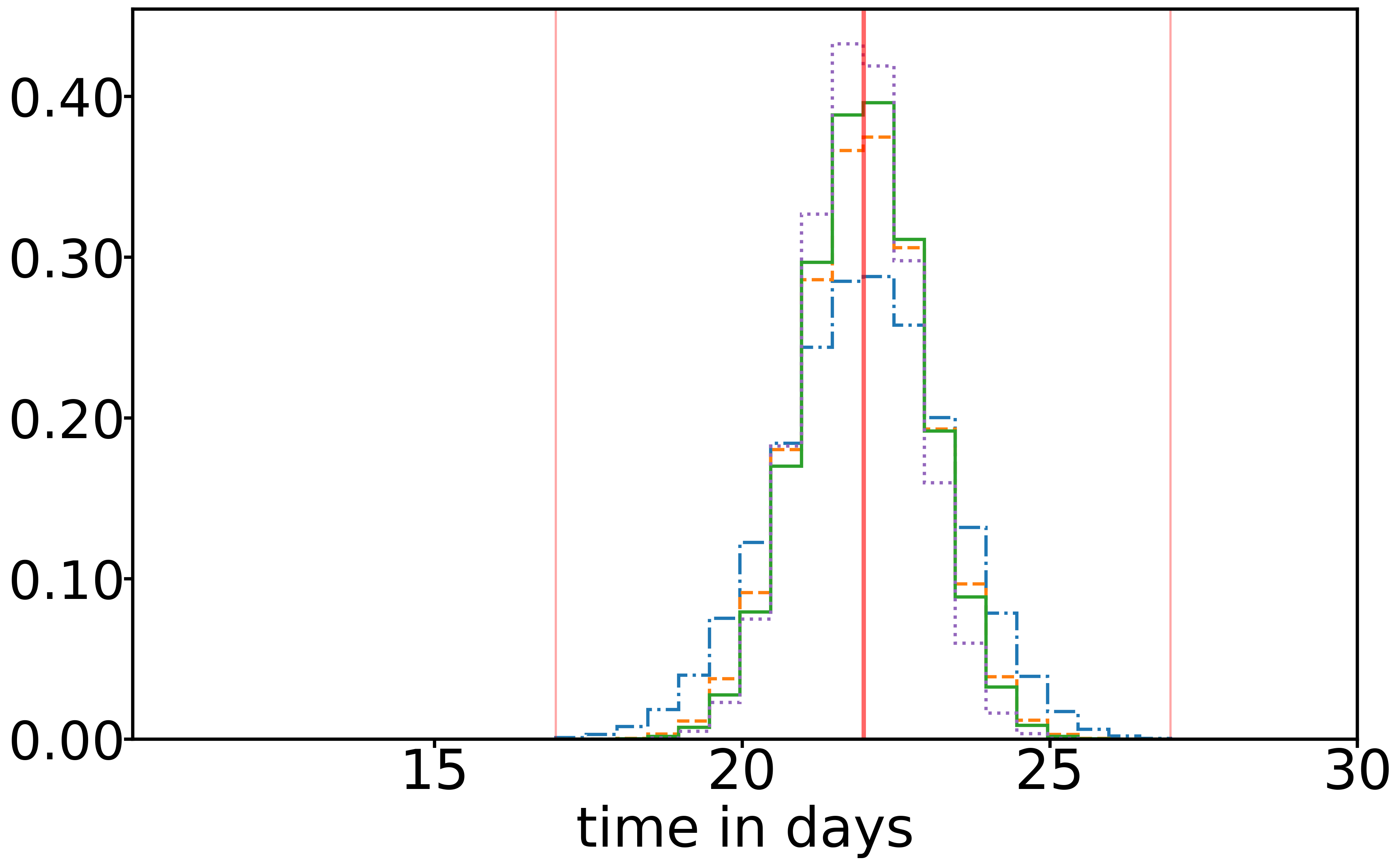}}\
\caption{\label{phases_short} Histograms of the retrieved phases of the
  second SN image using the absorption lines Fe\,\textsc{ii},
  H$\alpha$, and H$\beta$ and using only the third and fifth epoch of the first image. The phase inference
  is done for $S/N = 10$, 20, and 30 and for noiseless spectra. The thin red
  vertical lines in the histograms indicate the epochs of the two
  used spectra of the first image. The thick vertical line
  marks the input value of the second image.}
\end{figure*}

Furthermore, we consider a case where the fourth epoch that we want to retrieve with the second image is not measured within the first image to check how well the method works if the available data are limited. Another reason for this scenario is that with real measurements we might not have data of the second image covering the same epochs as for the first image.
For this case we only used the third and fifth epoch and interpolated linearly between them to create samples as explained previously in this section. 
We justify the linear interpolation by the linear behavior of the absorption minima of the last three epochs as shown in Fig. \ref{temp_wave_min}.
In particular, H$\alpha$ follows a linear trend to a high degree of precision. We thus expect H$\alpha$ to give the best results in this case. The resulting phase retrieval histograms are shown in Fig. \ref{phases_short} with the weighted average and the 1$\sigma$ uncertainties for the different absorption lines, their combination, and for the different S/N realizations shown in Table \ref{phase_uncertainties_short}. As expected, the recovered weighted average of the phase using the H$\alpha$ absorption line agrees well with the exact value of 22.0 days. Even though the precision is almost unchanged, the accuracy is strongly biased in the case of Fe\,\textsc{ii} and H$\beta$.
This most likely results from the nonlinear evolution of the last three bins, which is not as linear as for H$\alpha$.\\
\noindent\hspace*{4mm} In general we detect three sources for bias: microlensing, fitting ranges, and the cadence of the observations. The cadence is an especially important factor for the first image. At the moment we can only estimate the strength of each of these bias contributions from our results. How much each bias affects the measurements will depend on the details of actual data once they are available.
Combining the phase inference from multiple spectral features can potentially reduce the bias compared to the measurements for a single line. As illustrated in Fig. \ref{phases_short} and Table \ref{phase_uncertainties_short}, when we multiply the phase-retrieval histograms together, the bias cancels out since the underprediction of the phase from the H$\beta$ absorption line counterbalances the overprediction of the phase from Fe\,\textsc{ii}. Depending on the spectral features used, such a perfect cancelation of the biases might not occur.  Therefore, when applying our method to real data, we recommend investigating individual spectral features through simulations to quantify potential biases, as we have done here.\\
\noindent\hspace*{4mm} For accurate phase retrievals, we ideally need as many measured epochs of the first image as possible to ensure that we cover the range of phases where we might be able to measure the second image and use it in the phase retrieval. This further ensures that linear interpolations only cover short periods, which minimizes the bias induced by a long gap between epochs. 
From the example in Fig. \ref{phases} we can see that a cadence of around five days for the first image already reduces the bias significantly compared to a cadence of approximately ten days as used in Fig.\ref{phases_short}. 
Having as many epochs as possible for the first image also ensures that we do not need to extrapolate (which would lead to another source of bias) to cover an epoch of the second image that could be missing in the measurements of the first image.
As the combination of the phase retrieval of several different absorption lines can lead to the canceling of the bias, we recommend using at least three different absorption lines and also considering them individually, which gives a general hint of the amount of bias to expect.

\begin{table*}[hbt!]
\centering
\begin{tabular}{|c|c|c|c|c|}
\hline 
$S/N$ & 10 & 20 & 30 & $\infty$ \\ 
\hline 
Fe\,\textsc{ii} & 22.5 $\pm$ 2.3 days & 22.7 $\pm$ 1.9 days & 22.7 $\pm$ 1.8 days & 22.8 $\pm$ 1.7 days \\ 
\hline 
H$\alpha$ & 21.9 $\pm$ 2.0 days & 21.9 $\pm$ 1.5 days & 21.9 $\pm$ 1.4 days & 21.7 $\pm$ 1.3 days \\ 
\hline 
H$\beta$ & 21.6 $\pm$ 2.1 days & 21.5 $\pm$ 1.8 days & 21.5 $\pm$ 1.8 days & 21.4 $\pm$ 1.7 days \\ 
\hline 
Fe\,\textsc{ii} + H$\alpha$ + H$\beta$ & 22.0 $\pm$ 1.4 days & 22.0 $\pm$ 1.0 days & 22.0 $\pm$ 1.0 days & 21.9 $\pm$ 0.9 days \\ 
\hline 
\end{tabular}
\caption{\label{phase_uncertainties_short}Weighted average and the 1$\sigma$ uncertainties of the
  phase retrievals considering the third and the fifth epoch of the first image. We list the results for the three absorptions lines (Fe\,\textsc{ii},
  H$\alpha$, and H$\beta$) for the three
  different S/N values (10, 20, and 30) and for the noiseless case, indicated by
  $S/N = \infty$. The last row shows the values when
  using all three lines, assuming the noise and the microlensing in these lines are
  uncorrelated.}
\end{table*}

\section{Discussion}
\label{sec: Discussion}

As of today, only one strongly lensed SN II has been observed, namely
SN Refsdal, which was first detected by \cite{Kelly2016a, Kelly2016b}. With its
five detected and resolved images, time-delay cosmography with a SN II
is being conducted for the first time. The system has so far primarily
been used to crash test and validate lens modeling codes' predictions of the place and
time of the reappearance of the last trailing SN image.
The short time delays between the first four appearing images have been measured by \citet{Rodney2016} based on their light curves and by \citet{Baklanov2020} using model light curves, and there are ongoing efforts to measure the time delay of the last trailing SN image using light curves (P.~Kelly, private communications). 

Type Ia supernova iPTF16geu, discovered by
\cite{Goobar2017},
has four spatially resolved images from which
time-delay measurements have been conducted. The delays have been
calculated through two independent channels: from photometry of
consecutive image appearances by \cite{Dhawan2020} and from spectral features by
\cite{Johansson2020}. The authors mention that microlensing effects
\citep{More2017, Moertsell2020} on the light curves cause
inaccuracies in the time-delay measurements. Therefore, they used the
spectra for the first time to retrieve time delays. They fit the blended spectra of three of the four SN images and, separate from that group, the spectrum of the remaining SN image 
with two different sets of templates: the template
spectra of \cite{Hsiao2007} and the spectral energy distribution template of SN 2011fe
\citep{Amanullah2015}. Significant features of the fits
were then used to measure the time delay between the image groups. In the case of iPTF16geu, the
spectroscopic time delay is not as
precise as the photometric time delay, but the spectra seem to be less
sensitive to microlensing than the color curves \citep{Johansson2020}. In contrast to our new method, which takes advantage of the temporal evolution of the blue shift of absorption lines, the method by \cite{Johansson2020} needs very good sets of spectral templates. These only exist for SNe Ia. 

In the context of our new
method of retrieving time delays from absorption lines of SNe II, the studies of iPTF16geu indicate that light curves and color curves with strong features are in general more promising than spectral lines for the calculation of time delays. Microlensing has a
minor impact on the color curves, especially if the intensity
profiles appear to be mostly achromatic. The spectral approach is also highly
dependent on the observational data quality. If the measurements are dominated by noise, features and absorption lines may become
undetectable, making spectra unusable for time-delay measurements. Nevertheless, in cases where the color
curves do not show useful nonlinear structures for delay measurement,
which is the case in the early plateau phase of SNe II-P, we demonstrate
that the phase of a spectrum can be accurately inferred. The
uncertainties of the retrieved phase using a single absorption line lie within $\pm$2 days
(provided several spectra of one of the other LSN images in the system have been obtained to determine the evolution of the absorption line). Considering that delays of galaxy-scale LSNe typically
range from days to weeks, our new method is very promising for
obtaining delays with $<$10 \% uncertainty for systems with delays
longer than 20 days.

By comparing our LSNe II-P work to previous LSNe studies \cite[e.g.,][]{Goldstein2018,Foxley2018,Huber2019, Huber2020},
we find that LSNe Ia and II-P have separate advantages for measuring 
time delays. The homogeneity of SNe Ia facilitates the comparison with templates, making it easier to retrieve time delays with them. 
In contrast to SNe Ia, the plateau phase of SNe II-P lacks 
characteristic maxima or minima within the light and color curves, which makes 
this phase not ideal for time-delay measurements using color curves. 
However, this phase 
showed less ``chromatic'' evolution within the intensity profiles than SNe Ia \citep{Huber2019,Huber2020}, making LSNe II-P less susceptible to 
microlensing effects within the early plateau phase. 
Therefore, our newly
developed time-delay measurement method using the absorption lines of the spectra of LSNe II-P is a promising way of conducting cosmography in the LSST era
\citep{LSSTScienceCollaboration2009} as more LSNe II than LSNe Ia are expected to be detected \citep{Oguri2010,
Wojtak2019, Goldstein2018, Goldstein2016}. Methods of conducting time-delay measurements during the plateau phase of SNe II-P will be most useful as this is the time range where observations will be easiest to perform due to the brightness and duration of that phase.

\section{Conclusion}
\label{sec: Conclusion}

In this work we investigated strongly lensed SNe II and the possible
impact of microlensing on time-delay measurements. We used model
spectra of SN 1999em from the \textsc{tardis} code modified
by \cite{Vogl2019}. These were microlensed using magnification maps
generated with the code \textsc{gerlumph} \citep{Vernardos2015}. We
then investigated the effect of microlensing on light and color
curves. We find that the color curves are mostly unaffected by
microlensing since the specific intensity profiles in different bands are very similar during the plateau phase.
During
the investigated time span, the light curves, and therefore also the color
curves, are almost linear, with no strong characteristic features that could
be used as easily as for SNe Ia around the same epochs for time delay retrieval.
Further investigation of time-delay measurements using SNe II-P color curves is deferred to future studies.
We thus developed a new method for calculating the time
delay of two SN images from the absorption lines in their spectra. For both images,
we simulated spectra that were distorted by microlensing and had
noise with different S/N values. We discovered that the phase can
be determined from a single spectral feature with a 1$\sigma$ uncertainty of
$<$2\,days and the weighted average reproducing the input phase value within the 1$\sigma$ uncertainty, assuming a
realistic S/N of 20. If several measurements of different
absorption lines are combined, the uncertainty can be further
minimized and the accuracy improved as possible biases can cancel one another out. By neglecting any correlations between the errors of the
different absorption lines, the 1$\sigma$ uncertainty can be minimized
to $\pm$ 1.0 days for $S/N = 20$. Given the complementarity between the
spectral and photometric delay inference, we advocate for the
combination of both approaches in future measurements of LSNe.

\begin{acknowledgements}

We thank A.~Kozyreva, C.~McCully, and S.~Sim for discussions. We also thank the anonymous referee for helpful and detailed comments.
We thank the Max Planck Society for support through the Max Planck
Research Group for SHS.
This project has received funding from the
European Research Council (ERC) under the European Union’s Horizon
2020 research and innovation programme (LENSNOVA: grant agreement
No. 771776; COSMICLENS: grant agreement No. 787886).
This research is supported in part by the Excellence Cluster ORIGINS which is funded by the Deutsche Forschungsgemeinschaft (DFG, German Research Foundation) under Germany's Excellence Strategy -- EXC-2094 -- 390783311.
J.~H.~H.~Chan acknowledges support from the Swiss National Science Foundation (SNSF).

This research made use of \textsc{Tardis}, a community-developed software
package for spectral synthesis in supernovae
\citep{Kerzendorf2014, kerzendorf_wolfgang_2019_2590539}.
\textsc{Tardis} is fiscally sponsored by NumFOCUS, a nonprofit that
promotes open practices in research, data, and scientific computing.
The development of \textsc{Tardis} received support from the
Google Summer of Code initiative
and from ESA's Summer of Code in Space program. \textsc{Tardis} makes
extensive use of Astropy and PyNE.

\end{acknowledgements}

\bibliographystyle{aa}
\bibliography{MPA_SNII_microlensing}

\clearpage

\appendix

\FloatBarrier
\onecolumn
\section{Color curves}
\label{sec: Appendix A}

\begin{figure}[hbtp]
\centering
\subfigure[]{\label{colorcurve_u_r}\includegraphics[width=0.31\textwidth]{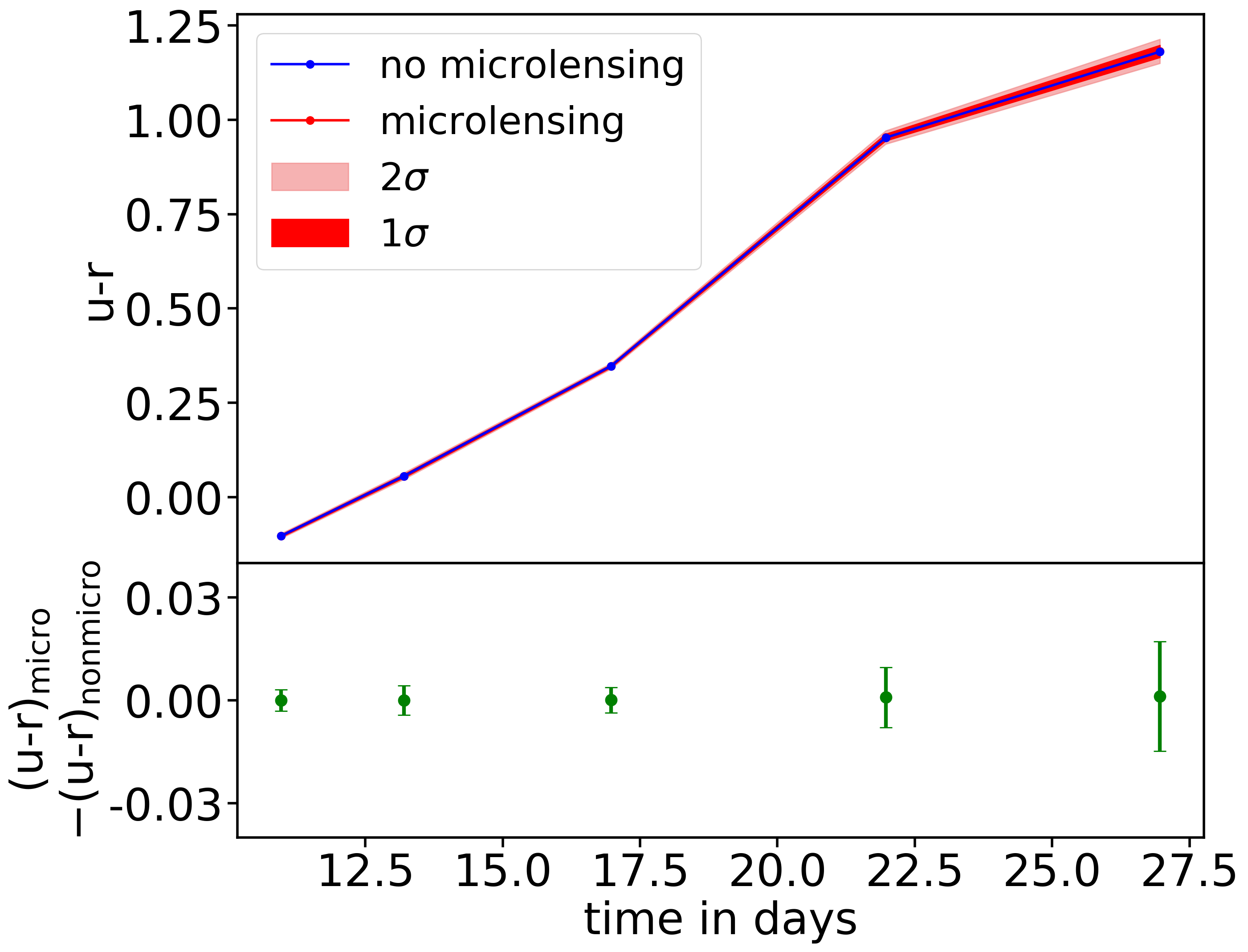}}
\subfigure[]{\label{colorcurve_u_i}\includegraphics[width=0.31\textwidth]{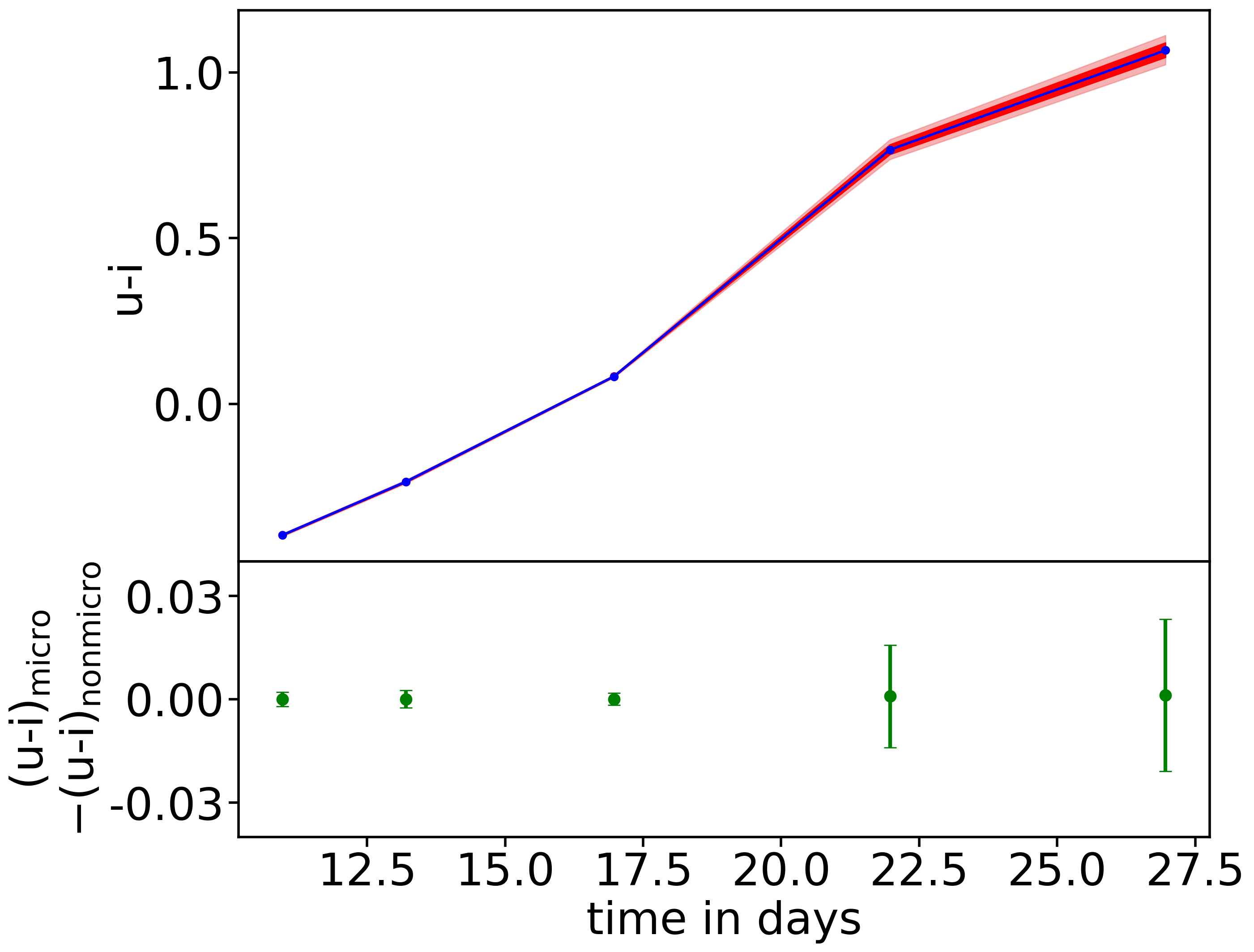}}
\subfigure[]{\label{colorcurve_u_z}\includegraphics[width=0.31\textwidth]{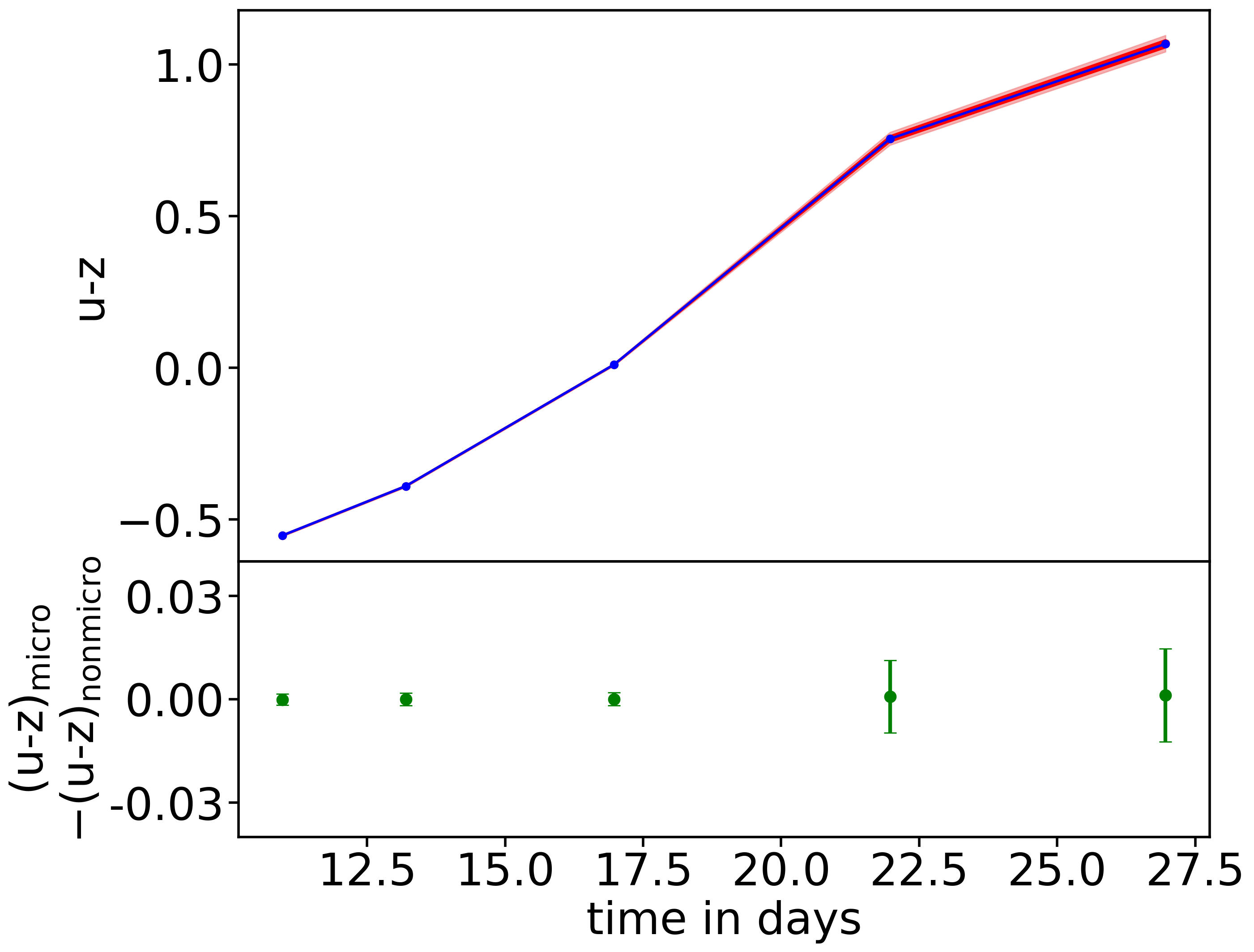}}\vskip3MM
\subfigure[]{\label{colorcurve_u_y}\includegraphics[width=0.31\textwidth]{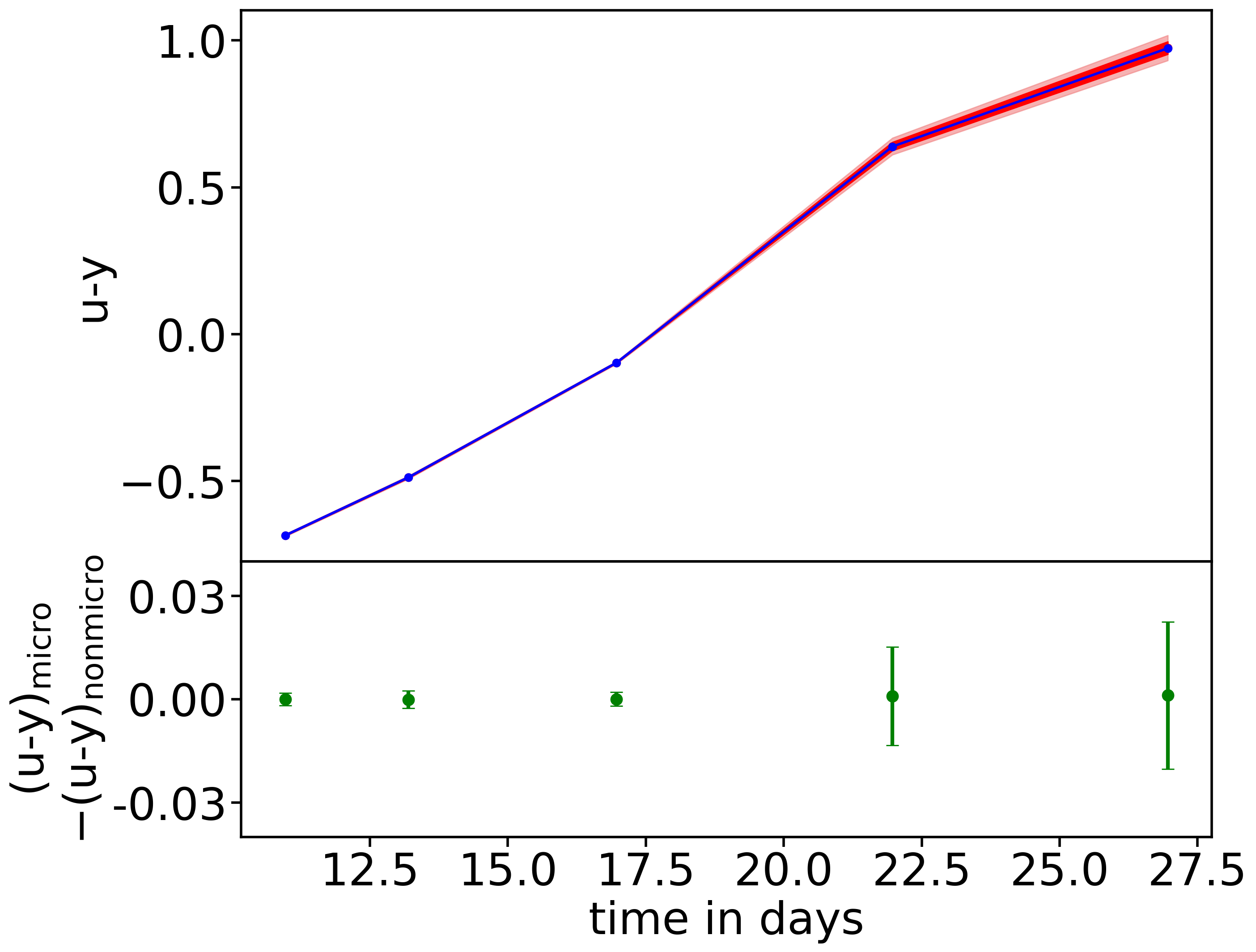}}
\subfigure[]{\label{colorcurve_g_r}\includegraphics[width=0.31\textwidth]{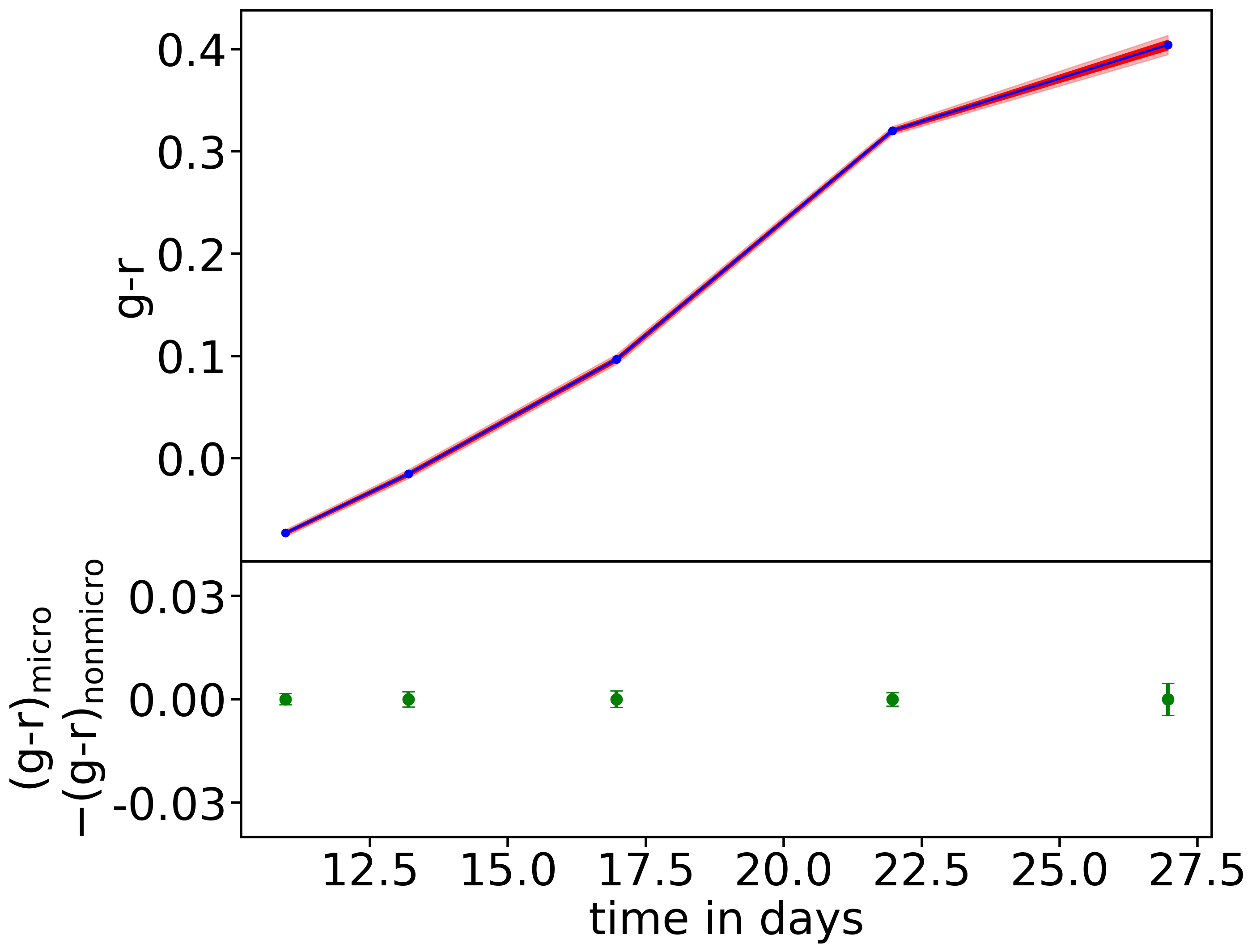}}
\subfigure[]{\label{colorcurve_g_i}\includegraphics[width=0.31\textwidth]{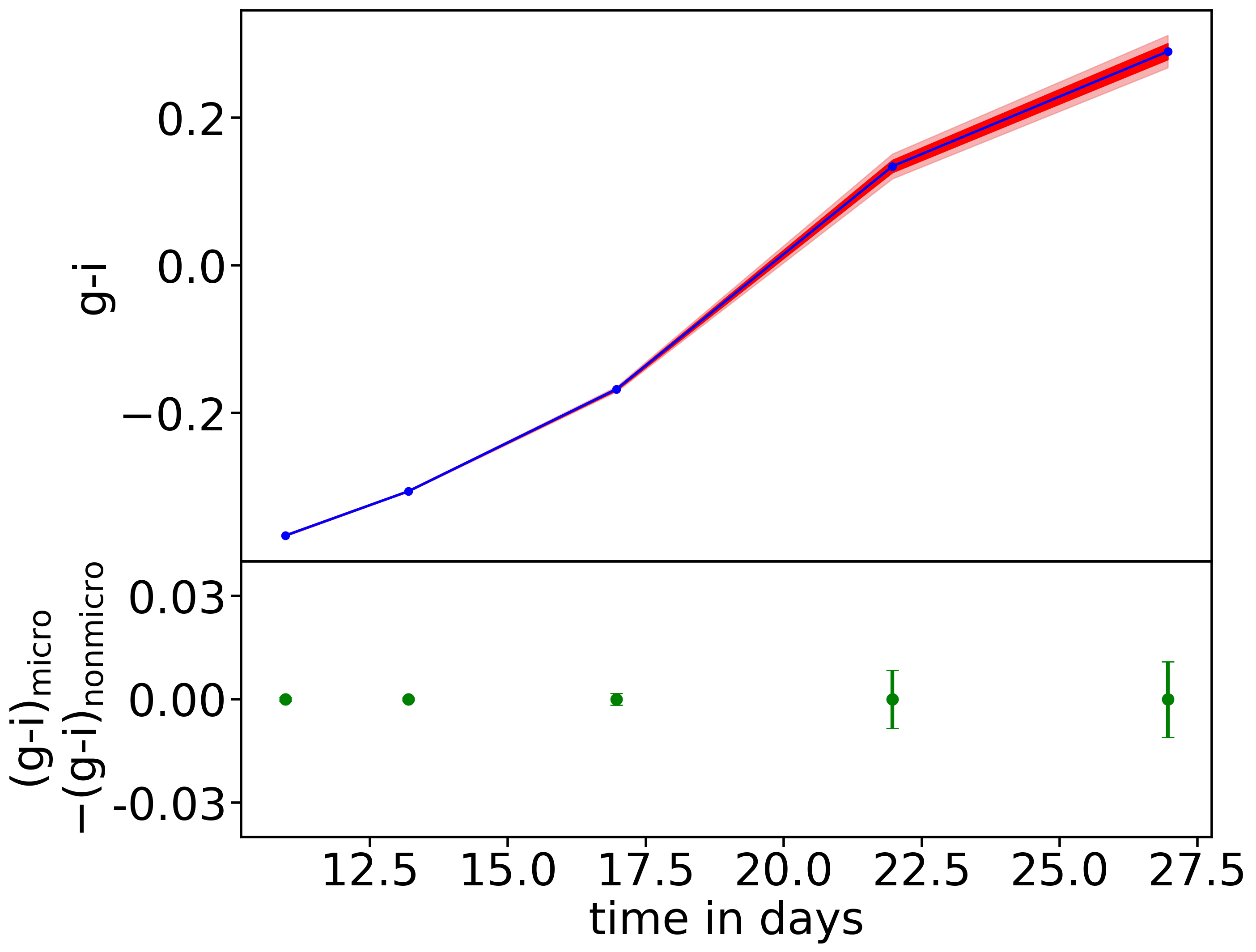}}\vskip3MM
\subfigure[]{\label{colorcurve_g_z}\includegraphics[width=0.31\textwidth]{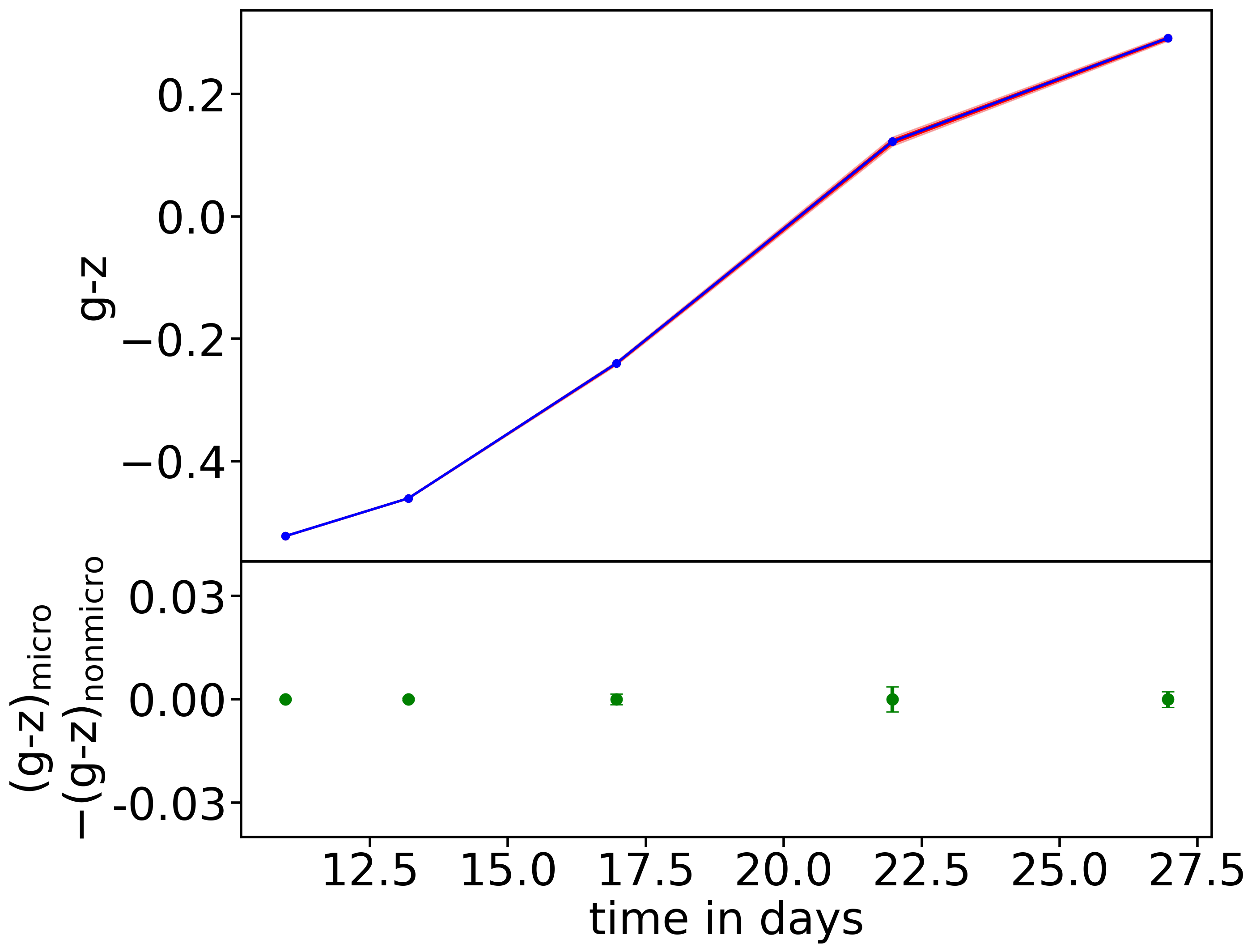}}
\subfigure[]{\label{colorcurve_g_y}\includegraphics[width=0.31\textwidth]{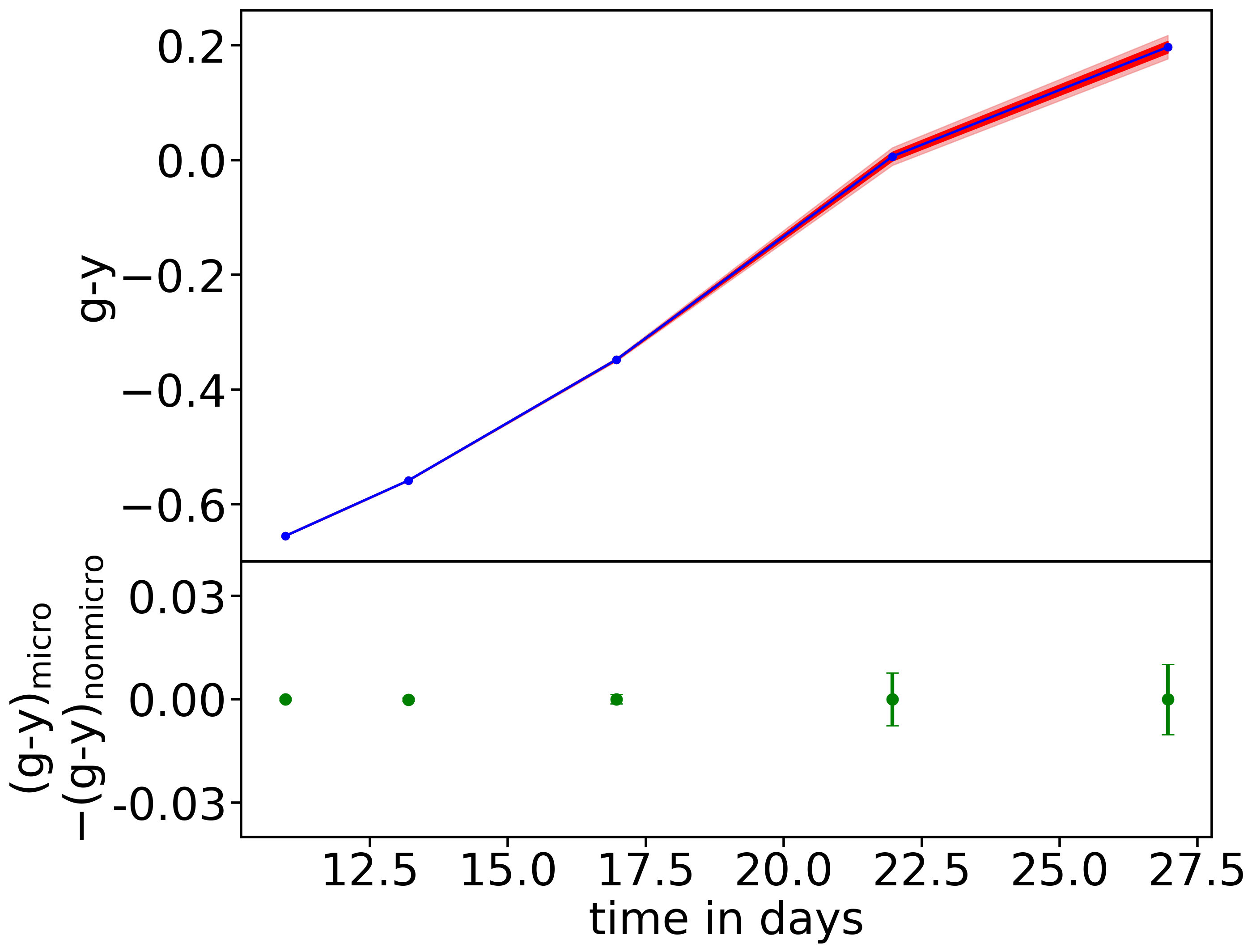}}
\subfigure[]{\label{colorcurve_r_z}\includegraphics[width=0.31\textwidth]{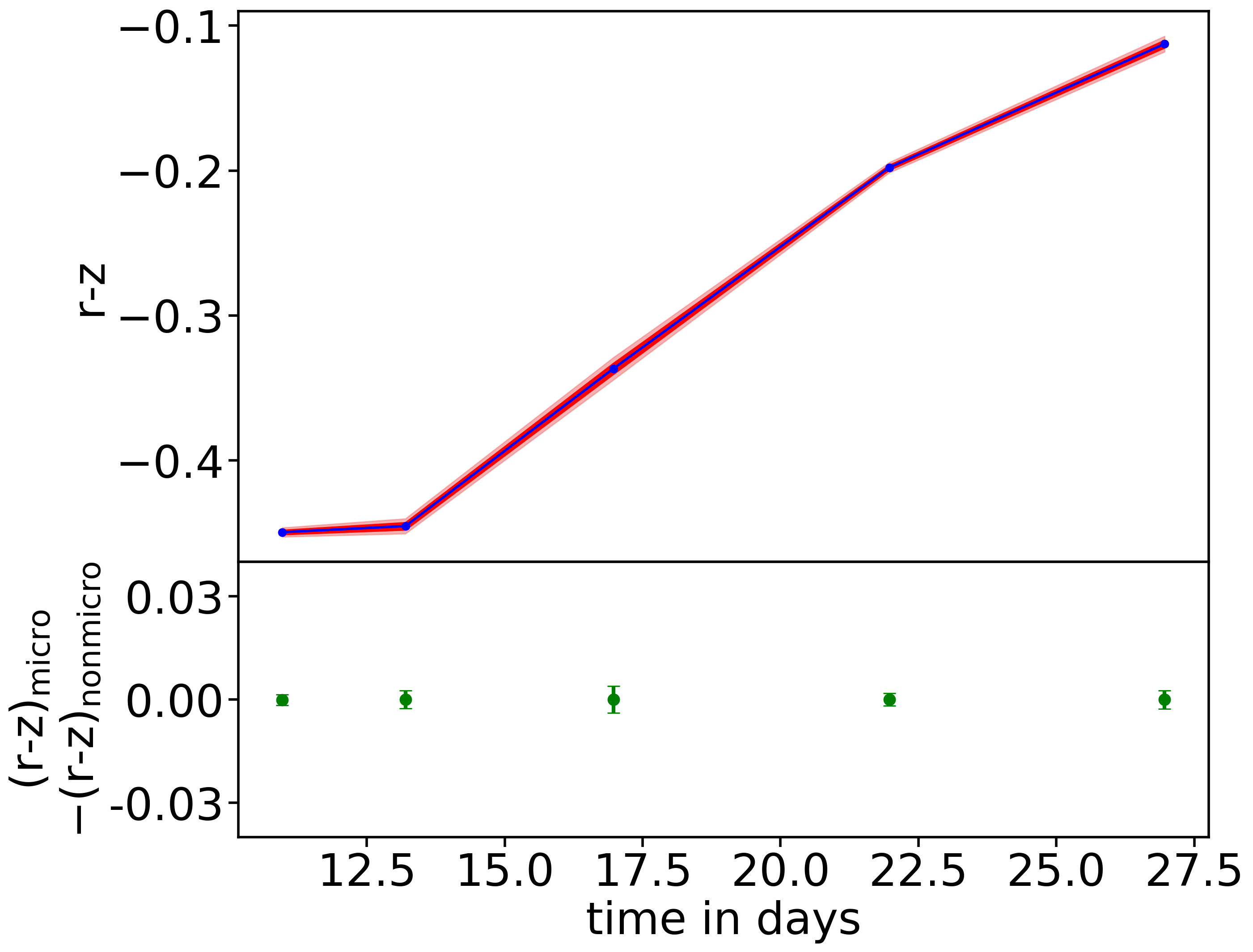}}\vskip3MM
\subfigure[]{\label{colorcurve_r_y}\includegraphics[width=0.31\textwidth]{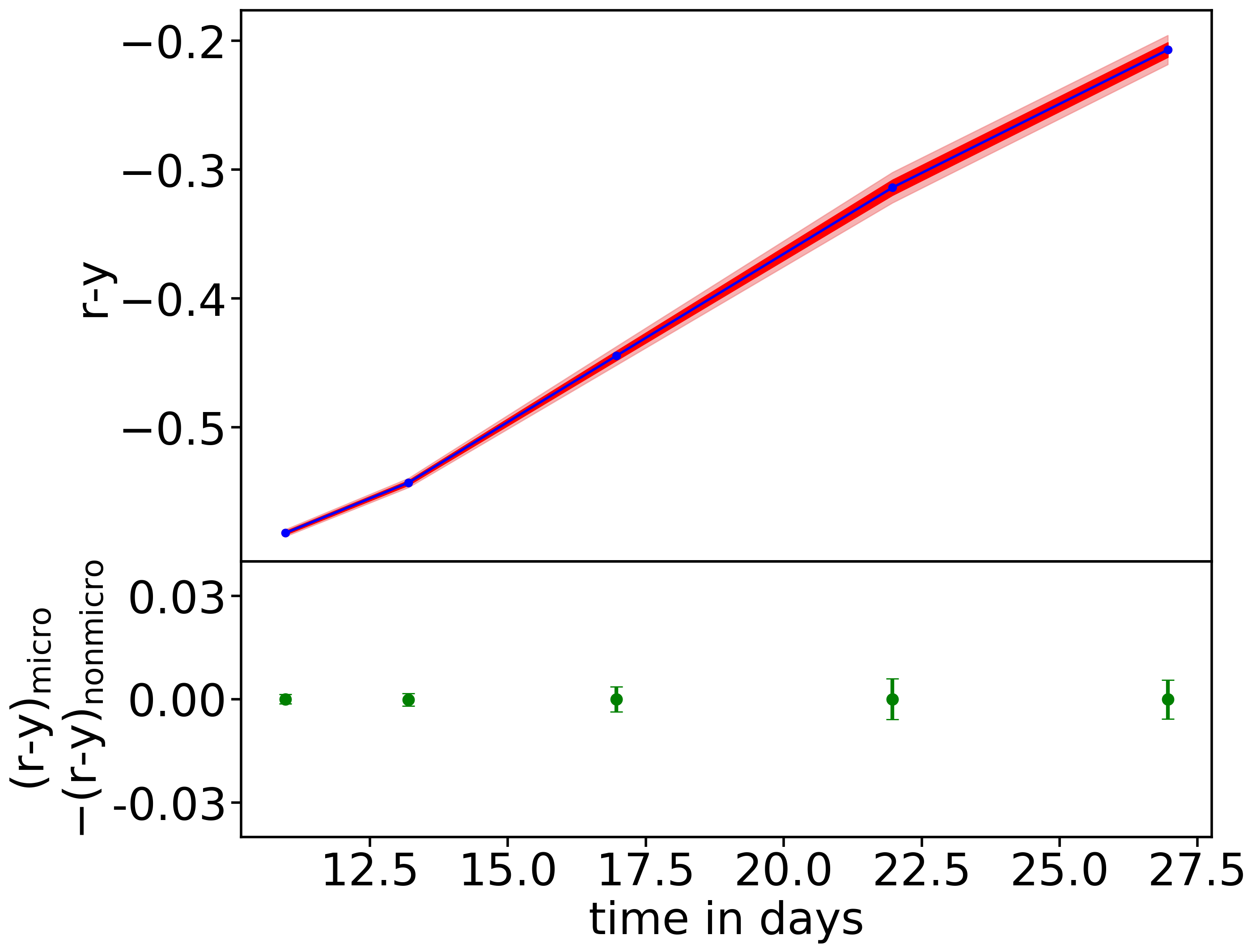}}
\subfigure[]{\label{colorcurve_i_y}\includegraphics[width=0.31\textwidth]{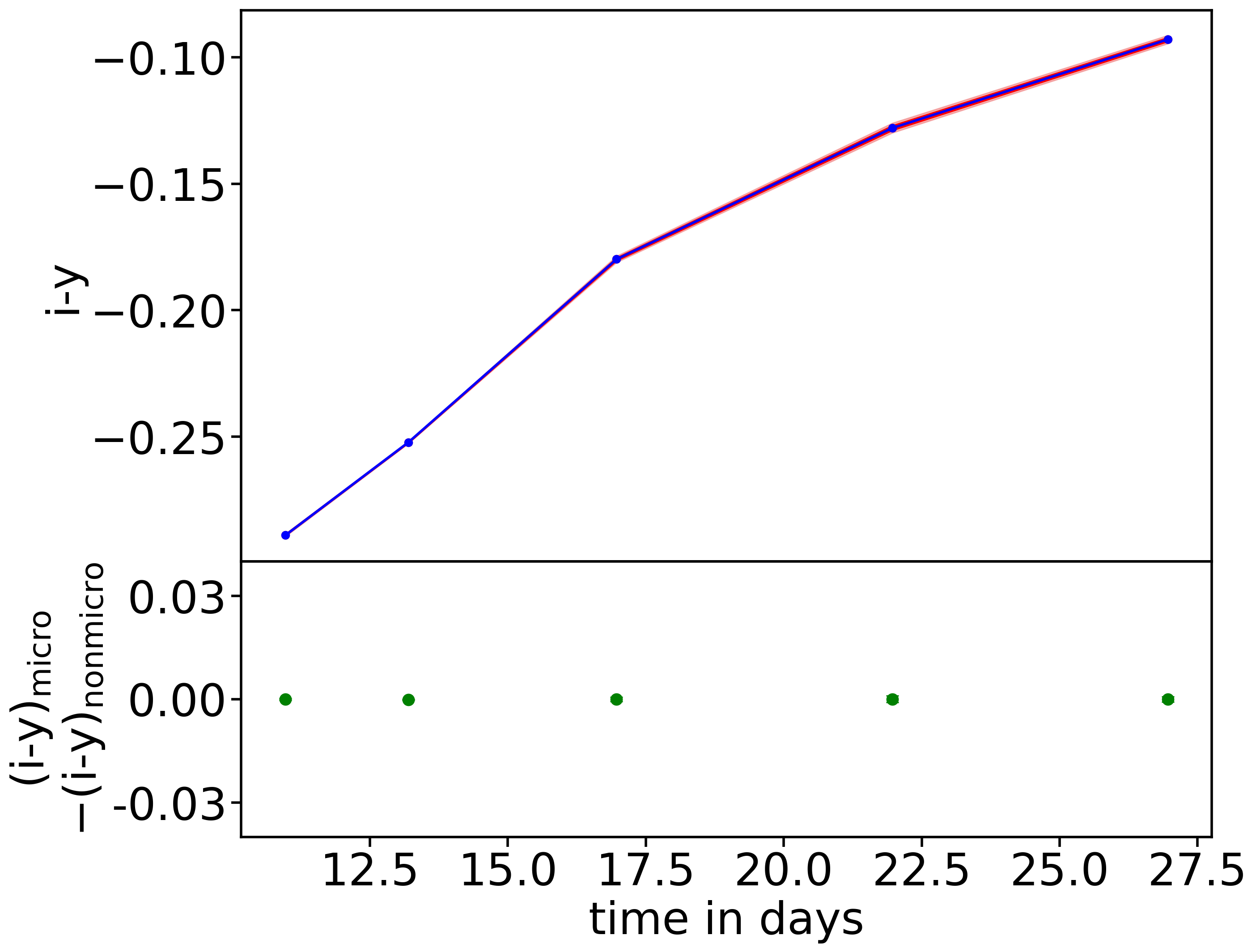}}\
\caption{\label{colorcurves_appendix} Color curves of the non-microlensed and
  microlensed spectra with median, 1$\sigma$, and 2$\sigma$ ranges obtained
  from 10000 random positions in the type I lensing magnification
  map with $\kappa$ = 0.36 and $\gamma$ = 0.35. In addition, the bottom part of each panel (\ref{colorcurve_u_r} to \ref{colorcurve_i_y}) shows the deviation of the microlensed color from the non-microlensed color with 1$\sigma$ uncertainties.}
\end{figure}

\FloatBarrier
\twocolumn

\section{Microlensing maps with high magnification}
\label{sec: Appendix B}

\begin{figure}[hbt!]
\centering
\subfigure[\hbox{Type I microlensing map}]{\label{micromap4}\includegraphics[width=0.21\textwidth]{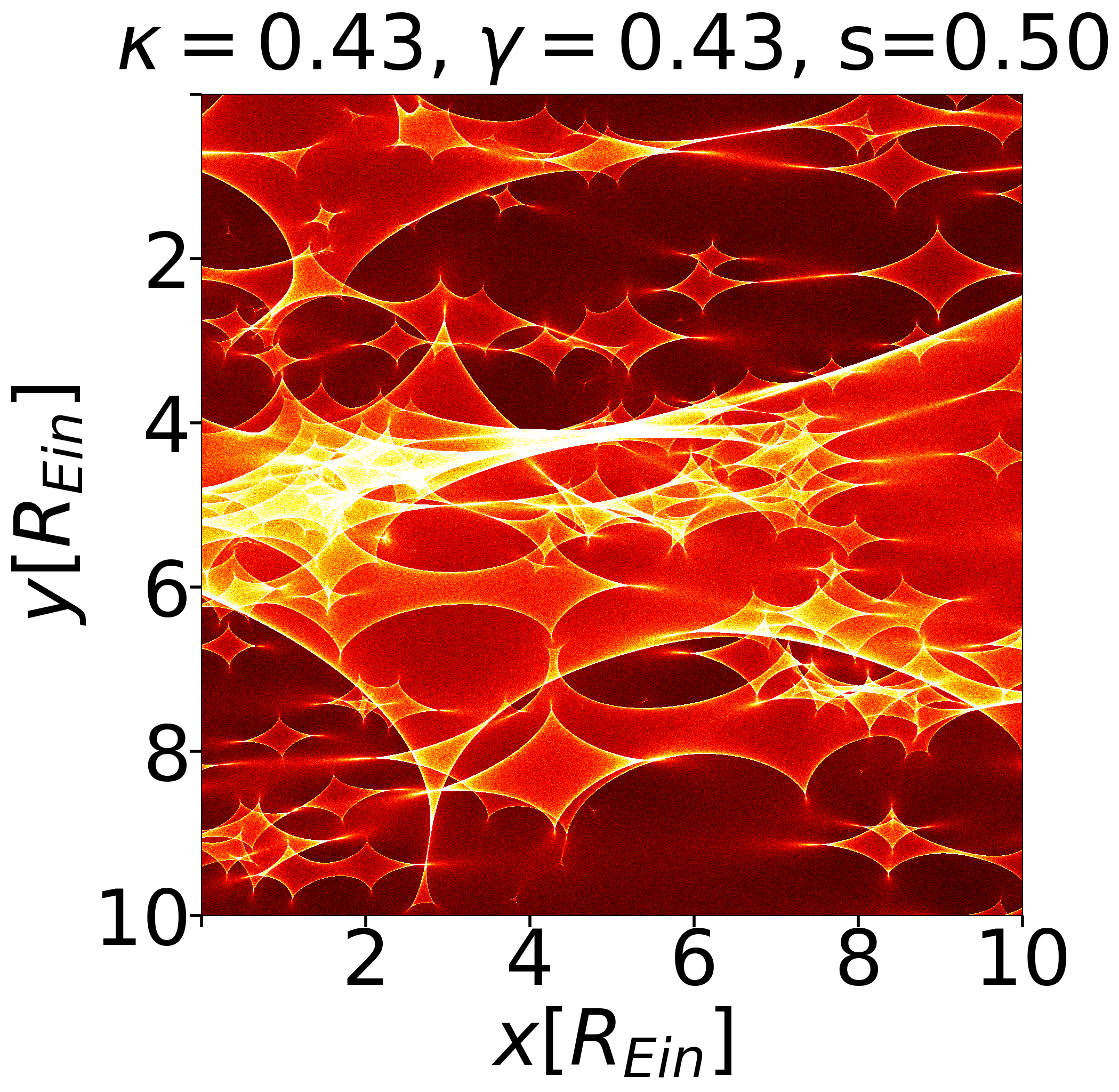}}
\subfigure[\hbox{Type II microlensing map}]{\label{micromap5}\includegraphics[width=0.21\textwidth]{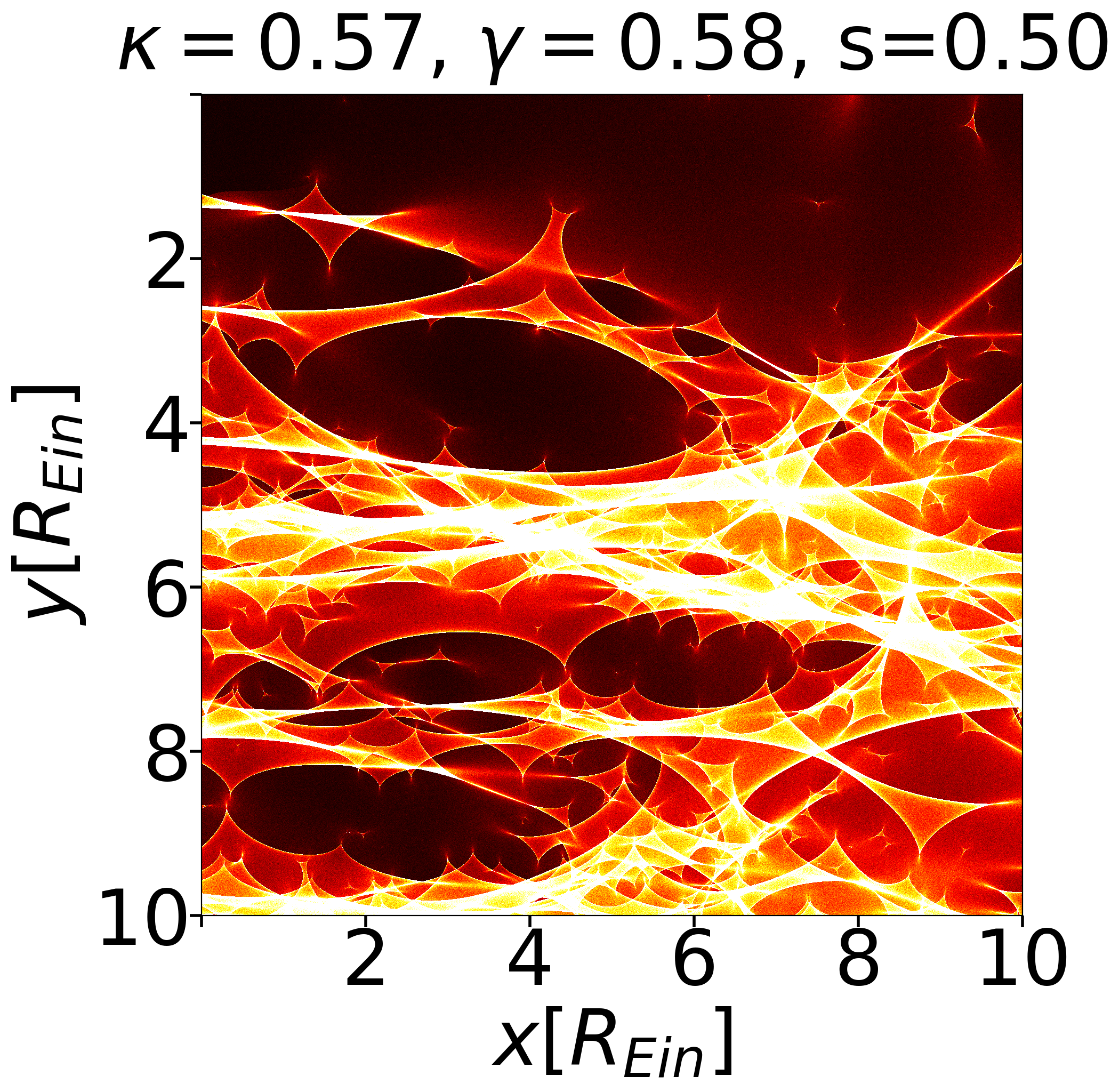}}
\subfigure{\label{micromap6}\includegraphics[width=0.059\textwidth]{plots/maps/colorbar.png}}\
\caption{\label{micromaps_2} Magnification maps for two different $\kappa$ and $\gamma$ values with smooth matter fraction $s$ = 0.50. The magnification $\mu(x, y)$ is indicated by the color scale on the right-hand side.}
\end{figure}

In addition to the already investigated microlensing case using the magnification maps with $\kappa = 0.36$ and $\gamma = 0.35$ (image I) and $\kappa = \gamma = 0.70$ (image II) from Fig. \ref{micromaps}, in this section we investigate a high magnification scenario. 
Therefore, we applied our method following the steps explained in Sects. \ref{sec: SNe II spectra and absorption features} and \ref{sec: SN phase inference from spectra} to \textsc{gerlumph} magnification maps with $\kappa = \gamma = 0.43$ (image I) and $\kappa = 0.57$ and $\gamma = 0.58$ (image II) as shown in Fig. \ref{micromaps_2}. 
These two microlensing maps correspond to the high magnification case of the 1$\sigma$ contours of LSNe in the OM10 catalog for images I and II, respectively \citep{Huber2020}.

\begin{figure*}[hbtp]
\centering
\subfigure{\label{temp_wave_legend_2}\includegraphics[width=\textwidth]{plots/wave_min/temp_evol_legend.png}}\hfill
\subfigure{\label{temp_wave_min_H_alpha_1_2}\includegraphics[width=0.49\textwidth]{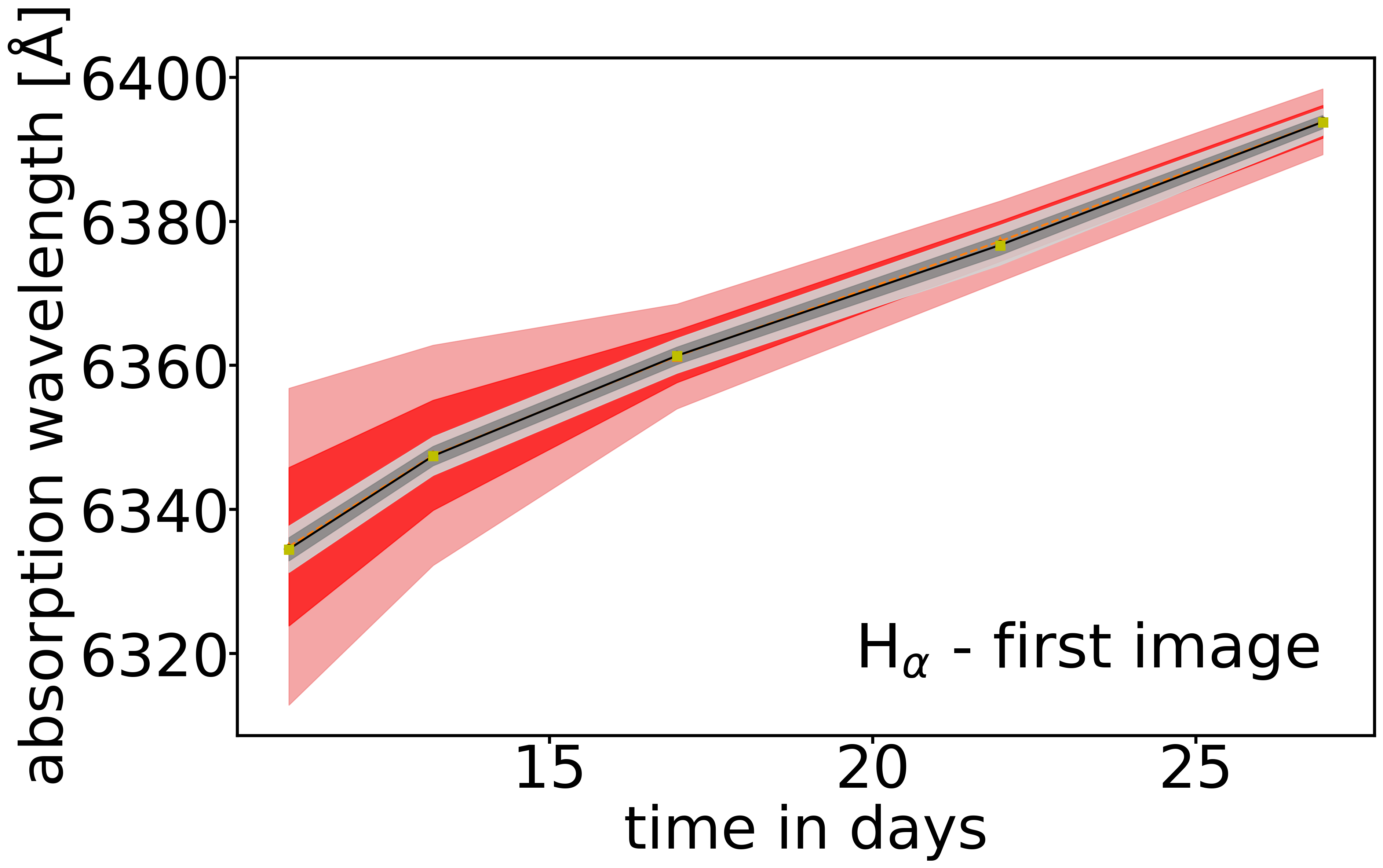}}\hfill
\subfigure{\label{temp_wave_min_H_alpha_2_2}\includegraphics[width=0.49\textwidth]{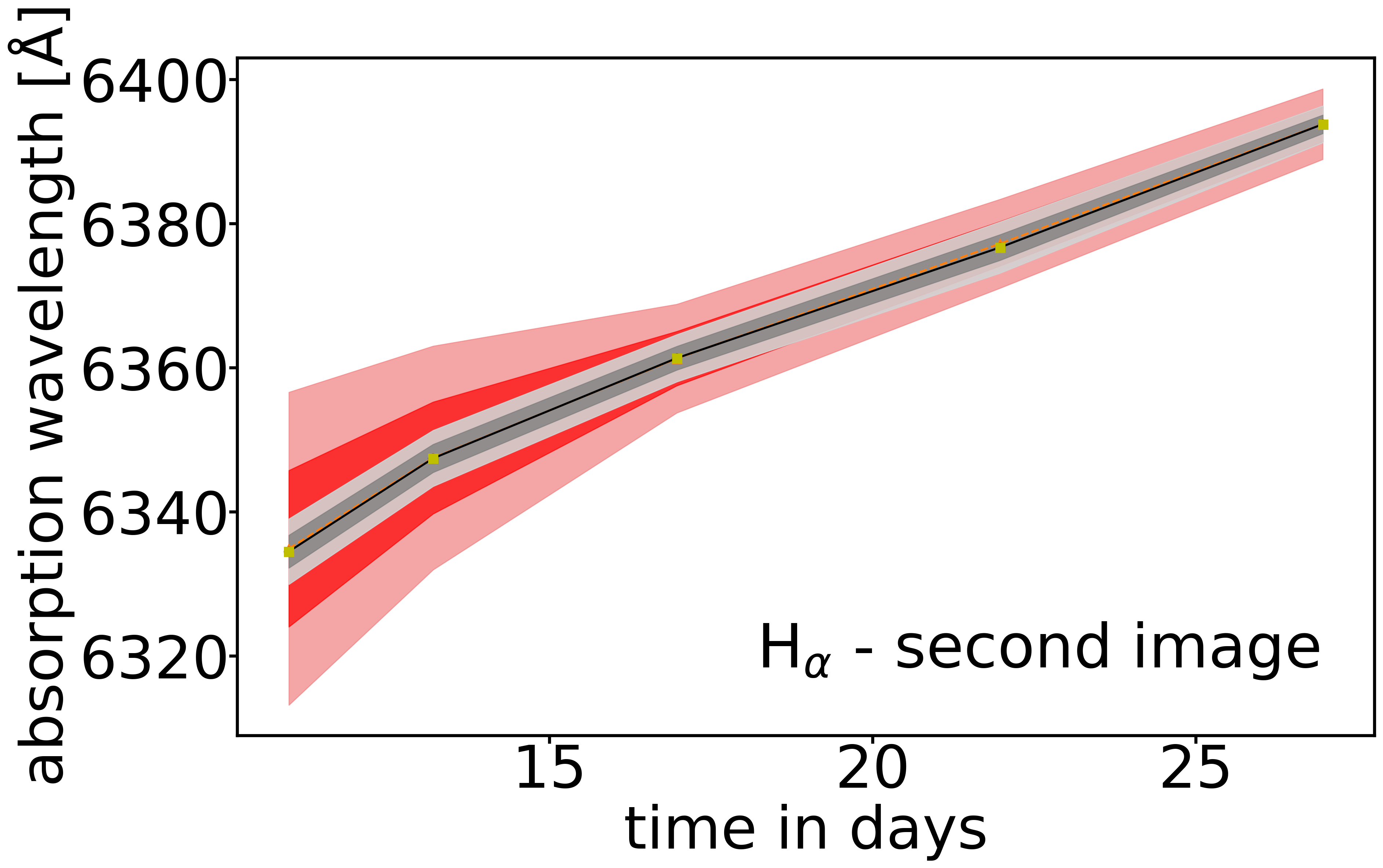}}\\
\subfigure{\label{temp_wave_min_H_beta_1_2}\includegraphics[width=0.49\textwidth]{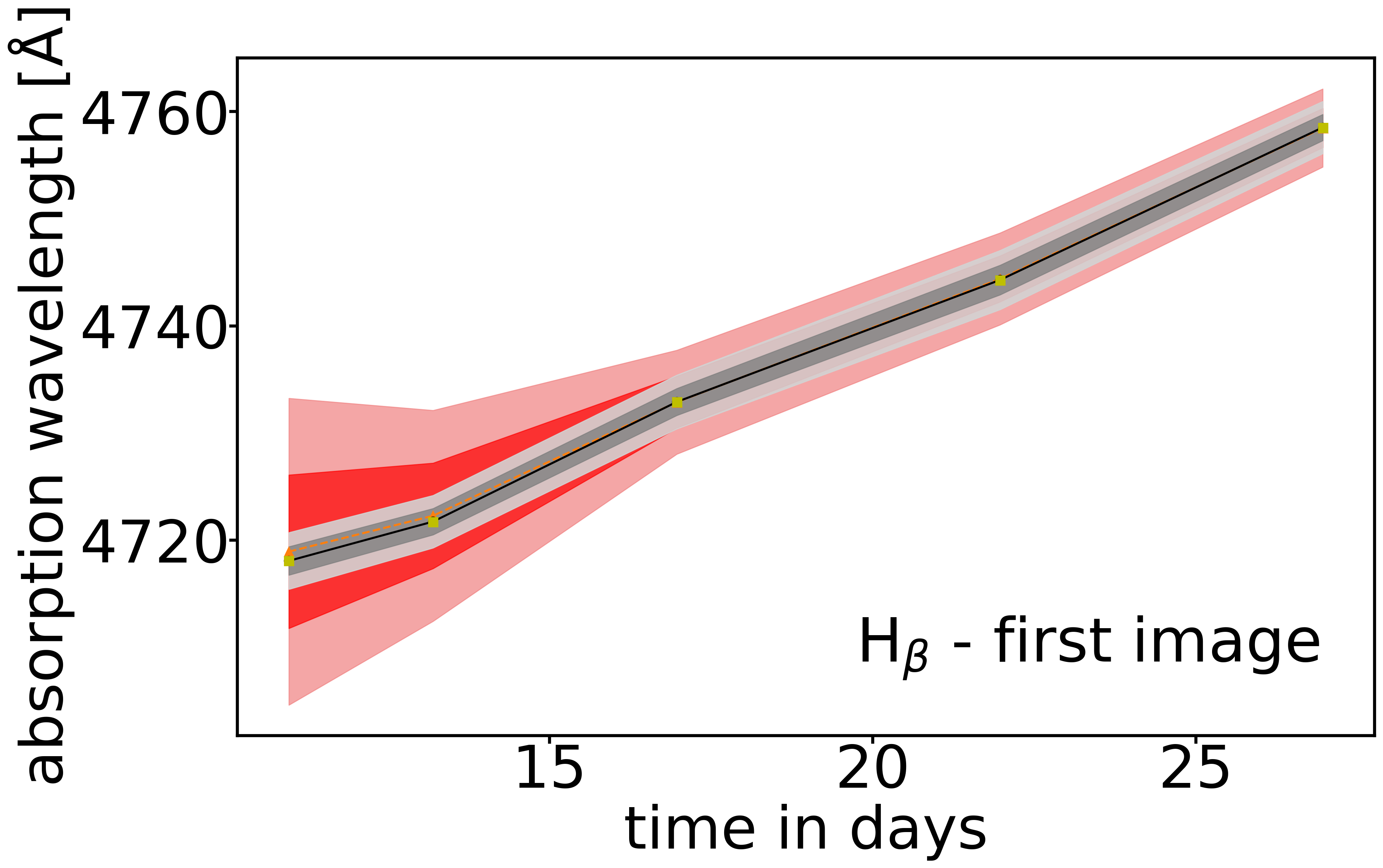}}\hfill
\subfigure{\label{temp_wave_min_H_beta_2_2}\includegraphics[width=0.49\textwidth]{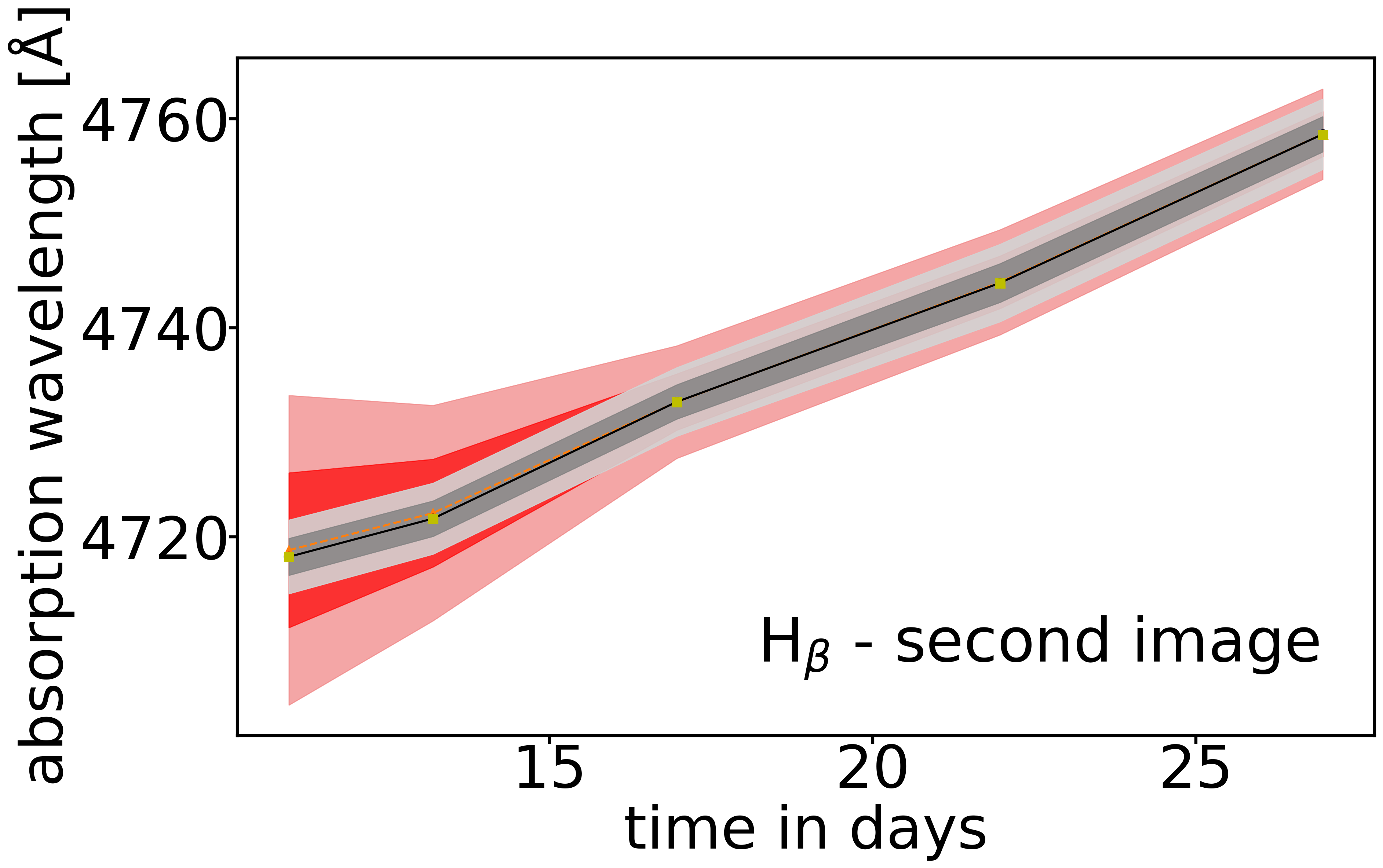}}\\
\subfigure{\label{temp_wave_min_FeII_1_2}\includegraphics[width=0.49\textwidth]{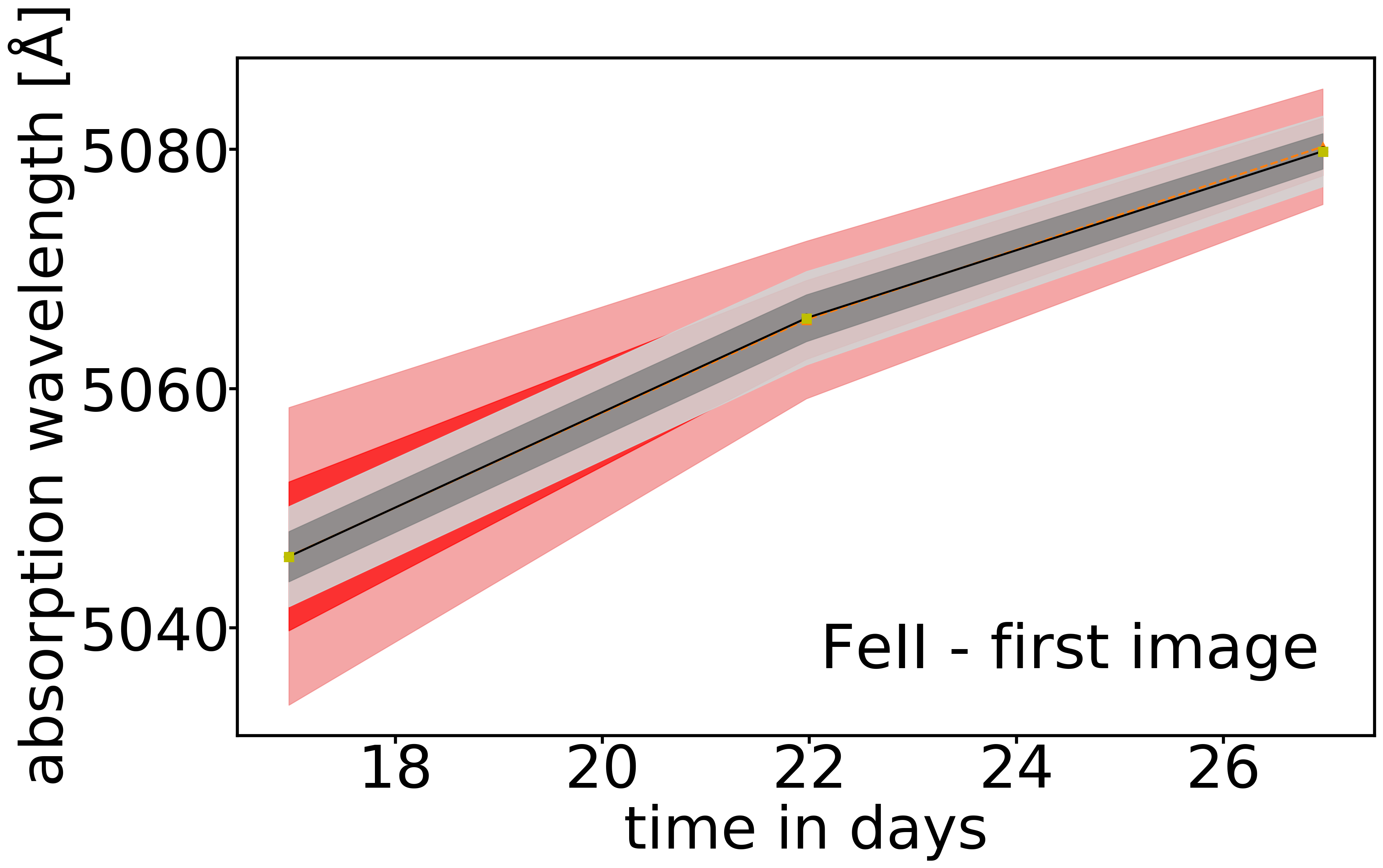}}\hfill
\subfigure{\label{temp_wave_min_FeII_2_2}\includegraphics[width=0.49\textwidth]{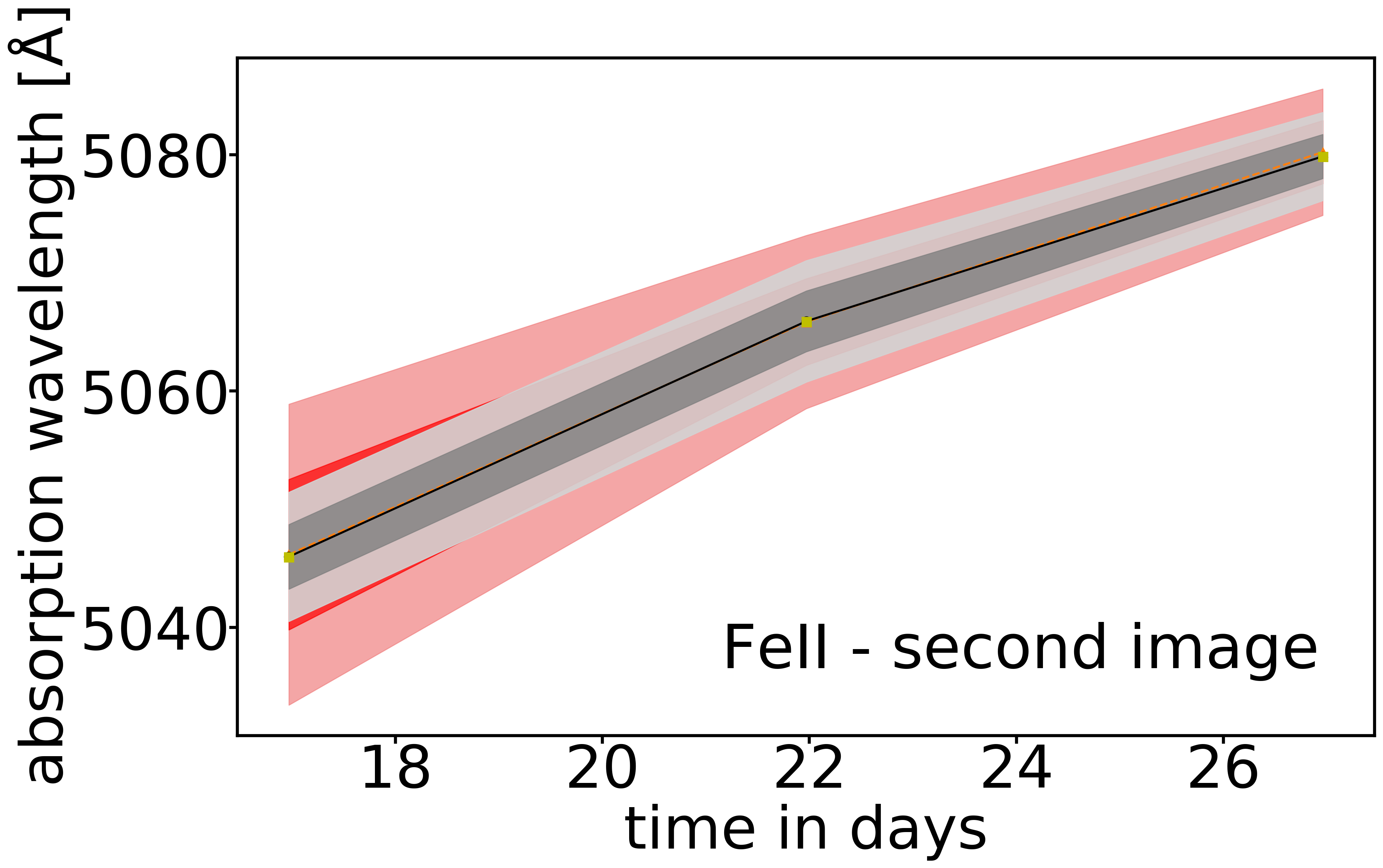}}\
\caption{\label{temp_wave_min_2} Temporal evolution of the absorption
  minima of the H$\alpha$, H$\beta$, and Fe\,\textsc{ii}
  lines for non-microlensed and microlensed spectra using the two
  different magnification maps of Fig. \ref{micromaps_2} (left: Fig. B.\ref{micromap4}; right: Fig. B.\ref{micromap5}).
}
\end{figure*}

We show the temporal evolution of the
absorption minima of H$\alpha$, H$\beta$, and Fe\,\textsc{ii} of the high magnification case in Fig. \ref{temp_wave_min_2}.
Compared to the case discussed in Sect. \ref{sec: SNe II spectra and absorption features}, there is barely any visible difference.

\begin{figure*}[hbtp]
\centering
\subfigure[\hbox{Retrieved phase using the absorption line of Fe\,\textsc{ii}}]{\label{phase_FeII_2}\includegraphics[width=0.48\textwidth]{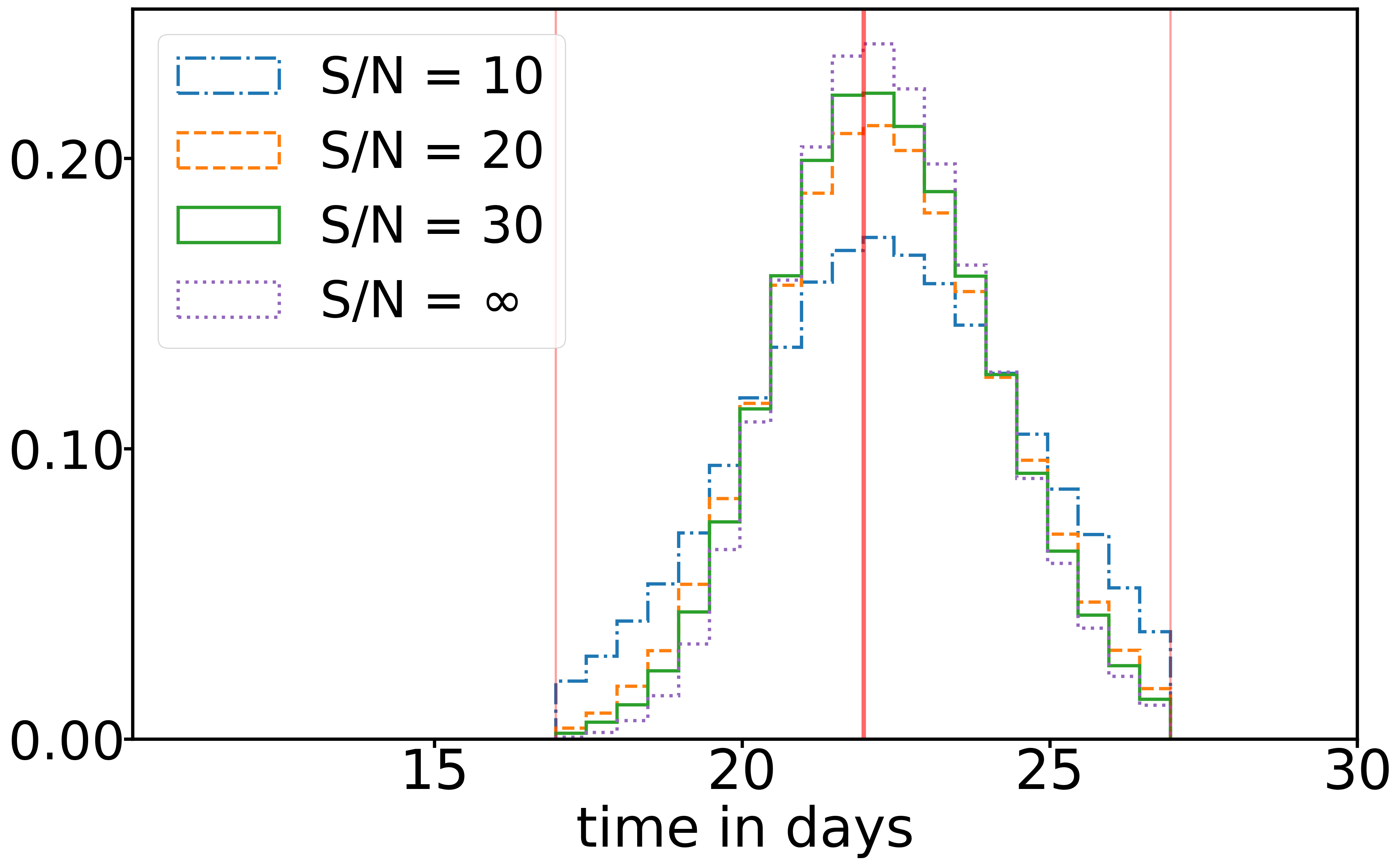}}
\subfigure[\hbox{Retrieved phase using the absorption line of H$\alpha$}]{\label{phase_H_alpha_2}\includegraphics[width=0.48\textwidth]{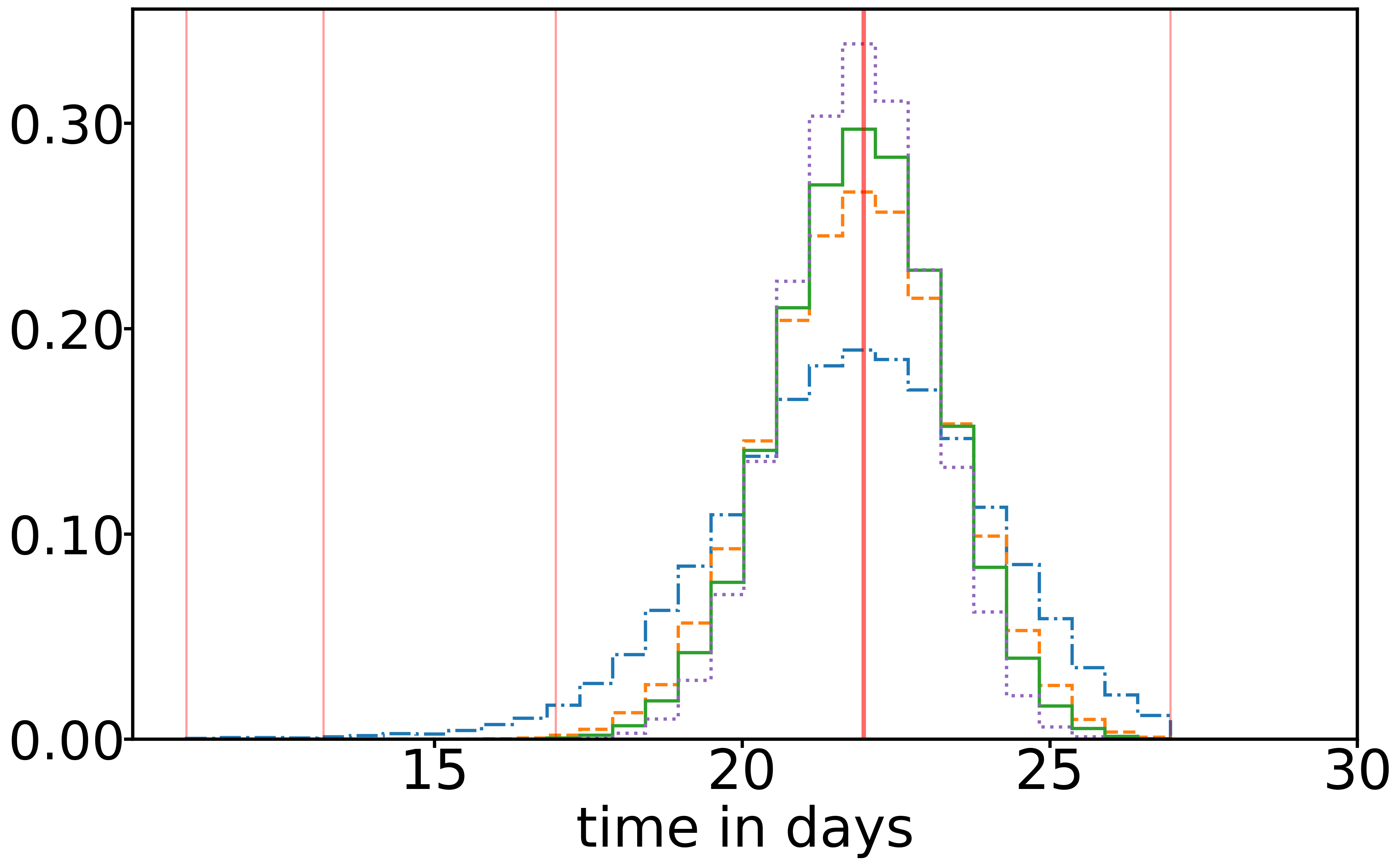}}\\
\subfigure[\hbox{Retrieved phase using the absorption line of H$\beta$}]{\label{phase_H_beta_2}\includegraphics[width=0.48\textwidth]{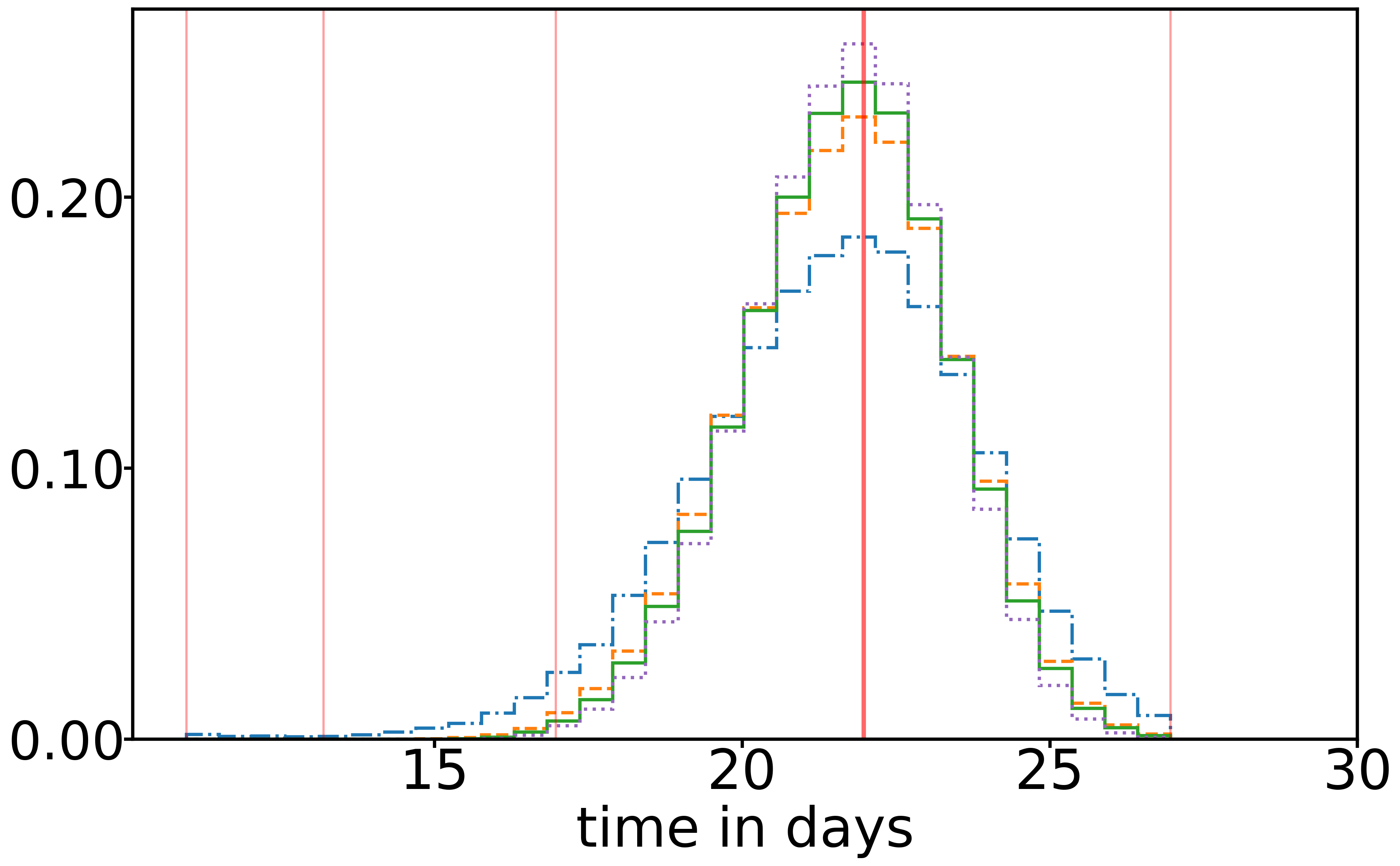}}
\subfigure[\hbox{Retrieved phase combining all three absorption lines}]{\label{phase_combined_2}\includegraphics[width=0.48\textwidth]{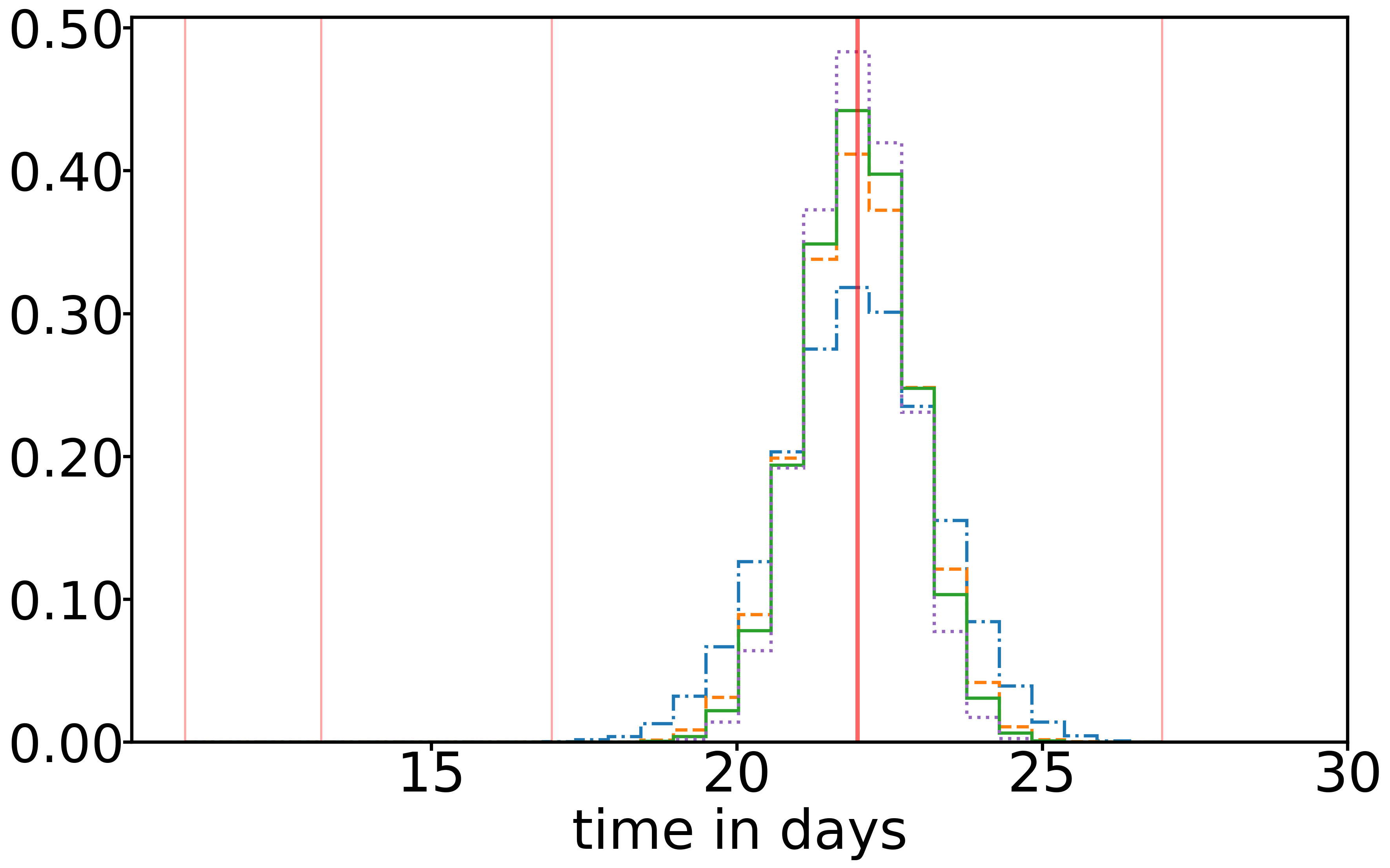}}\
\caption{\label{phases_2}  Histograms of the retrieved phases under high microlensing magnifications of the
  second SN image and second microlensing scenario using the absorption lines Fe\,\textsc{ii},
  H$\alpha$, and H$\beta$. By using the wavelengths of the absorption
  features, the phase of the second SN image of the second microlensing scenario is correctly recovered.}
\end{figure*}

The histograms of the phase retrieval using this second microlensing scenario are shown in Fig. \ref{phases_2}. In this scenario the input epoch is also recovered very well, especially for $S/N = 20$ and higher S/N cases.

In Table \ref{phase_uncertainties_2} we list the weighted average and the 1$\sigma$ deviations of
the retrieved phases. Compared to the averages and uncertainties of the originally investigated microlensed spectra in Table \ref{phase_uncertainties}, we see that this second case has the same or smaller uncertainties and retrieves the phase with the same accuracy. As the second scenario microlensing maps correspond to a high magnification case, we expected higher or at least equal uncertainties. The smaller uncertainties can be explained by fluctuations caused by the spectral fitting procedures. As the absorption lines have the asymmetric P-Cygni profile, the fitting ranges influence the phase retrieval uncertainties by $\lesssim 0.2$ days. The chosen fitting ranges discussed in Sect. \ref{sec: SNe II spectra and absorption features} appear to work better for the second scenario than for the first scenario, especially for H$\beta$, resulting in smaller phase retrieval uncertainties in some cases. In general, both investigated microlensing cases have comparable phase retrieval results.

\onecolumn

\begin{table*}[hbtp]
\centering
\begin{tabular}{|c|c|c|c|c|}
\hline 
$S/N$ & 10 & 20 & 30 & $\infty$ \\ 
\hline 
Fe\,\textsc{ii} &22.3 $\pm$ 2.2 days&22.3 $\pm$ 1.8 days&22.3 $\pm$ 1.7 days&22.4 $\pm$1.6 days\\ 
\hline
H$\alpha$ &21.8 $\pm$ 2.2 days&21.9 $\pm$ 1.5 days&21.9 $\pm$ 1.3 days&21.9 $\pm$ 1.2 days\\ 
\hline 
H$\beta$ &21.5 $\pm$ 2.2 days&21.7 $\pm$ 1.7 days&21.7 $\pm$ 1.6 days&21.7 $\pm$ 1.6 days\\ 
\hline
Fe\,\textsc{ii} + H$\alpha$ + H$\beta$&22.0 $\pm$ 1.3 days&22.0 $\pm$ 1.0 days&22.0 $\pm$ 0.9 days&22.0 $\pm$ 0.8 days\\ 
\hline 
\end{tabular}
\caption{\label{phase_uncertainties_2}Phase retrieval under high microlensing magnifications. We list the weighted average and the 1$\sigma$ uncertainties of the
  phase retrievals using the three absorption lines (Fe\,\textsc{ii},
  H$\alpha$, and H$\beta$) for the three
  different S/N values (10, 20, and 30) and for the noiseless case, indicated by
  $S/N = \infty$. The last row shows the uncertainty on the phase when
  using all three lines, assuming the noise and the microlensing in these lines are
  uncorrelated.}
\end{table*}

\end{document}